# Roadmap for Animate Matter

**Giorgio Volpe[1], Nuno A. M. Araújo[2,3], Maria Guix[4], Mark Miodownik[5]**, Nicolas Martin[6], Laura Alvarez[6], Juliane Simmchen[7], Roberto Di Leonardo[8,9], Nicola Pellicciotta[9,8], Quentin Martinet[10], Jérémie Palacci[10], Wai Kit Ng[11], Dhruv Saxena[11], Riccardo Sapienza[11], Sara Nadine[12], João F. Mano[12], Reza Mahdavi[13], Caroline Beck Adiels[13], Joe Forth[14,15], Christian Santangelo[16], Stefano Palagi[17], Ji Min Seok[18], Victoria A. Webster-Wood[19], Shuhong Wang[20,21], Lining Yao[20], Amirreza Aghakhani[22], Thomas Barois[23], Hamid Kellay[23], Corentin Coulais[24], Martin van Hecke[25,26], Christopher J. Pierce[27], Tianyu Wang[27], Baxi Chong[27], Daniel I. Goldman[27], Andreagiovanni Reina[28,29], Vito Trianni[30], Giovanni Volpe[13], Richard Beckett[31], Sean P. Nair[32], Rachel Armstrong[33]

1 – Department of Chemistry, University College London, 20 Gordon Street, London WC1H 0AJ, United Kingdom
2 – Departamento de Física, Faculdade de Ciências, Universidade de Lisboa, 1749-016 Lisboa, Portugal
3 – Centro de Física Teórica e Computacional, Faculdade de Ciências, Universidade de Lisboa, 1749-016 Lisboa, Portugal
4 – Department of Materials Science and Physical Chemistry, Institute of Theoretical and Computational Chemistry, University of Barcelona, 08028, Barcelona, Catalonia, Spain
5 – Department of Mechanical Engineering, University College London, London, United Kingdom
6 – University of Bordeaux-CNRS, Centre de Recherche Paul Pascal, UMR 5031, 115 Avenue du Dr. Albert Schweitzer, 33600 Pessac, France
7 – Pure and Applied Chemistry, University of Strathclyde, United Kingdom
8 – Department of Physics, "Sapienza" University of Rome, Rome 00185, Italy
9 – NANOTEC-CNR, Institute of Nanotechnology, Soft and Living Matter Laboratory, Rome 00185, Italy
10 – Institute of Science and Technology Austria (ISTA), Klosterneuburg, Austria
11 – Blackett Laboratory, Department of Physics, Imperial College London, London, UK
12 – CICECO - Aveiro Institute of Materials, Department of Chemistry, University of Aveiro, Campus Universitário de Santiago, 3810-193 Aveiro, Portugal
13 – Department of Physics, University of Gothenburg, SE-41296 Gothenburg, Sweden
14 – Department of Physics, University of Liverpool, Crown Street, Liverpool L69 7ZD, United Kingdom
15 – Department of Chemistry, University of Liverpool, Oxford Street, Liverpool L69 7ZE, United Kingdom
16 – Physics Department, Syracuse University, Syracuse, NY 13244, USA
17 – The BioRobotics Institute and Department of Excellence in Robotics & AI, Sant'Anna School of Advanced Studies, Pisa, Tuscany, 56025, Italy
18 – Department of Mechanical Engineering, Carnegie Mellon University, USA
19 – Department of Mechanical Engineering, Department of Biomedical Engineering, The Robotics Institute, Carnegie Mellon University, USA
20 – Morphing Matter Lab, Mechanical Engineering, UC Berkeley, USA
21 – Zhejiang University, China
22 – Institute of Biomaterials and Biomolecular Systems, University of Stuttgart, Pfaffenwaldring 57, 70569 Stuttgart, Germany
23 – University of Bordeaux, CNRS, LOMA, UMR 5798, F-33400 Talence, France
24 – University of Amsterdam, Institute of Physics, Science Park 904, 1098 XH, Amsterdam, the Netherlands
25 – Huygens-Kamerlingh Onnes Laboratory, Universiteit Leiden, PO Box 9504, Leiden, 2300 RA, the Netherlands
26 – AMOLF, Science Park 104, 1098 XG, Amsterdam, the Netherlands
27 – School of Physics, Georgia Institute of Technology, Atlanta, GA 30332, USA
28 – Centre for the Advanced Study of Collective Behaviour, Universität Konstanz, Germany
29 – Department of Collective Behaviour, Max Planck Institute of Animal Behavior, Konstanz, Germany
30 – Institute of Cognitive Sciences and Technologies, CNR, Via San Martino della Battaglia 44, Rome, Italy
31 – The Bartlett School of Architecture, University College London, United Kingdom
32 – Eastman Dental Institute, University College London, United Kingdom
33 – Department of Architecture, KU Leuven, Ghent, Campus Sint-Lucas, Belgium.

**Guest editors:**
Giorgio Volpe: g.volpe@ucl.ac.uk
Nuno A. M. Araújo: nmaraujo@fc.ul.pt
Maria Guix: maria.guix@ub.edu
Mark Miodownik: m.miodownik@ucl.ac.uk

**Abstract**

Humanity has long sought inspiration from nature to innovate materials and devices. As science advances, nature-inspired materials are becoming part of our lives. Animate materials, characterized by their activity, adaptability, and autonomy, emulate properties of living systems. While only biological materials fully embody these principles, artificial versions are advancing rapidly, promising transformative impacts across various sectors. This roadmap presents authoritative perspectives on animate materials across different disciplines and scales, highlighting their interdisciplinary nature and potential applications in diverse fields including nanotechnology, robotics and the built environment. It underscores the need for concerted efforts to address shared challenges such as complexity management, scalability, evolvability, interdisciplinary collaboration, and ethical and environmental considerations. The framework defined by classifying materials based on their level of animacy can guide this emerging field encouraging cooperation and responsible development. By unravelling the mysteries of living matter and leveraging its principles, we can design materials and systems that will transform our world in a more sustainable manner.

## Introduction

Humankind has always looked at nature for inspiration to design new devices and materials. As science and technology progress, nature-inspired artificial materials that previously could only exist in the realms of myths and fantasy [1] are now part of everyday life, including self-healing materials [2, 3], energy and water harvesting systems [4, 5], and autonomous devices mimicking cognition [6, 7].

Animate materials are a novel class of potentially transformative artificial materials reproducing key properties of living systems [8]. They are not alive but are defined according to the **three principles of animacy**, being active, adaptive and autonomous [8]. By being **active**, they can use energy available to them in the environment to perform tasks and work, such as motion, growth, communication, etc. Their **adaptive** nature means that they can sense, process and respond to environmental changes and stimuli, often exhibiting an emerging behaviour. Finally, being **autonomous**, they can initiate behavioural changes based on a certain internal degree of information processing (which may or may not include memory) without external monitoring or control by human agents. The combination of these three properties sketches a future where materials are more resilient and sustainable with the capabilities for, e.g., multitasking, self-regulation, self-healing, and even evolution. Currently, examples of these materials are being developed at all human-relevant length scales, from molecular aggregates to urban ecosystems (**Figure 1**): some materials are fully artificial while others hybridize biological materials with synthetic ones, hitchhiking on natural properties to improve animacy of human-made systems.

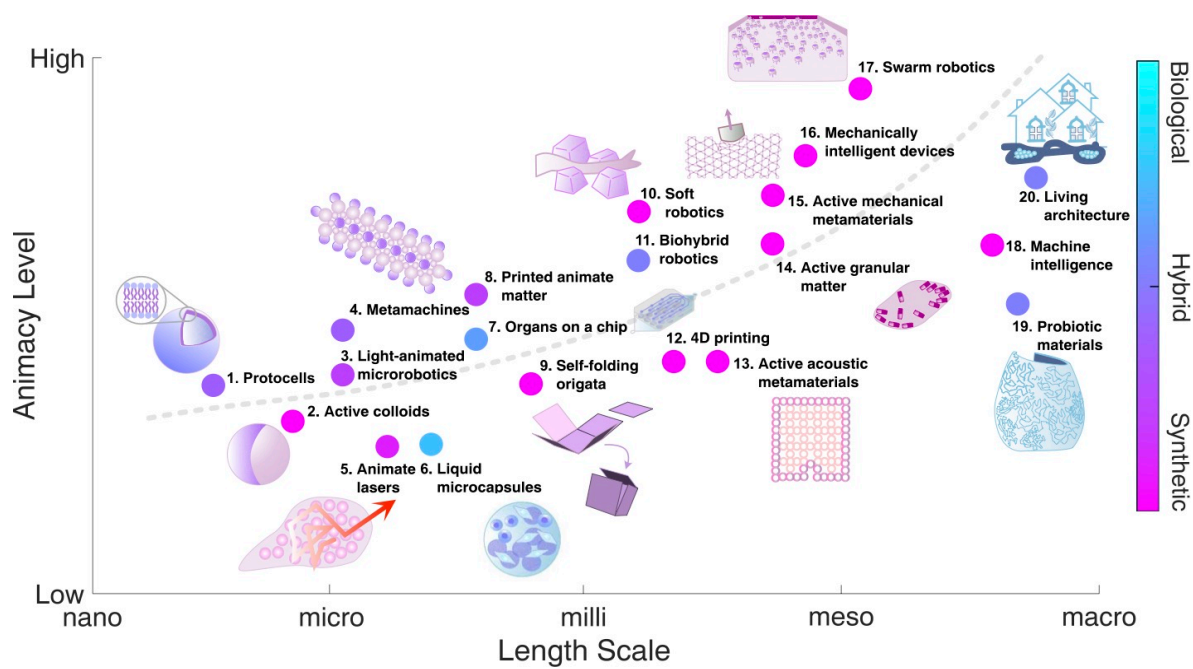

**Figure 1 | Efforts towards realizing animate materials to date**. A plot showing the overall animacy level (vertical axis, from low to high) versus length scale (horizontal axis, from nanoscopic to macroscopic) for the different systems discussed in this roadmap numbered based on how they appear in the table of contents (roughly as a function of scale with the caveat that some systems can span multiple scales, e.g., printed animate matter, self-folding orikata, soft robotics, biohybrid robots, 4D printed systems). Animacy here estimates the cumulatively level of activity, adaptiveness and autonomy based on our joint perception. Only biological materials (not represented here) can be considered fully animated to date. Most artificial systems to date show an average level of overall animacy lower than biological systems, with mesoscopic and robotic systems typically scoring higher due to the possibility of programming them directly. As can be expected, due to the miniaturisation problem, overall animacy levels tend to increase going from the microscopic scale to the larger scales (grey dashed trend

line as a guide for the eyes). The colour code highlights the material origin: most systems are fully artificial (synthetic), while some incorporate biological elements to different extents (hybrid).

To date, indeed, only biological materials score highly under all three principles as these typically led to competitive evolutive advantages, while materials created through human agency typically lag behind their natural counterparts in at least one core principle (Figs. 1-2) [8]. As can be seen in Fig. 1, miniaturising animacy also present challenges due to issues in scaling components and functionalities to smaller scales. Nonetheless, worldwide research effort on the topic of animate materials is ever growing driven by both the fundamental and applied challenges they represent. On the more fundamental aspect, developing and studying animate materials can shed light on the physics of living systems. On the more applied side, these materials hold great promise to revolutionize our technology in diverse sectors, including medicine, robotics, and infrastructure towards a more sustainable future. It is therefore a matter of time until artificial, fully animate materials will leave the realm of curiosity-driven research to start contributing to human technology.

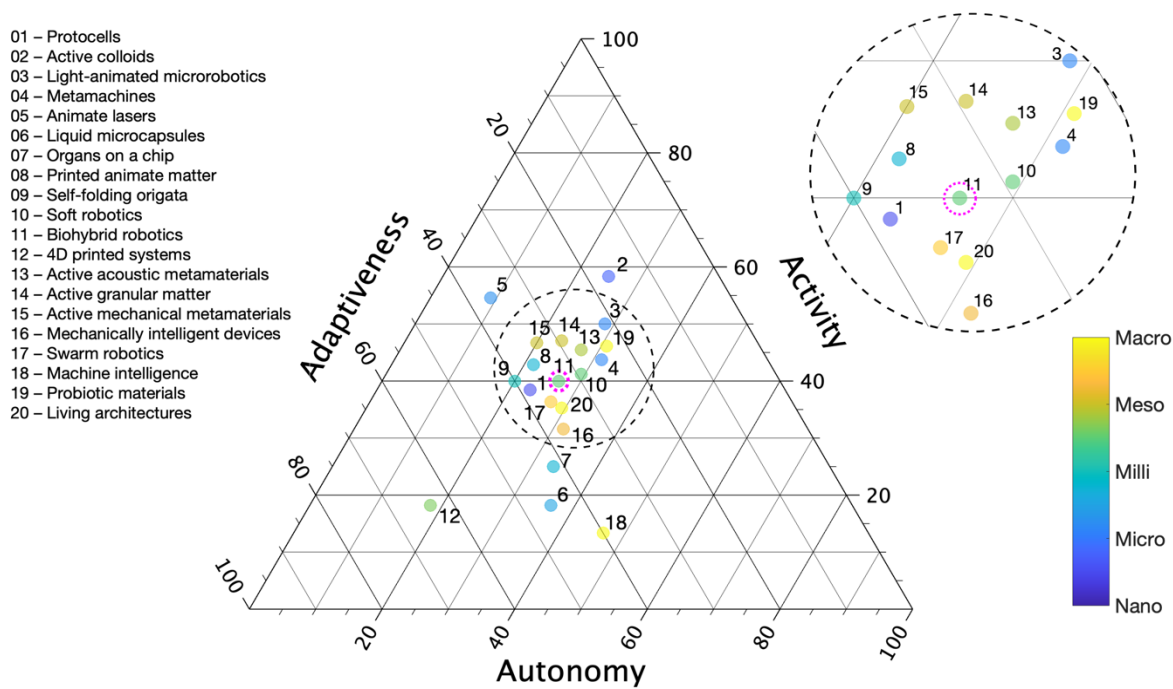

**Figure 2 | Relative contribution of each principle to the overall animacy level**. This ternary diagram sketches the relative importance of each property of animacy (i.e., activity, adaptiveness and autonomy) for each system discussed in this roadmap based on our joint perception. The overall level of animacy for each system (cumulatively given by all three properties) is provided in Fig. 1. Interestingly, most systems appear to cluster roughly around the centre of the diagram. This highlights the fact that efforts from the research community have been equally divided to improve the three properties of animacy, although the overall level of animacy in many systems is still far from that of their biological counterparts (Fig. 1). Despite roughly equal efforts, of the three properties, activity and adaptiveness are those that have seen more development on average compared to autonomy that lags behind the other two properties. The magenta dashed circle represents the current average level of relative development in the three properties of animacy of all systems considered, being more skewed towards higher relative levels of activity and adaptiveness. While activity and adaptiveness have improved thanks to material engineering, autonomy is more difficult to achieve as it requires, e.g., computation to be implemented. This task is notoriously difficult when moving from macroscopic systems to smaller length scales due to current limitations in miniaturization. Interestingly, some systems discussed in this roadmap have mainly seen development in one property of animacy and score relatively lower on others: for example, active colloids (2) and animate lasers (5) score relatively high on activity; 4D printing (1w) stands out for adaptiveness while machine intelligence (18) jointly for adaptiveness and autonomy. The colour code highlights the scale of the various systems.

This Roadmap consists of a series of twenty authoritative perspectives that capture the current state of research in animate materials in different disciplines (Figs. 1-2). We roughly organised them by length scale, with the caveat that some sections (e.g., printed animate matter, self-folding origata, soft robotics, biohybrid robots, 4D printed systems) can span across several scales. By no means intending to be an exhaustive list, we highlight representative areas where the concept of animate materials has found (consciously or not) application by driving advancements in materials science and has helped our understanding of living matter. Our roadmap embarks on a journey through this dynamic landscape, charting pathways across disciplines at different length scales relevant to human experience, from nanotechnology to robotics, from synthetic biology to artificial intelligence, from metamaterials to living architectures. We aim to highlight common challenges and opportunities emerging in the design of animate materials at all scales without being dismissive of the unique challenges and opportunities emerging in each discipline and length scale.

What clearly emerges from these contributions is how animate materials represent a convergence of physics curiosity, chemical and biological principles and engineering expertise, fostering the creation of materials with unprecedented functionalities and potential applications. To achieve this overarching goal, a concerted effort across disciplines and scales is needed. So far, efforts to develop animate materials have often been siloed, focusing mainly on specific systems and scales in isolation, such as protocells, robots, or living architectures. However, there are many analogous questions and technical challenges found across multiple scales, systems and disciplines, which need to be addressed systematically before the full potential of animacy is harnessed to design materials and systems. Shared open challenges include understanding, modelling and controlling complexity emerging from animacy, miniaturizing its components, scaling up and deploying laboratory prototypes to real-life applications, defining a unified language to discuss animate materials, fostering opportunities for interdisciplinarity, promoting cross-fertilization of ideas across systems, and weighting the ethical, environmental and societal implications of developing animacy in artificial materials.

Key for the advancement of the field is the coordination of multiscale research efforts aimed at integrating and combining concepts and approaches from different systems and scales. The emergence of complex functionality in biological systems relies on the existence of intercommunicating hierarchical structures, from molecules to entire ecosystems via the formation of, e.g., macromolecules, cells, tissue and organisms [9]. Reproducing this multiscale approach in synthetic animate materials requires the development of interconnected hierarchical levels that can communicate and, if needed, store information effectively across scales. Succeeding in this task can enable the creation of synthetic animate materials presenting a hierarchy of structures and functions, thus unleashing the full scientific and technological potential of this class of materials and fulfilling their promise for animacy in terms of, e.g., self-regulation, self-healing, and even evolution.

The animacy framework of this roadmap focusing on active, adaptive and autonomous materials is not meant to be prescriptive. We hope it will serve as a flexible framework for this emerging field defining a common language that can guide development across disciplines and length scales. As we embark on this multidisciplinary journey, we invite researchers, engineers, and visionaries alike from diverse backgrounds to join us in shaping the future of animate materials in a socially responsible and sustainable manner. Together, we can unravel the mysteries of living matter, harnessing its principles to create a new generation of materials and systems that are not merely passive objects, but active symbiotic participants in the vast fabric of human life and society.

**Acknowledgements**

N.A.M.A. and G.V. (Giorgio) acknowledge support from the UCL MAPS Faculty Visiting Fellowship programme. N.A.M.A. acknowledges financial support from the Portuguese Foundation for Science and Technology (FCT) under Contracts no. UIDB/00618/2020 (https://doi.org/10.54499/UIDB/00618/2020) and UIDP/00618/2020 (https://doi.org/10.54499/UIDP/00618/2020).

## Contents

# 01 – Protocells

Nicolas Martin, Laura Alvarez

University of Bordeaux-CNRS, Centre de Recherche Paul Pascal, France

### Status

The mystery of the origins of life involves understanding and harnessing the structural and functional requirements to transition from non-living to living matter. Protocells are primitive compartments (assembled from prebiotic components) considered precursors to living cells [1]. Synthetic protocells provide a unique and versatile platform to investigate fundamental aspects (e.g., interactions between building blocks, environmental requirements) and design simple life-like functionalities.

Protocells are spatially confined nano- or micro-compartments emerging from the self-assembly of different molecules. Compartmentalization establishes chemical environments separated from their surroundings (**Figure 1.1A**). The simplest protocell models are membrane-less droplets (e.g., coacervates and emulsions), forming distinct microenvironments through liquid-liquid phase separations, akin to intracellular organelles known as biomolecular condensates. Coacervates formed by attractive interactions within or between polymers, surfactants or biomolecules (e.g., nucleotides, peptides) [2] can encode chemical functionalities specific to their composition and serve as dynamic reactors when formed and dissolved under external stimuli (e.g., temperature, pH, light). The minimalistic synthetic analogues of cell membranes are membrane-bound protocells [3], believed to have evolved from coacervates [4]. These vesicle-like structures, typically formed by bilayers of amphiphilic lipids or polymers, allow for compartmentalization and encapsulation of biomolecules with selective permeability for environmental exchange.

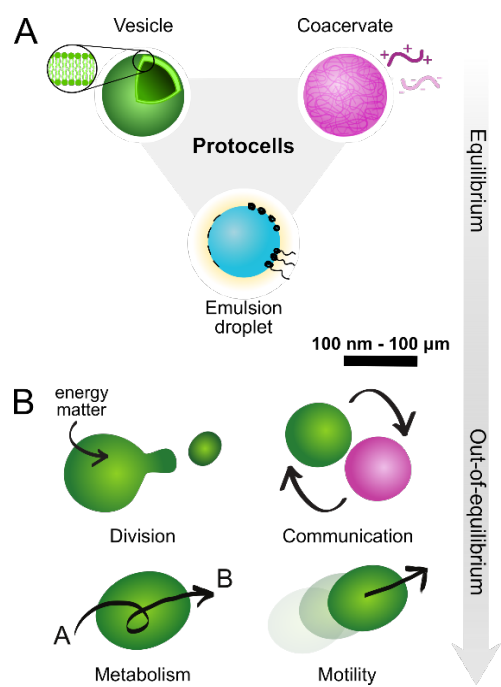

**Figure 1.1 | Incorporating out-of-equilibrium properties in protocell design**. (**A**) Schematic representation of the main models for protocellular compartments, including coacervates, emulsion droplets and vesicles. (**B**) Out-of-equilibrium processes characteristic of life-like protocells, including self-replication, communication, metabolic activities and motility.

The primary focus of protocell research is understanding and predicting the assembly and stability of the required minimal building blocks. However, recent efforts have expanded to out-of-equilibrium processes (**Figure 1.1B**), driven by intake of environmental energy and matter – a crucial step in transitioning to autonomous, self-sustainable units [5]. While reaction cycles, growth, self-replication and chemical communication are achievable through materials design approaches, physical transformations (e.g., self-propulsion or morphogenesis) still mostly rely on external actuation. The development of self-propelled protocells (e.g., active coacervates [6] or lipid vesicles [7,8]) mirrors cellular dynamics and their interactions with environmental cues, potentially offering insights into survival strategies of early pre-cellular units.

Advances in hybrid protocells, combining lipid vesicles with proteinosomes or coacervates, mimic the complex hierarchy of high-evolved living cells. This structural organization enhances spatially localized biochemical processes but also allows the combination of incompatible reactions within the same vessel together with their temporal orchestration. Sophisticated protocells have recently shown morphogenetic capabilities [9], especially when integrated with energy-sustaining bacterial components.

**Current and Future Challenges**

Creating life-like protocells using simple architectures poses a significant challenge, requiring solutions to key issues in their adaptive, active, and autonomous nature:

- **Stability.** Despite their structural simplicity, environmental perturbations (e.g., pH, light, temperature) might affect protocells' long-term stability, limiting their use in experiments and applications (e.g., oxidation of phospholipids, coalescence of coacervates). Yet, environmental changes are crucial to trigger certain prebiotic processes. Understanding the relationship between stability and spatiotemporal evolution is vital, especially for implementing functionalities in non-equilibrium conditions.
- **Autonomous vs actuated.** The formula for self-regulation remains a complex puzzle driven by thermal forces and external stimuli. Unlike living cells that move, divide and continuously replace components to maintain integrity, protocells might reach a static state after self-assembly without external actuation. The challenge lies in designing self-regulating 'factories' via either continuous matter intake or internal machineries that produce their components. The design of non-equilibrium processes based on dissipative (or transient) self-assembly and -disassembly remains an open challenge.
- **Chemical programming.** Integrating multiple (bio)chemical processes to achieve autonomous behaviours and developing a transcription-like machinery pose major challenges. This requires chemical compatibility amongst multiple components and managing cross-reactivity. Compartmentalization is essential to regulate these parameters, by allowing selective permeability and solubility. Yet, spatial localization can affect the reactivity of chemical species (e.g., due to crowding or interactions with soft interfaces). Moreover, linking molecular content with structural characteristics might mimic genotype-phenotype correlations, but specificity and variety of building blocks and reactions are currently limited.
- **Motility.** Motion is ubiquitous to all living systems as a survival tool. Self-directed motion is related to crucial intercellular migration or protocell interactions. These autonomous and spatial-dependent dynamics involve intricate exchanges of matter and energy (see also Section 2). Designing efficient motility mechanisms and understanding how softness and structural adaptation to dynamical stress affect protocells remains untapped.
- **Protocell interactions**. Understanding and replicating the communication and physical interactions between protocells, such as chemical signalling or fusion, involve high levels of physicochemical understanding of all elements involved. This includes studying competition for resources and predatory/cooperative behaviours among protocells. In this context, interactions between membrane-bound and membrane-less droplets are gaining increasing attention.

- **Reproducibility.** While nanoscale compartment production is rather optimised, cell-sized protocell fabrication methods are not achieving uniform composition, size, and geometry. This precision is essential for robust statistical studies across large samples. On one hand, microfluidic techniques – despite their popularity – suffer from their sensitivity to chip fabrication. On the other hand, template-assisted methods are not fully reliable to control size or encapsulation of colloidal or molecular units. Consequently, the variability in experimental approaches complicates reproducibility and hinders scaling protocells reliably for advanced applications.

**Advances in Science and Technology to Meet Challenges**

The rational design of novel building blocks is crucial for integrating advanced chemical networks into protocells to achieve life-like processes (e.g., chain reactions in membrane-bound compartments, molecular self-replication, motion, dissipative self-assembly). Key is the development of new molecules and building blocks that can initiate and maintain out-of-equilibrium processes. This involves accepting that prebiotic compositions are not fully replicable and the need to explore alternative systems. The use of enzymes presents a powerful chemical tool in that direction. Enhancing the performance and compatibility of enzymes through directed evolution could expand the range of functional chemical modules that can be integrated within protocells. Additionally, organic chemistry is crucial in developing unique and novel molecules that perform similar functions to prebiotic ingredients. Assembly Theory [10] and its predecessors quantify the degree of selection in an ensemble of evolved units and identify molecular biosignatures that could lead to life-like compartments. Integrating such models into experimental design might help identify and create new biochemical building blocks for engineering self-regulated protocells.

Tool development is also critical towards achieving robust experimental setups. Microfluidics is a well-established, high-throughput production method and environmental control platform (e.g., through light or local pressure). Intrinsically, this technique is capable of extensive parameter screening and optimization, which can be time-consuming and beyond current manpower capabilities. Developing microfluidics-automated reactions mimicking 'digitalization chemistry' approaches (i.e., based on chemical programming languages and automatized robotic platforms) might help screen an otherwise impossible condition range. Artificial intelligence is a valuable emerging tool to analyse vast data amounts and predict possible outcomes from experimental information on protocell design.

Due to the sensitivity of protocells to environmental changes and protocol variations, reproducibility and standardization protocols are needed. Online platforms for sharing data and protocols could facilitate the unification of results, leading to parameter standardization otherwise difficult to control in individual laboratories (e.g., humidity and solvent quality).

Extending experimental efforts to alternative environments (e.g., high-pressure settings or microgravity in sounding rockets or parabolic flights) can offer novel approaches to understanding the impact of gravity and pressure on the formation and evolution of protocells.

Finally, studying the diverse and complex nature of protocells requires a multidisciplinary approach, integrating knowledge from soft matter, chemical engineering, biology, and computational science.

This need could be met by establishing graduate programs targeting this topic and promoting interdisciplinary training.

**Concluding Remarks**

The study of protocells is pivotal in understanding the transition from non-living to living matter, marking a significant advance in bioinspired materials, especially by integrating in- and out-of-equilibrium processes. Protocells, considered as precursors to complex cells, offer insights into the fundamental interactions of molecular building blocks within controlled environments. This research is key to understanding the assembly, stability, and evolution of these minimal units towards autonomous systems. Challenges in protocell research include achieving stability under environmental changes, designing self-regulating systems, and integrating multiple chemical functionalities. These efforts aim to replicate life-like behaviours, such as motility, communication, growth, and replication, mirroring the dynamics of living cells and offering insights into early life forms. The development of sophisticated protocells, capable of morphogenesis and integrated with energy components, represents a step towards complex architectures and life-like behaviours. This field extends beyond understanding the origins of life, by also contributing to the development of microscale materials with unique, self-sustaining features across synthetic biology, soft matter, systems chemistry, and biophysics. While significant advances have been made, challenges in sustainability, self-regulation, and evolutionary capabilities remain, making protocell research a rich area for future scientific exploration, with potential applications in energy harvesting, environmental remediation, catalysis and biomedicine.

**Acknowledgements**

L. Alvarez and N. Martin acknowledge funding from IdEx Bordeaux and the French Agence Nationale de la Recherche (ANR-23-CE06-0007-01 and ANR-21-CE06-0022-01), and support by the Univ. Bordeaux Research Network Frontiers of Life. N. Martin also thanks Région Nouvelle-Aquitaine (AAPR 2020-2019-8330510) for funding. We want to thank Prof. J.C. Baret and Prof. S. Mann for stimulating, fruitful, and enjoyable collaborations and V. Willems for her contributions to some of the work cited here. L. Alvarez thanks her peers at the ESA FTD Soft Matter and Biophysics group for fruitful discussions.

## 02 – Active colloids

Juliane Simmchen

University of Strathclyde

### Status

Biological microswimmers have evolved over millions of years to perfectly adapt their swimming mechanisms to the low Reynolds number conditions dictated by their small size. Physicists have been fascinated by various aspects of them over the last century, leading to seminal work such as the postulation of the Taylor sheet [1], Purcell's interpretation of "low Reynolds number life" [2] and Anderson's work on phoresis [3]. In the early 2000s, researchers were intrigued by the idea of artificially recreating the ability of these biological micro-entities to move at the microscale.

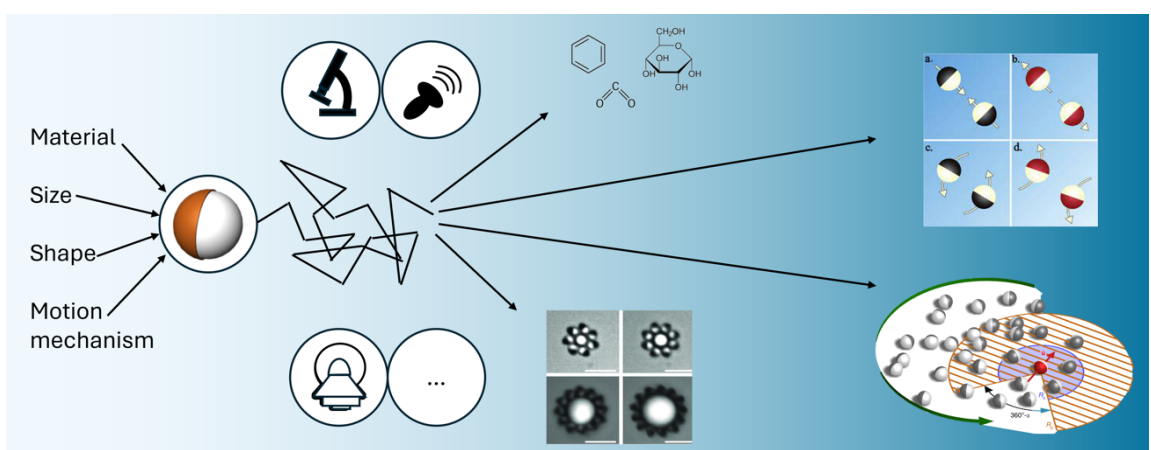

**Figure 2.1 | Influencing and detecting emerging behaviours in active colloids**. (Left) The swimming behaviour of active colloids is influenced by several factors such as material, shape, size and motion mechanism. (Middle) Detection is largely dependent on the environment, with optical detection being the simplest although it relies on transparent environments; ultrasonic or magnetic properties extend the range of imageable environments, and new alternatives are constantly being developed. (Right) Beyond controlling the motion properties (from diffusion to ballistic), the various fuel sources (first arrow from the top), the interactions between individuals (second and third arrow from the top) and larger groups (third and fourth arrow from the top) can lead to promising emerging behaviours. Insets (clockwise from the top): fuel sources; pair interactions reproduced from [11], use permitted under the CC-BY 4.0 licence; colloidal swarm reproduced from [12], use permitted under the CC-BY 4.0 licence; cogwheels reproduced from [13], use permitted under the CC-BY 4.0 licence.

Along with pioneering experiments with bimetallic rods [4], a theoretical paper postulated the active motion of a sphere with a catalytic patch [5]. Subsequently, the first self-propelling colloidal spheres were proposed using an inert particle half coated with platinum that moved autonomously when placed in a hydrogen peroxide solution [6]. This spherical, reliably moving model system (known as a Janus particle) allowed closer interaction between theory and experiment, leading to the exploration of different behaviours and the generation of a number of different modelling approaches able to capture different aspects of experimental observations. In addition to many proof-of-concept applications in fields as diverse as sensing, biomedicine and environmental remediation, the active matter community began to explore different material and fuel combinations for artificial microswimmers, providing a broader insight into a range of swimming mechanisms and their limitations. These investigations, progressing in parallel with theoretical advances, have not only demonstrated the versatility of these experimental systems but have also deepened our understanding of self-propulsion at the microscale.

In more recent years, active colloids have been challenged by a variety of stimuli and have been integrated with digital technologies. For example, an artificial computer vision model for active colloids has been proposed to observe the consequences of different visual cones on collective interactions [7]. Looking towards applications beyond colloids and the obvious drug delivery, biomedical uses depend on imaging independent of optical visibility. If visibility is impeded by particle size, dynamics light scattering (DLS) and other scattering techniques have become essential. For larger microrobots alternative imaging strategies using ultrasound, acoustic techniques or magnetic resonance can be used to extract information about positions, behaviours and speed of the active particles [8]. However, real-world industrial applications of active colloids, including microscale sensors, advanced control systems, and artificial intelligence, are yet to be developed.

## Current and Future Challenges

One of the major challenges that will help move the field of active colloids from isolated knowledge to full understanding is to perform proper **comparisons between different active colloidal systems**. Changing propulsion mechanism, shape, and material affects many individual and collective behaviours (**Figure 2.1**, see also Sections 3 and 4) but experimental studies that evaluate these effects are sparse because unfortunately, this type of study is not highly valued in the scientific community due to a lack of innovation. As a result, such studies are much more difficult to fund and publish, and the field is still far from reaching out to industrial partners who would contribute to a broader body of knowledge. Approaches such as interlaboratory comparisons, which would lead to greater standardisation and cross-comparability of results, have been discussed but not yet implemented. Similarly, at the level of understanding certain phenomena that have often been observed for one type of microswimmer, studies across the spectrum comparing shapes and swimming mechanisms are largely lacking.

A major challenge, which has gained momentum as the community has grown and expanded beyond physics, is **the integration of 'new chemistry'** and a wider range of materials to increase the flexibility and options available to experimental systems.

Furthermore, although the first steps towards digital control have been taken, the integration of **digital features and computational capabilities** is still in its infancy. Due to the small size, mostly non-planar morphology and frequent imperfections of colloidal particles compared to traditional *in silico* components, the integration of traditional transistor-based computing facilities on colloids has been very limited. While alternative computational modes have begun to take shape for biologically active materials, their artificial analogues have only been theoretically postulated.

A future challenge that seems to be very closely related to applications is to look at **larger collectives** and model behaviours across scales (see also Sections 3 and 4). While a single micromotor might provide enough signal for sensing, most cargo delivery or remediation applications will require more than an individual microswimmer to perform a given task. Therein, not only the emergence of complexity is posing a challenge, but also the sheer volume of data to handle from large numbers of individuals, their trajectories, spatial distribution, etc.

## Advances in Science and Technology to Meet Challenges

To elevate research on active colloids from a curiosity-driven domain to a technology contributing to our economies and unlocking potential to solve global problems using microrobotics, a few advances are still required. While advances in microscopy have and continue to make tremendous impact for visualisation of small features, improvement in processes and techniques such as advanced particle image velocimetry to detect flows and innovative markers and signalling can continue to enable new levels of understanding, providing more detailed and dynamic insights into the behaviour of microswimmers (biological and artificial ones).

Better and more extended integration of technology (including moving stages and smart detection mechanisms, online-coupling, feedback and remote control) can improve the digital control and processing of active matter at the interface with computation and allow for real-time manipulation, data collection and evaluation. The combination of this fascinating area of research and digital requirements will contribute to increase the digital literacy of traditionally less exposed STEM areas such as biology and chemistry.

Improved synthetic and fabrication capabilities in nanotechnology will allow for advanced design of materials with highly specific functionalities. Additive manufacturing techniques, including stop flow lithography and a broad range of 3D printing techniques, can contribute to broadening the available materials and enhance their functionalities and capabilities [9,10].

A significant challenge that is less technological and more cultural is the collaboration between different disciplines. While research on active colloids and microswimmers has risen as a trans-disciplinary field of research, enabling and facilitating further knowledge transfer among disciplines to avoid 're-inventing the proverbial wheel' is of crucial importance and will ensure a more holistic approach to solving cross-disciplinary research problems.

**Concluding Remarks**

In conclusion, the fascination of creating artificial or robotic capabilities on the microscale, on par with nature's toolbox on this scale, is currently inspiring many scientists from different disciplines around the world. To fully realize the potential of active matter technology, a few steps towards routine studies, standardisation and enhanced reproducibility are necessary to create not just amazement but trust and acceptance among the different industries. Closer interactions with advanced techniques for AI (Section 18), computation, enhanced manipulation, control strategies and image analysis will require effective knowledge transfer mechanisms but have the potential to unlock precision microrobots acting as a collective to perform the most intricate tasks on the microscale.

**Acknowledgements**

JS acknowledges a fellowship from the German Fulbright Cottrell foundation.

### 03 – Light-animated microrobotics

Roberto Di Leonardo, Nicola Pellicciotta
Dipartimento di Fisica, Sapienza Università di Roma
NANOTEC-CNR, Soft and Living Matter Laboratory, Institute of Nanotechnology

**Status**

Robots are man-made machines that can be programmed to perform tasks autonomously. Usually, a robot needs two main components: a decision-making system informed by external data and actuators to execute planned actions. On the microscale, microrobots can bring automation to laboratories built on microscope slides (lab-on-chip) or operate within the human body [1]. Due to their small sizes, synthetic microrobots cannot yet incorporate complex decision-making systems and rely on external computers for sensing their environment and computing responses. As for actuation, fuel and controls should be provided wirelessly to enable remote control of the microrobots while operating in physically inaccessible environments. A solution is direct manipulation with non-contact forces (e.g., magnetic, electrical, optical or acoustic). Although these systems are usually referred to as microrobots, micromanipulation is probably a more appropriate definition when both control and force generation are external. Conversely, self-propelled microrobots integrate remotely controllable active components that can generate forces using locally available energy (see also Section 2). If microrobots are to execute complex and precise tasks, such as lab-on-a-chip sorting and cargo delivery, their movements must be directed accurately and independently. Light is an ideal carrier of both energy and control instructions delivered with high resolution in space and time [2]. Here we focus on a class of microrobots that can be remotely controlled by light using a variety of physical and biological mechanisms to convert photon energy into motion (**Figure 3.1**):

- **Optical and optoelectronic micromanipulation.** Optical actuation can be direct, i.e. through radiation pressure exerted by focused laser beams (optical tweezers). Multiple holographic traps can be used to grab, move, and rotate complex 3D-printed polymeric structures, e.g., for indirect live-cell manipulation [3]. Alternatively, by using photoconductive substrates as light-reconfigurable electrodes, strong electric field gradients with controlled spatiotemporal profiles can manipulate microscopic objects by dielectrophoretic forces [4].
- **Light-deformable materials.** Light-driven microrobots can be also built from softer polymeric materials with strong mechanical response to light. These are typically hydrogels or liquid crystal polymers that swell or deform upon light-induced local temperature changes. These microrobots can perform a variety of mechanical actions, like gripping and swimming, through shape deformations dictated by a dynamically structured illumination [5].
- **Light-controlled biological motors**. Motile cells can be used as biological propellers within microfabricated synthetic chassis. The use of light-driven proton pumps enables the selective powering of spatially separated bacteria. Modulation of bacterial thrust on different parts of a micromachine enables the execution of complex manoeuvres (see also Section 4) [6].

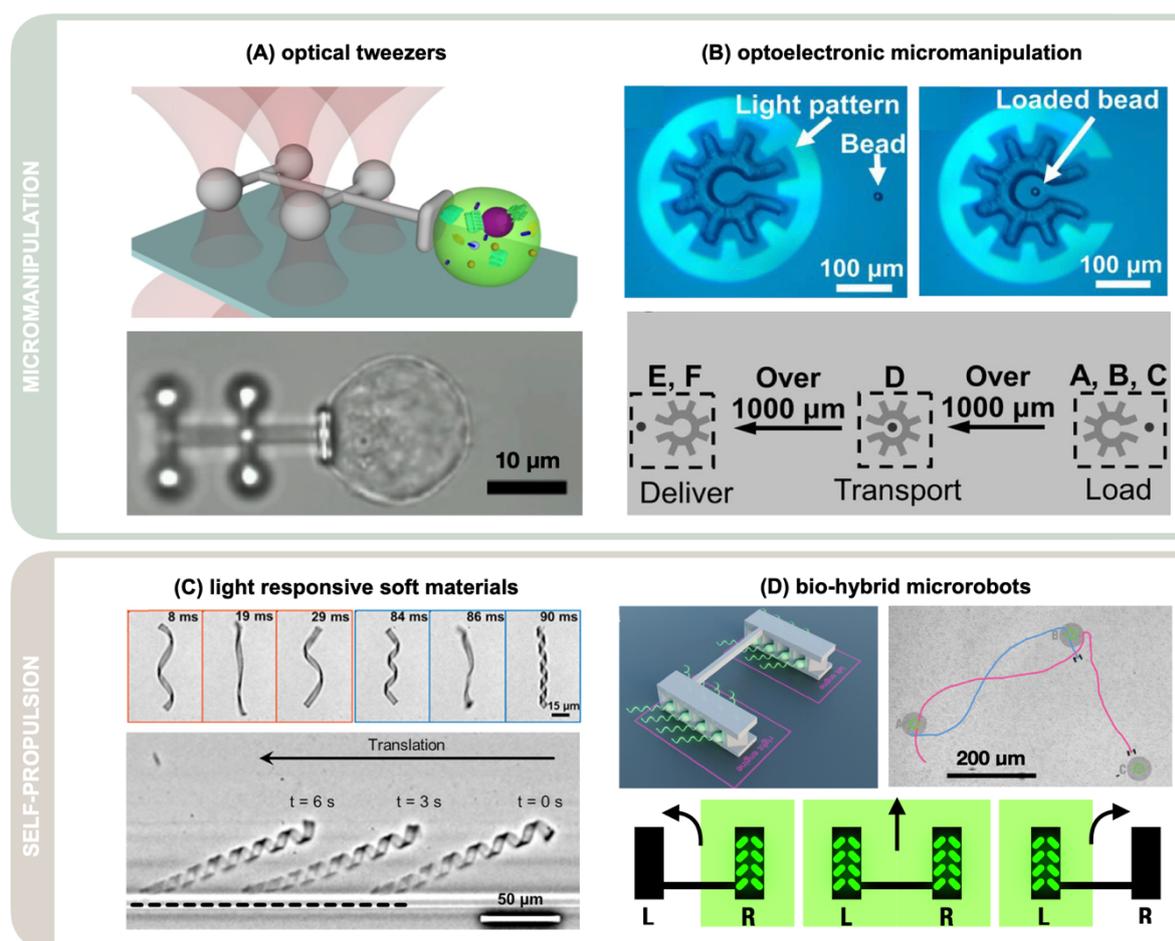

**Figure 3.1 | Examples of light-actuated microrobots**. (**A**) A 3D printed microrobot operated by holographic optical tweezers enables indirect manipulation of live cells. Adapted with permission from [3], Copyright © 2016, Optical Society of America. (**B**) A cogwheel shaped microrobot operated by optoelectronic tweezers delivers a colloidal cargo. Reproduced with permission from [4], Copyright © 2019, published by National Academy of Sciences. (**C**) A soft hydrogel micro-robot undergoes large conformational changes triggered by laser light. Under periodic stimulation the helix can translate over a confining wall with a speed of 20 μm s⁻¹. Reproduced from [5], use permitted under the CC-BY 4.0 licence. (**D**) Biohybrid microbot propelled by light-guided bacteria. A self-assembled microrobot consisting of a synthetic chassis and *E. coli* bacteria as propellers. A feedback loop running on an external computer navigates two microrobots through distributed checkpoints. Reproduced from [6], use permitted under the CC-BY 4.0 licence. Copyright © 2023, The Authors, published by Wiley-VCH GmbH.

## Current and Future Challenges

Translating light-guided microrobots from the research laboratory to real-world applications poses several challenges:

- **Biocompatibility.** Probably one of the most stringent demands is to use technologies that are versatile and compatible with cellular environments. In this context utilizing electrical currents in dielectrophoretic actuation or high-power laser light may cause significant perturbations to cell physiology. On the other hand, using bacteria as propellers poses potential contamination risks.

- **Friction.** Cell media are often salty, so that electrostatic repulsion is screened and strong nonspecific adhesions (e.g., van der Waals) give rise to irreversible sticking. In addition, an often poorly controlled macromolecular composition of the environment causes tethering and increased friction. Even under ideal chemical conditions, sliding on a flat substrate involves a significant amount of shear that can result in high viscous drag and low velocities.

- **Noise resilient navigation strategies.** Brownian noise, combined with the inherent randomness of any microscopic propulsion mechanism, results in stochastic dynamics and the need for constant feedback to counteract deviations from planned trajectories.

- **Cooperation.** Implementing interactions between multiple microrobots is also important to amplify functional throughput or even to create higher-order functionality that exceeds the capabilities of a single microrobot [7] (see also Section 17). In biological communities, such as microorganisms or flocks of birds, physical interactions and chemical signalling between individual entities are critical to facilitate complex tasks such as navigation, predation and survival. Swarms of microrobots will need interactions to avoid collisions during parallel control and to be programmed for seamless collaboration on a common task.

- **3D navigation.** Most of the light-driven microrobotics developed so far is limited to 2D motions over a substrate while the possibility of navigating 3D environments could really help move the first steps out of the microfluidic chip and into real-world microenvironments. 3D navigation would also avoid sliding on the substrate, reducing friction and sticking risk.

- **Embedded control.** The main limitation of light driven microrobots is that they require a clear optical access to be operated. Their 'sensing' capabilities are actually based on global imaging and processing by a central control computer. It will be a major challenge to equip these microrobots with integrated sensors that can detect physical or chemical signals from the environment and calculate a mechanical response.

- **Scalability.** Many microrobots with 3D structures are fabricated with two-photon polymerization techniques, where the fabrication time increases linearly with the number of microrobots and that are therefore not particularly suitable for mass production.

**Advances in Science and Technology to Meet Challenges**
Meeting these challenges requires scientific and technological advances:

- **Improve light-to-force conversion efficiency.** Improving actuation efficiency is crucial for successfully implementing collaborative schemes that involve simultaneously controlling large numbers of microrobots with relatively low optical power. This is essential to avoid compromising the viability of biological samples and the accuracy of manipulation. Biological motors show superior efficiency when compared to synthetic systems. With only a few nanowatts of optical power, bacteria expressing the light-driven proton pump proteorhodopsin can generate piconewton forces through flagellar propulsion. Synthetic systems, however, can generate stronger forces that are best suited to overcome friction, sticking and noise.

- **Understand and exploit friction.** A clear analysis of friction and sticking problems of microfabricated structures sliding on solid substrates is still lacking. A comprehensive study that incorporates both molecular and hydrodynamic interaction aspects could provide the necessary elements to select the best materials and geometries to build microrobots that maintain a high mobility in any biologically relevant media. Alternatively, friction could be exploited for new locomotion strategies such as walking [5].

- **Implement cooperativity and autonomy.** Optical control allows the implementation of effective interactions through feedback control loops. In light-activated colloids, effective interactions can be attractive or repulsive and even mimic animals' visual perception [8]. This approach could prove valuable in shaping collective behaviours within swarms of self-propelled microrobots. However, precise knowledge of the location of the microrobots and their surroundings by the central computer may not always be guaranteed, as in the case of *in vivo* applications. Advancing towards the full autonomy of microrobots requires integrated sensors. One potential solution could come from an interdisciplinary collaborative effort involving synthetic biology. Within biohybrid microrobots, engineered cells can serve not only as actuators but also as sensors and processors of external signals [9].

- **Machine learning for navigation.** Machine learning developments can help devise more robust navigation strategies for complex biological environments characterized by spatially and temporally inhomogeneous conditions such as obstacles and flows (see also Section 18). Approaches based on reinforcement learning seem to be particularly promising, although most studies are only theoretical or conducted on simplified experimental settings [10].
- **Advanced 2D microfabrication.** Scalable solutions could come from alternative ways of using 2D lithography, which can parallelize microfabrication much more easily than two-photon polymerization. Grayscale lithography offers a faster solution for the creation of embossed (2.5D) microstructures. Alternatively, soft 2D microstructures can be designed to fold into a 3D shape in response to specific stimuli [5].

**Concluding Remarks**

Capable of generating stronger forces and with a higher level of control, optical and optoelectronic micromanipulation techniques are still more effective and reliable than semi-autonomous microrobots. However, the realization of autonomous units, which are minimally dependent on external control and power supply, offers more opportunities for innovation and applications that can go beyond the use in microfluidic chips. The path to the realization of autonomous light-driven micro-robots presents exciting but daunting challenges, and the key to overcoming them lies in fostering an interdisciplinary collaboration that brings together experts in microfabrication, surface chemistry, synthetic biology, and control theory for noisy systems.

**Acknowledgements**

R.D.L. acknowledges funding from the European Research Council under the ERC Grant Agreement No. 834615.

## 04 – Metamachines

Quentin Martinet, Jérémie Palacci

ISTA

### Status

Machines are assemblies of parts that transmit forces, motion and energy to one another in a predetermined manner. Their development thrust the industrial revolution, unleashing technical progress with important societal and environmental impact. By converting energy sources into work, machines expand human abilities beyond physiological limits, such as lifting enormous weighs or covering large distances at rapid speed. Specialized machines (or mechanisms) are combined and directed by a computational unit to form a machine of variable complexity, whether a blender or a self-driving car. This approach shows great impact in the engineering of macroscale machines but hits conceptual bottlenecks when scaling down to the microscale. There, parts experience surrounding noise that can render robust placement challenging or the reading of signal needed for sensing elusive. In addition, micromachines carry limited resources that make conventional actuation and *in silico* computation beyond reach.

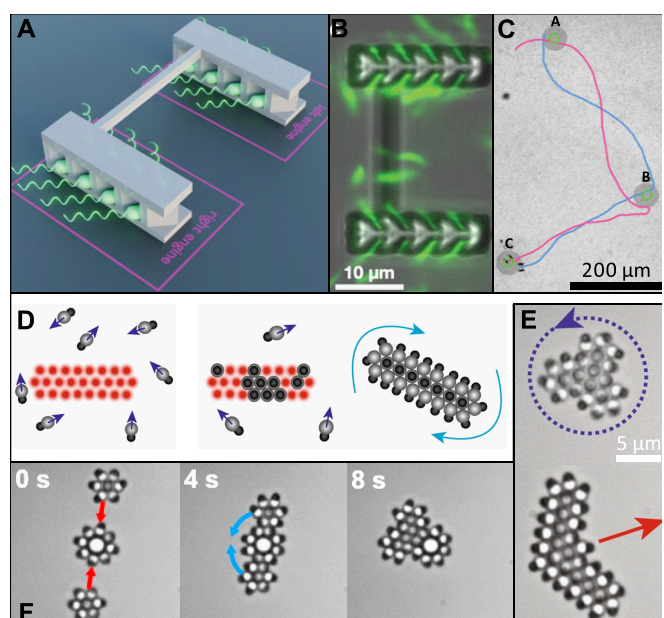

**Figure 4.1 | Examples of hybrid and fully synthetic metamachines. (A-C)** Hybrid metamachines with light-powered bacteria. (**A**) Schematic representation of the hybrid metamachine powered by light-activated bacteria. Adapted from N. Pellicciotta *et al.* [3], use permitted under the CC-BY 4.0 licence. (**B**) Brightfield image of the microfabricated chassis superposed with green fluorescence light-powered bacteria. Adapted from N. Pellicciotta *et al.* [3], use permitted under the CC-BY 4.0 licence. (**C**) Control of the trajectories of the hybrid metamachines along a predetermined pathway, using computer vision and externally controlled light patterns. Adapted from N. Pellicciotta *et al.* [3], use permitted under the CC-BY 4.0 licence. (**D-E**) Optically templated colloidal metamachines. (**D**) A pattern of optical traps constitutes a programmable 2D template to assemble a metamachine. Colloidal microswimmers position on the traps forming a metamachine that remains stable after removal of the optical traps. The metamachine is an autonomous machine with prescribed dynamics. Adapted from A. Aubret *et al.* [2], use permitted under the CC-BY 4.0 licence. Copyright © 2021, The Authors. (**E**) Top: Chiral metamachine rotating counterclockwise. Bottom: Axisymmetric metamachine displaying translational motion. Adapted from A. Aubret *et al.* [2], use permitted under the CC-BY 4.0 licence. Copyright © 2021, The Authors. (**F**) Parts of a machine self-position and lock to form a metamachine, showing elementary features of configurability. Adapted from Q. Martinet *et al.* [6], use permitted under the CC-BY 4.0 licence. Copyright © 2022, The Authors.

The inspiration from nature, which exhibits prototypical examples of functional and autonomous machines at the microscale, funds the concepts of metamachines, or machines made of machines [1]. *Metamachines are modular and functional machines made of machines with the ability to compute,* *e.g. sense and respond to their environment autonomously.* As a result, metamachines have constitutive elements that are internally driven. They can be active colloids (Section 2) for purely synthetic metamachines [2] or responsive bacteria for hybrid metamachines (see also Section 3) [3] (**Figure 4.1**). The field of metamachines therefore appears as a complement to the field of conventional microrobotics. There, major advances were achieved, using monolithic structures that were powered externally. Microscopic walking robots [4] or reconfigurable micromachines [5] show important progress towards Feynman's original vision of "swallowing the surgeon". They however remain large (50-100 µm) and present limited modularity or autonomy. At the microscale, machines that respond to their environment require a design that departs from conventional robots, leveraging physical phenomena relevant at this scale, e.g. interfacial phenomena, to power mechanisms and sense. Practically, metamachines require microscale building blocks that consume fuel. They demand novel engineering approaches to assemble them in structures with meaningful functions and embedded computation capability. Ultimately, metamachines will rival the versatility of biological machines. They will present the emergent dynamics seen in non-equilibrium systems and accomplish various tasks and evolve in complex environments, especially when presenting collective motion.

**Current and Future Challenges**

Important progress remains to be achieved and the field is in its infancy. To date, experimental realizations of metamachines are scarce, with shortcomings to the ultimate goal of an autonomous and functional machine. We will briefly present here pioneering results towards this goal, highlighting their advantages and the drawbacks to motivate future challenges.

Light-powered bacteria enable the design of a programmable micro-vehicle, successfully integrating micrometric actuators made by light-powered *E. coli* bacterial cells with a microfabricated chassis (see also Section 3) [3]. A computer vision software coupled to an external light source to control the power of each actuator (i.e., the light-powered bacteria) allows the micro-vehicle to follow predetermined pathways. This hybrid metamachine is small (sub-10 µm) and makes a nifty use of light-powered bacteria as tuneable actuators at the microscale, typically challenging at small scales. This approach relies on the self-positioning of the bacteria in the micro-chambers constituting the chassis. This step of self-assembly is central but typically challenging with conventional colloidal building blocks, hindering progress. Advances will require tailoring interactions between building blocks to form specific architectures, potentially leveraging recent developments in patchy particles or DNA-origami [7].

In an alternative approach, the propulsion forces of self-propelled colloids are delicately balanced by the optical forces of an optical trap to position and trap active colloids onto a controllable 2D template. Such templated structures remain stable after the removal of the optical traps thanks to non-equilibrium interactions providing cohesion to the overall structure. Those non-equilibrium interactions lead to propulsive forces responsible for the dynamics of the metamachines into prescribed dynamics. They can form small colloidal machines, typically below 10 µm, with embedded dynamics: self-spinning microgears, whose chirality can be set by optical vortices, or translational rods. Those metamachines show elementary features of configurability, e.g. self-positioning parts into larger and dynamical structures [6] and respond to light gradients. They, however, lack internal dynamics and emergent features, e.g. oscillations or self-replication, that are landmarks of non-equilibrium systems. Similarly, they lack the ability to reconfigure and adapt their shape or function to a varying environment.

These advances would constitute a significant step towards the realization of metamachines that compete with living organisms. Next steps towards this overarching goal will be to embed interactions between metamachines in order to build larger architectures, relying on the concerted input of functional machines to achieve emergent behaviour. For example, we could think hypothetically of self-oscillating metamachines that would self-replicate to form a synchronized collection of cilia able to transport fluid at large scales!

**Advances in Science and Technology to Meet Challenges**

Meeting these goals requires stepping up the current state of basic science and technology to (**Figure 4.2**). Metamachines and their ability to reconfigure would benefit from flexible and responsive building blocks, that replace the current monolithic chassis. Similar blocks have been recently made available by progress in micro- and nanofabrication techniques [8]. Composite metamachines, made of different materials, would benefit from the combination of their different physical properties to realize multiple and complex functions by decoupling parts for propulsion, actuation and sensing. Compliant structures will provide additional degrees of freedom, that can be leveraged to compute and respond to environmental constraints [9]. Self-organization and reconfiguration of metamachines with deformable parts can enable complex tasks such as going through a constriction or moving a load.

To date, metamachines and their functions have been limited to two dimensions, so has their navigation. For either medical or environmental applications, metamachines will have to navigate in a three-dimensional complex environment and be biocompatible. A collection of ferrofluidic iron colloidal particles self-assembled into a dynamic colloidal collective thanks to the application of a rotating magnetic field. This fluid architecture was then controlled using the convective flows arising from heating by a focused laser beam [10], These convective flows allowed the structure to overcome gravity and navigate in three dimensions, by lifting the colloidal collective. This example of metamachine is however millimetric and a comparable implementation at microscale remains elusive. It nonetheless highlights how soft and reconfigurable machines could displace in a three-dimensional environment, responding to external control.

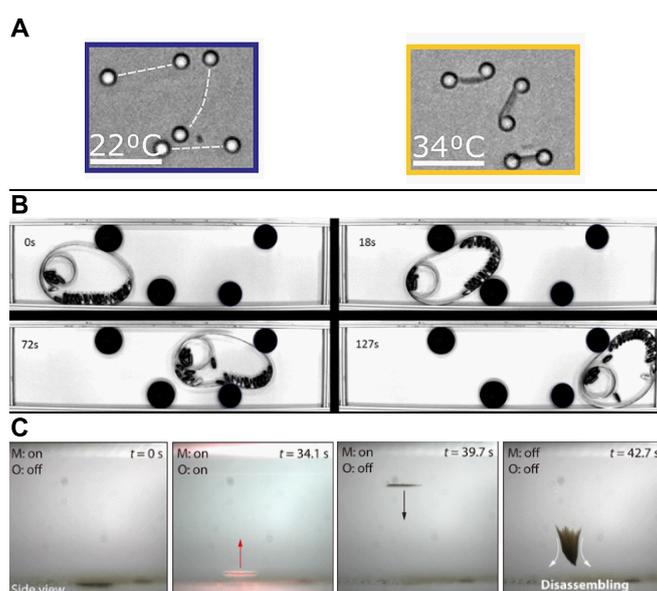

**Figure 4.2 | The future of metamachines. (A)** Composite colloids with responsive components. Silica colloidal dimer linked by a thermoresponsive hydrogel. Brightfield images of the colloidal dimers at 22°C and 35°C. Scale bar: 10 μm. Adapted from S van Kesteren *et al.* [8], use permitted under the CC-BY 4.0 licence. **(B)** Smart robots from flexible scaffolds. Photographs of the superstructure resulting from the self-organization of centimetre-sized robots enclosed scaffolds moving through obstacles and carrying a load. From J. F. Boudet *et al.* [9]. Reprinted with permission from AAAS. **(C)** Fluid dynamic colloidal

collective navigates in 3D. Driven by a rotating magnetic field, sedimented ferrofluid iron colloids self-assemble into a dynamic colloidal collective. The optical field triggers the generation of convective flow through photothermal effect, allowing the 3D navigation of the colloidal collective. The M and O labels indicate respectively the status of the magnetic and optical field. The red and black arrows show the direction of displacement of the colloidal assembly. Scale bar: 1 mm. Adapted from M. Sun *et al.* [10], use permitted under the CC-BY 4.0 licence.

## Concluding Remarks

Metamachines is a nascent field aimed to complement the current developments in microrobotics with autonomous and responsive colloidal machines. Its success will hinge on the use of smart(er) materials in nano- and microfabrication, allowing responsiveness to the environment and flexibility. In coming years, we envision that metamachines will be endowed with advanced properties, e.g. reconfigurability, self-replication or oscillatory behaviour, ultimately providing a flexible platform for bottom-up assembly of hierarchical structures.

## Acknowledgements

JP and QM acknowledge support from the US Army Research Office under award W911NF2310008.

## 05 – Towards animate lasers: Bioinspired lasers with active matter

Wai Kit Ng, Dhruv Saxena, Riccardo Sapienza
Imperial College London

### Status

The invention of new optical materials and lasers has always been sought after for technological progress and the creation of new applications. Small lasers, for example, made from single nanodisks [1] and microspheres [2] led to the development of 'biolasers' that could attach and function inside biological tissues/cells, spurring research activity in cell tagging, imaging and even targeted treatment inside the body. However, developing lasers that can reliably function in complex dynamic environments, like in an organism, is challenging, as it requires precise control over the position and orientation of individual components.

One solution is to build lasers out of disordered structures, utilising multiple scattering for light trapping and active manipulation (**Figure 5.1**). Such type of lasers, termed random lasers, are free from rigid cavity designs and, hence, can better maintain their performance in dynamic environments [3]. Random lasers have been made using strong scattering optical materials such as photonic glasses, semiconductor powders, and colloidal particles. These are typically static and hard to tune. In contrast, the recent use of animate matter – that can respond autonomously to its environment [4] – allowed for the dynamic synthesis and control of a random laser. This is an exciting development, as the self-organisation of animate matter opens new possibilities to build complex artificial optical metamaterials and devices.

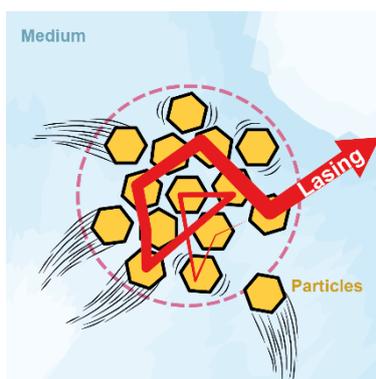

**Figure 5.1 | Self-assembled random lasers.** Schematic showing light amplification in a colloidal solution with scattering particles and laser dye. Dense clusters formed by self-assembly enable random lasing to occur.

Lasing is an important functionality for animate materials, whose potential is yet to be exploited. The first steps towards realising active lasers were focussed on manipulating the properties of the structure during fabrication, with fixed architectures afterwards. For instance, the dynamic assembly of semiconductor nanorod lasers was demonstrated under an external electric field [5]. With the tendency of the nanorods to align with the electric field in the solution, nanorod superstructures with tuneable orderliness could be deposited on the substrate with corresponding tuneable lasing or amplified spontaneous emission (ASE) thresholds.

Recently, active colloidal random lasers built via self-assembly with additional functional controllability were demonstrated [4]. By incorporating Janus particles (see also Section 2) for thermo-osmotic attraction, colloids in a dye solution were assembled into clusters surrounding a Janus particle. The threshold for random lasing was reduced by the enhancement of scattering in the cluster, thus enabling random lasing under optical excitation. Together with the active control of particle dissipation, the laser was controlled adaptively with reversible and reconfigurable lasing behaviours.

This recent advancement in active random lasers suggests the possibility of achieving programmable animate lasers that could autonomously adapt to their environment.

**Current and Future Challenges**

Many challenges remain to turn active matter lasers and biolasers into practical technologies. Materials development for active lasers in complex systems requires consideration of many factors, including viscosity, size, density, solubility, hydrophilicity, and surface functionality. Furthermore, pH, toxicity, and carcinogenicity must also be considered for use in a biological environment. All these considerations significantly constrain the material choices and make it challenging to realise low-threshold, small-footprint active lasers.

Future active lasers will require rapid manipulation of their components to respond faster to the stimuli from the environment. The current active lasers constructed using thermal gradients are limited to an assembling speed of tens of minutes [4], which is far slower than changes in a dynamic environment. Important laser functionalities, such as pulsed emission through mode-locking and frequency comb for wavelength multiplexing, are also still missing in active lasers and will require substantial development.

One of the major targets in the active laser community is to realise programmable and autonomous lasers that can self-organise, react, cooperate, and adapt to the environment like biological materials. This would require responsiveness from the system to sense environmental changes and feedback mechanisms to reconfigure its structures. This is an extraordinary function which has not yet been implemented in any photonic structure. Realising self-adapted lasers capable of changing their photonic/lasing properties in response to environmental stimuli is a big challenge in the active laser community that awaits to be addressed.

**Advances in Science and Technology to Meet Challenges**

The future development of active lasers will require advances in the gain material as well as in the active control of the self-assembly process and of the interaction between the elements in the swarm. Specifically,

- **Lasing materials:** Materials play a crucial role in active laser performance. Low-index contrast materials hamper the lasing process by reducing light confinement. Recent advancements in nanophotonics have achieved nano/micro-particles with various high refractive indices, such as perovskite-$SiO_2$ nanocomposites [6] and inorganic semiconductors [7]. These nanoparticles are biocompatible and intrinsically combine high optical gain and strong-scattering properties, which are desirable for assembling random lasing swarms in dynamic environments. Lasing systems can also implement gain both in its units (e.g., fluorescent beads) and environment (e.g., perovskite nanocrystals in solution [8]) and incorporate materials with ultra-high refractive indices (e.g., Germanium or artificial metamaterials [9]) to achieve highly confined, small-volume active lasers.

- **Active control methods:** Additional driving forces could be exploited to realize faster responding active lasers. Optical, electric or magnetic fields could improve material assembly as well as swarm formation, rapidity and flexibility. Photo-induced accumulation and depletion of a colloidal assembly on a conductive substrate shows the potential to achieve faster responding laser swarms with assembly times under a minute [10]. In addition, collective structures with varying shapes and sizes can be formed, for example, by exploiting electric fields in metal or dielectric colloids [11]. Implementing these faster and more flexible assembly schemes in active laser systems could lead to programmable swarming that can not only adapt and operate in a fast-changing environment, but also actively change the lasing properties in the environment.

- **Swarm interactions:** To realise truly autonomous and life-like swarming lasers, achieving self-monitoring of the particle motions is crucial. If the swarm motion is driven externally, possibilities for control and mobility remain limited. For an ideal laser swarm, individual units should be able to "communicate" by sensing other units' changes and adapt to the overall swarm motion. **Figure 5.2** illustrates this concept. With the target set for the laser swarm (e.g., by introducing chemical or temperature gradients), individual laser units assemble while simultaneously sensing their environment to guide the whole laser swarm to the target collectively and maintain the lasing properties and functionalities on the way. The interactions between the laser swarm and its environment are a critical part for enabling its automation. Sensing and signalling could be boosted by optical antennas with the capability to emit and receive directional signals from neighbouring swarms or obstacles. Although many implementation challenges are still awaiting to be addressed, learning how to engineer the driving mechanisms of the laser units and their coordination could enable active lasers to adapt to their environment and become truly programmable and autonomous.

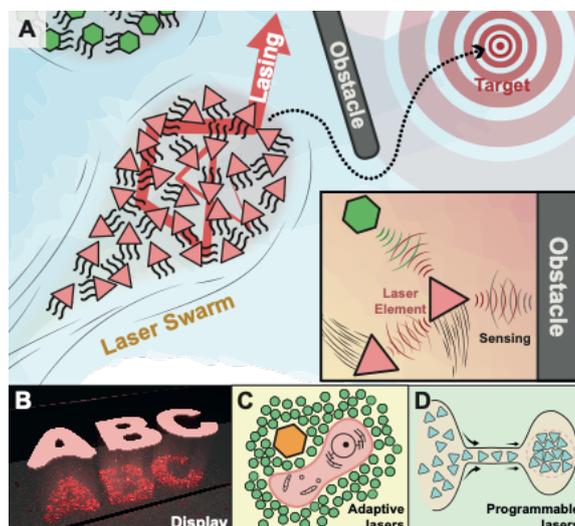

**Figure 5.2 | Swarming lasers**. (**A**) Illustration of an autonomous active laser. Laser units form a laser swarm through sensing and signalling to generate collective motion with lasing functionalities. The laser swarm can respond to environmental changes and be guided towards the target by introducing potentials. The examples of the laser swarms' possible applications, including (**B**) displays, (**C**) adaptive biological lasers, and (**D**) programmable lasers.

## Concluding Remarks

The development of active lasers is still in its infancy, with only a few architectures blending active matter concepts with nano-/micro-laser systems. However, the first success in manipulating active laser properties suggests new laser functionalities that can be realised in dynamic systems. With further development in tackling challenges such as laser materials, swarm manipulation, and environment interactions, active programmable lasers in dynamic systems could be achieved. This suggests exciting routes for realising novel functionalities in complex and biological environments, such as biological sensing, signal processing, and non-conventional computing.

## Acknowledgements

W.K.N. acknowledges the research support funded by the President's PhD Scholarships from Imperial College London.

## 06 - Liquid microcapsules: Life within confined liquid microenvironments

Sara Nadine, João F. Mano

CICECO – Aveiro Institute of Materials, University of Aveiro, Aveiro, Portugal

### Status

Life finds expression through a diverse array of liquid compartments. This phenomenon extends across various length scales, from the micro- to the macroscale, featuring entities such as lysosomes, animal cells, embryos, zebrafish eggs, or the mammalian placenta. Cells also utilize membranes to sculpt liquid and functional compartments, mirroring the sophisticated organizational principles observed in the human body, where organs meticulously fulfil specific functions. Within tissue engineering, researchers aspire to replicate the intricacies of natural biological systems to fabricate complex 3D microtissues. For example, researchers have successfully compartmentalized basic components to develop artificial cells [1] (see also Section 1). Cells have also been encapsulated within hydrogels to obtain tissue-like constructs such as myocardium or epithelium organoids. [2-3]. Inspired by nature, a cell encapsulation system was realized, serving as an authentic bioreactor supporting the autonomous development of microtissues [4]. These liquid capsules contain (i) a permselective membrane, wrapping the liquid content, (ii) microcarriers for cell adhesion alongside (iii) the cells themselves (**Figure 6.1**). The membrane facilitates the exchange of vital biomolecules (e.g., nutrients, oxygen, growth factors, and cytokines) throughout the 3D construct. Microcarriers play a pivotal role in supporting cell adhesion, proliferation, and differentiation, particularly beneficial for adherent cell types. They also support the encapsulation of suspension cells, faithfully recreating an environment that closely mimics their natural surroundings. The unique fluidic environment enables the orchestrated development of dynamic flows, optimizing the interaction of cells with microparticles for the creation of living matter. Besides working as a specialized bioreactor, these liquid microcapsules, by enhancing the precision of *in vitro* tissue-engineered constructs, are envisioned also as a promising alternative to conventional scaffolds. The membrane works as a selective barrier against the intrusion of immunoglobulins and immune cells and prevents the dispersion of the core contents into peripheral regions of the body upon implantation. Furthermore, liquid capsules demonstrate excellent injectability, facilitating their delivery by minimally invasive procedures. This pioneering system, already tested *in vivo*, emerged as a versatile cell encapsulation solution for diverse applications, including bone, cartilage, and bone marrow tissue engineering [5-7].

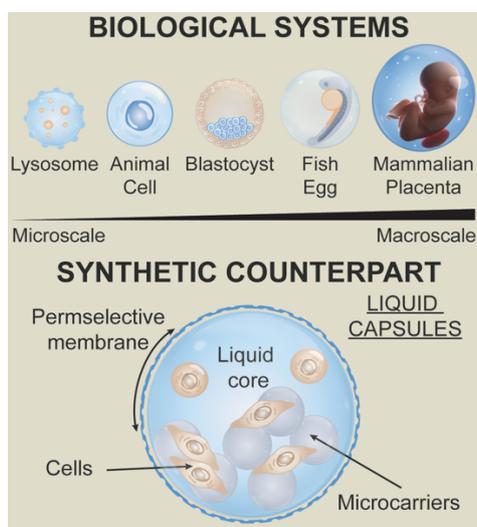

**Figure 6.1 | Biological and synthetic liquid compartments**. Liquid capsules as intricate synthetic constructs, drawing inspiration from the diverse liquid compartments prevalent in nature.

**Current and Future Challenges**

Developing the next generation of bioreactors and scaffolds requires careful consideration. Prioritizing *in vitro* vascularization prepares tissues for successful *in vivo* integration. Remarkably, liquid capsules (Fig. 6.1) enable the self-regulated organization of cells, primarily orchestrated by the cells themselves, presenting a compelling advantage in directing the differentiation of stem cells without external biochemical factors. Nevertheless, a significant obstacle persists within the liquid capsule strategy, namely the establishment of vascularization. Simultaneously, the introduction of an inflammatory environment alongside growing tissue is essential to mimic mammalian host responses. To mitigate these challenges, various co-culture systems have been conceptualized, including the encapsulation of mesenchymal-derived stem cells, macrophages, and endothelial cells. This approach not only enhanced the release of angiogenic factors but also elicited a regenerative immunomodulatory response [8]. Despite noteworthy advancements, establishing a fully functional vascular network within these capsules remains a central focus and a persistent challenge in the field [8]. A pivotal step for advancing tissue engineering outcomes involves the continuous monitoring of the bioreactor. Utilizing advanced imaging and sensing techniques facilitates the real-time, non-destructive evaluation of cell fate and tissue growth in the complex 3D environment, establishing the foundation for the automated control of the bioreactor. One of the advantages of liquid capsules lies in the semi-permeable membrane, which permits the passage of proteins released by cells, enabling their detection and measurement in the surrounding culture medium without compromising the integrity of the capsules. Scaling up these micro-units is another aspect to bear in mind. One approach is to incorporate these micro-units within hydrogels. Alternatively, the outer membrane of capsules can be functionalized, introducing adhesion sites conducive to the formation of larger structures [9]. This innovative approach involves a dual building block system, where microcapsules demonstrate the capability to generate microaggregates of cells and microparticles within the liquefied core, and simultaneously, these microcapsules can be assembled by cells to create macroscopic, more complex 3D structures [9].

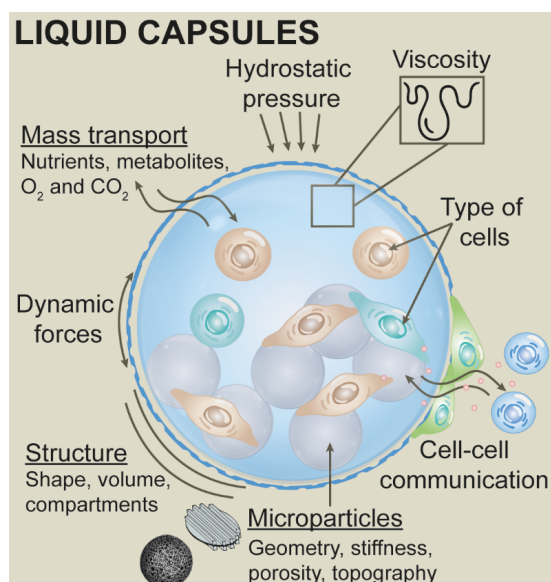

**Figure 6.2 | Liquid microcapsules**. Liquid capsules function as specialized bioreactors, offering the potential for the independent manipulation of various parameters tailored to specific applications and objectives.

**Advances in Science and Technology to Meet Challenges**

Various strategies can be explored to address the primary challenges of the field. While current liquid capsules demonstrate remarkable potential in orchestrating the development of microtissues, future advancements must steer towards more sophisticated and forward-looking solutions. Each liquid unit possesses the versatility to harbour distinct living matter, including both suspension and adherent cells, derived from multiple sources, such as human-based and animal-based cells, or even microorganisms. Combining different cell phenotypes in varying ratios could offer a promising approach to enhancing vascularization and reducing heterotypic tissue formation. Another avenue worth exploring is the integration of engineered living materials for applications such as biosensing or sustained drug release, effectively managing the regulation of differentiation and vascularization [10]. Moreover, the quality of the generated tissues may also be upgraded by the independent manipulation of various parameters (**Figure 6.2**). For instance, precise control over hydrostatic pressure is possible, allowing the replication of native mechanical microenvironments. This control is an essential factor influencing the regulation of the differentiation of stem cells. Additionally, modifying the core viscosity by adjusting fluid composition, may enable the modulation of shear stress magnitude.

Another notable possibility of guiding stem cell differentiation and improving vascularization could be the customization of microparticle characteristics, offering flexibility in the type of material, quantity, geometry, stiffness, porosity, and topography, or even incorporating specialized features for controlled drug release. The encapsulation of microparticles not only provides precise mechanical stimuli to trigger the initial differentiation of stem cells towards controllable lineages but also expands the range of stiffness compared to hydrogels. Taking advantage of the liquid core, these capsules can be cultivated in dynamic environments like spinner flasks, providing mechanical stimulation that profoundly influences cell behaviour and response. Cells can sense the mechanical stimuli resulting from the agitation of the compartmentalized liquid within capsules, a characteristic easily customized by adjusting the rotations per minute of the culture flask. Moreover, the adjustment of other parameters, including pH, temperature, and oxygen levels, can be easily accomplished by altering the conditions of the culture medium, also owing to the properties of the membrane (see also Section 7).

In addressing the challenge of scaling up, liquid capsules offer a promising avenue as modular building blocks. One viable strategy involves leveraging bioprinting techniques to integrate these capsules into larger, more intricate structures. Due to the absence of a core, it is hypothesized that incorporating liquid capsules within hydrogel precursors can address the limitations associated with direct cell embedding, such as lack of direct cell-cell contact or limited diffusion of nutrients, while also reducing the amount of bioink required for such processes. Additionally, the application of magnetic and acoustic fields presents an opportunity to precisely arrange capsules, facilitating the creation of customized designs. The adaptability of this system extends to the modification of external features of the capsule, such as shape, compartments, or volume, easily achievable based on the chosen fabrication technique. While spherical systems are commonly utilized, there is a growing exploration of generating multi-shaped, intricate 3D structures to emulate the complexity of native tissues.

**Concluding Remarks**

Inspired by the intricate liquid compartments observed in nature, liquid capsules stand as a ground-breaking advancement in tissue engineering. Serving as animate matter and specialized bioreactors, these units facilitate reproducible and controlled variations of specific environments during the *in vitro* development of living tissues. Moreover, liquid capsules enable the simultaneous application of multiple stimuli, spanning from mechanical to biochemical cues. This unique structural design permits systematic investigations into the effects of these stimuli on cell functions and tissue morphogenesis, elevating the precision of *in vitro* tissue-engineered constructs. Like conventional bioreactors, liquid capsules allow independent manipulation of multiple parameters, being adaptable to sizes ranging from approximately 50 μm to 2 mm in diameter. This micro-scale versatility enables tailored mimicry of both the geometry and nature of mechanical and biochemical stimuli specific to the target tissue. Overall, liquid capsules represent a transformative approach, offering unprecedented control and adaptability in tissue engineering that holds great promise for advancing the field.

**Acknowledgements**

This work was financed by the European Research Council Advanced Grant ''REBORN'' (grant agreement n. ERC-2019-ADG-883370. This work was developed within the scope of the project CICECO-Aveiro Institute of Materials, UIDB/50011/2020 (DOI 10.54499/UIDB/50011/2020), UIDP/50011/2020 (DOI 10.54499/UIDP/50011/2020) & LA/P/0006/2020 (DOI 10.54499/LA/P/0006/2020), financed by national funds through the FCT/MCTES (PIDDAC).

## 07 – Organs on a chip: Guiding cellular specialization with environmental cues in advanced cell cultures

Reza Mahdavi, Caroline Beck Adiels

Department of Physics, University of Gothenburg

**Status**

Organ-on-a-chip technology has advanced significantly since its inception in the early 2000's, reshaping how we study human biology and disease. These microfluidic devices house living, miniaturized tissues that mimic human organ physiology and pathology. The field's rapid growth is fuelled by the integration of tissue engineering and microfabrication techniques, allowing precise control over tissue microenvironments and spatiotemporal dynamics [1] (see also Section 6).

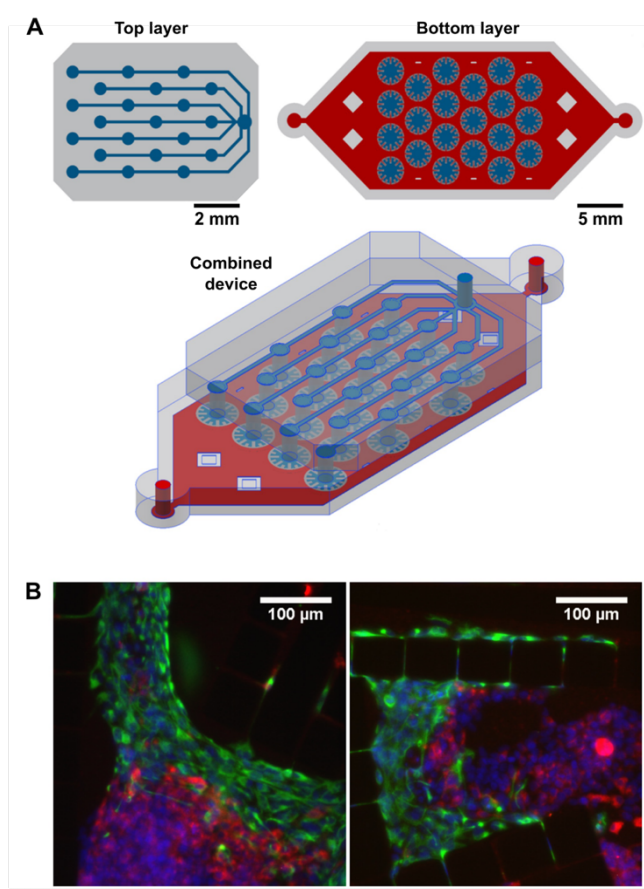

**Figure 7.1 | An organ-on-a-chip platform reproducing the architecture and fluid dynamics of human liver lobules and incorporating co-cultures of two liver-specific cells.** (**A**) The rendered figure showcases the top, and bottom layers, respectively, of the combined device (bottom). The medium channel is highlighted in red, while the chambers, seeding ports, and top layer channels are depicted in blue. The flow in the finalized polydimethylsiloxane (PDMS) device can be redirected through the top layer outlet (blue), mimicking the portal vein. Scale bars indicate 2 mm in the top layer image (top left) and 5 mm in the bottom layer (top right). (**B**) Presents on-chip staining of HepG2 (hepatocytes) and LX-2 (stellate cells) cell co-cultures in the device under nonadherent conditions. The cultures are stained with antibodies targeting albumin (HepG2, red) and vimentin (LX-2, green), and counterstained with DAPI (nuclei, blue). Adapted with permission from [2], use permitted by author of original publication.

Innovations in organ-on-a-chip technology have evolved from basic 2D monocultures to sophisticated 3D co-culture systems, manipulating cellular microenvironments and geometrical arrangements to enhance organ-like functions (**Figure 7.1**). This manipulation results in self-organization of cells driven by concentration gradients and physiologically relevant architectures. Cells, sensing their environment, undergo self-regulation, leading to specialized gene expression and cellular functions within these microfluidic devices. This dynamic interplay between cells and microenvironments advances organ-on-a-chip technology towards more accurate representations of human biology and disease states. Integrated with primary cell sources and induced pluripotent stem cell technology, organ-on-a-chip devices enable personalized studies of disease phenotypes and drug responses on a patient-specific level. On-chip sensors further enhance these platforms, enabling dynamic data acquisition and feedback systems for complex tissue architectures [2, 3].

Looking ahead, organ chips have the potential to transform drug screening by replacing animal models with miniaturized systems that replicate human responses more accurately, possibly evolving into body-on-a-chip platforms. This shift aligns with the principles of 3Rs: refining research models to align with human biology, reducing reliance on animal testing, and replacing animal testing wherever feasible. As organ-on-a-chip technology gains wider acceptance, increased funding, and undergoes continuous technological advancements, researchers foresee the era of personalized medicine facilitated by patient-derived organ chips. This progress promises to accelerate our comprehension of human diseases and accelerate discoveries in therapeutic interventions [1,3].

**Current and Future Challenges**

Organ-on-a-chip models present a promising avenue for *in-vitro* systems, yet their widespread application encounters several limitations. Current organ-on-a-chip devices suffer from limitations in the standardization of multi-chip systems (including fabrication techniques, materials, sensor incorporation) and in the spatiotemporal resolution necessary for dynamic studies [3].

Many organ-on-a-chip systems are still simple compared to human organs. This limitation is exacerbated by the utilization of cell sources which often relies on immortalized cancer cell lines or cells of animal origin. Even with the utilization of human primary cells, their inclusion is typically limited to a small number of cell types. Further, the cell source is finite, resulting in large heterogeneity and outcomes hard to interpret. Consequently, a broader array of natural cell types is essential to faithfully emulate human organ physiology. Ongoing investigations are directing efforts towards the integration of various cell types and the interconnection of organ-on-a-chip models, forming more comprehensive systems known as multi-organ-on-a-chips or, more ambitiously, body-on-a-chips. Despite these endeavours, fully replicating human physiology *in vitro* remains a formidable challenge (see also Section 6). Furthermore, the absence of standardization in tubing and liquid handling within multi-organ-on-a-chip systems introduces an additional hurdle, necessitating considerable time and resource investments for setup and customization [3, 4].

Further engineering challenges complicate the replication of tissue structural complexity and spatial organization at the microscale level. Soft lithography using polydimethylsiloxane (PDMS), a leading material in organ-on-a-chip studies, offers desirable features such as transparency, biocompatibility, gas permeability and ease of fabrication but is limited in drug study research due to its hydrophobicity and porosity. Advanced microfabrication techniques and biocompatible substitutes are proposed to overcome these challenges and enable robust and reproducible microfluidic chips, scaling up their throughput [5].

Integrating sensors for real-time monitoring during cell culture poses another challenge. Traditionally, analyses relied on endpoint assays, disrupting dynamic culture conditions. However, to elicit the full capabilities of organ-on-a- chip devices, the incorporation of different sensors is crucial. Often, current sensors are based on optical or electrochemical readouts, which face challenges such as low resolution, short lifespans, or constrains to 2D culture models. The development of microsensors tailored to organ chips holds promise for enabling sophisticated cellular measurements, contributing to precision testing. The potential of organ-on-a-chip models in *in-vitro* systems is hampered by various challenges, and addressing these complexities requires a multidisciplinary strategy. This approach involves advancements in fabrication techniques, materials, and sensor technologies to propel organ-on-a-chip systems towards a more mature and widely applicable platform [3, 6].

**Advances in Science and Technology to Meet Challenges**

As previously noted, expanding the utilization of induced pluripotent stem cells (iPSCs) in organ-on-a-chip devices can overcome several shortcomings associated with traditional cell sources. iPSCs offer increased availability compared to primary cells, as they can be generated in large quantities from diverse donors, thereby reducing heterogeneity in outcomes. Additionally, iPSCs enable patient-specificity, facilitating the customization of organ-on-a-chip devices tailored to individual genetic background, including ethnicity, sex, age, and health status. The integration of patient-specific parenchyma, specialized cells, and immune cells into versatile on-chip tissues supports more accurate and personalized disease modelling and treatment testing (**Figure 7.2**). Successful incorporation of iPSCs in organ chips requires robust differentiation and maturation protocols to acquire cell phenotypes that closely resemble their *in-vivo* counterpart. Achieving comprehensive on-chip recapitulation with multi-organ communication relies heavily on precise vascularization to connect different tissues and ensure perfusion of individual tissue grown on the chip [3, 7].

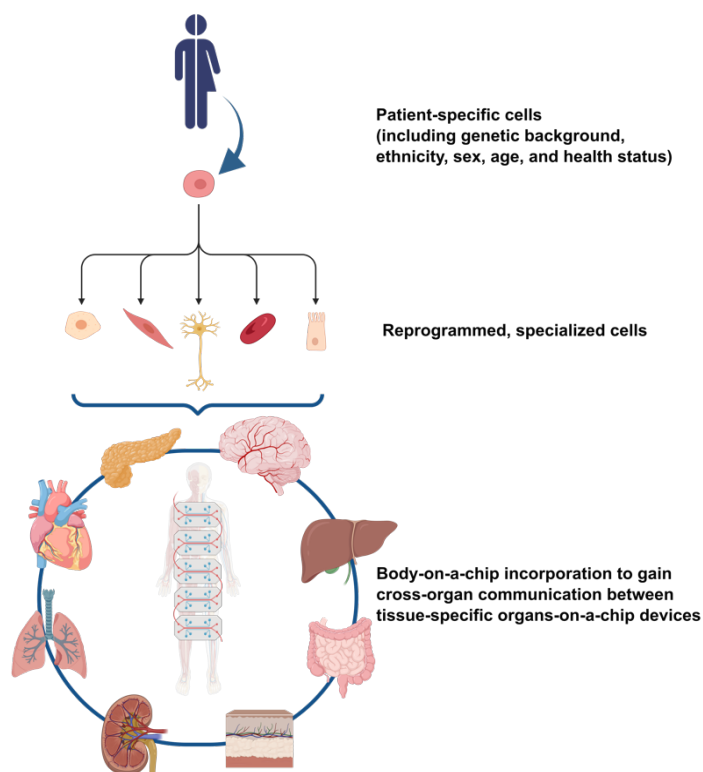

**Figure 7.2 | Illustration of the potential of combining microfluidics with iPSC technology**. Somatic cells, such as fibroblasts, are reprogrammed, karyotyped, and differentiated into a preferred specialized cell type through various protocols. Ideally, these cells are then cultured in physiologically relevant 3D environments, perfused, and interconnected with other organ-specific cell types. Together, they form a 'body-on-chip' platform for conducting patient-specific drug development and testing on a model representing both healthy and diseased bodies.

Concurrently, it is essential to focus on selecting biocompatible materials and balancing fabrication accessibility with cost considerations. Recent advancements in laser pyrolysis have made significant strides in high-speed, high-quality PDMS 3D processing, addressing issues associated with soft lithography. However, PDMS remains susceptible to absorbing lipophilic materials, prompting exploration of alternatives like glass and thermoplastics, each presenting unique trade-offs in terms of cost and fabrication efficiency. With the ongoing advancements in material science and the potential utilization of biologically-derived materials such as hydrogels or extracellular matrix components, i.e. bio-inks, along with advancements in 3D printing technology, both speed, precision and accuracy in creating organ-on-a-chip models are increasingly achievable [8].

Analytical biosensors integrated into organ-on-a-chip models enable real-time monitoring of cells and their microenvironment, providing invaluable insights. The urgent demand for continuous, non-invasive, real-time monitoring of tissue architectures requires the direct integration of biosensors. Currently, various sensors are employed to measure physiological and pathological conditions. Sensors are being improved to overcome their limitations. A recent study implemented an enhanced spatial transepithelial electrical resistance-based sensor in an organ-on-a-chip that incorporates scanning electrodes, offering increased precision in localized measurements of permeability [9]. In another instance, the application of 3D-printing technology has improved the efficiency and cost-effectiveness of producing microelectrode arrays [10]. Multisensor integration is crucial for body-on-a-chip platforms to capture complex interactions. For instance, a four-organ system combined electrical recordings, mechanical sensors, and optical analysis to continuously monitor liver, cardiac, neural, and muscle tissues. Future organs-on-chips will benefit from microscale, flexible, biocompatible sensors that provide multiparametric analysis unlocking the full potential of body-on-a-chip devices [6, 11].

### Concluding Remarks
Organ-on-a-chip technology represents a transformative shift in experimental approaches within biomedical research and pharmaceutical development pipelines. At the technological core, the microfluidic devices utilize specialized fabrication techniques to embed human living cells into an engineered environment mimicking key architectural and physiological features of organs and tissues. The result is an in-vitro model exhibiting complexity not achievable in conventional 2D cell cultures. As organ-on-a-chip platforms continue maturing through advances in biomaterials, microfluidics, and cell culture techniques, so does their ability to revolutionize medicine. Within academic labs, organ-on-a-chip adoption enables researchers to visualize dynamics of infection, toxicity, and treatment response with direct species-relevant observations at the cellular level. In parallel, biopharma integration of organ-on-a-chip technology for elevating compound screening and efficacy testing heralds the emergence of personalized therapies to deliver precision treatments. With strong pharmaceutical backing now accelerating commercial translation, organ-on-a-chip systems are poised to transform preclinical R&D over the next decade by enhancing probability of success for drug candidates prior to costly late-stage failures. For these reasons, organ-on-a-chips represent one of the most disruptive and transformative technologies that will shape the future of healthcare in the 21$^{st}$ century.

### Acknowledgements
This project received funding from the Swedish Foundation for Strategic Research (ITM17-0384).

## 08 – Towards printed animate matter

Joe Forth

Department of Physics and Department of Chemistry, University of Liverpool, Liverpool, United Kingdom

**Status**

The overwhelming majority of work on printed 'active' matter falls under the umbrella of bioprinting, which aims to recapitulate the physiological form and function of an organ for use in regenerative medicine or *in-vitro* modelling (see also Section 6). A more niche, albeit rapidly growing area applies the advanced fabrication techniques used in bioprinting to incorporate active components into complex structures as working parts, producing a distinct class of materials: Printed Animate Matter ('PAM'). To be classified as PAM, systems must satisfy three criteria. First, they must have some sort of architected structure, either through fabrication or controlled self-assembly. Second, they must incorporate embodied energy or the means to convert energy from their surrounding environment; they do not simply relax to equilibrium. Third, they must consist of discrete, potentially autonomous building blocks that can collectively give rise to complex emergent behaviour. Beyond obvious outcomes like actuation, PAM has also found use in stem cell culture, electronics, and ecological complexity modelling.

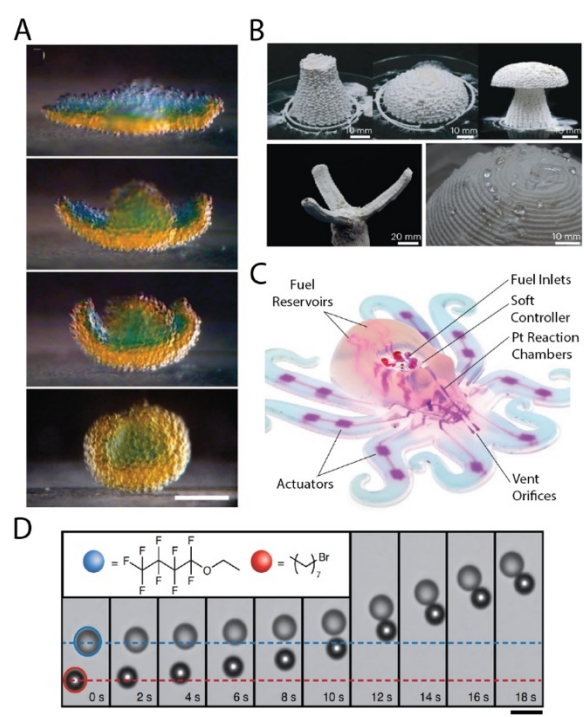

**Figure 8.1 | Examples of printed animate matter or its building blocks**. (**A**) Reconfigurable droplet-based constructs that change shape due to liquid flow between interconnected compartments. Scale bar, 200 µm. From Villar *et al.* [1]. Reprinted with permission from AAAS. Copyright © 2013, AAAS. (**B**) Living structures of mycelium made from inoculated, printed scaffolds. Reproduced with permission from Gantenbein *et al.* [2]. Copyright © 2022, Springer Nature. (**C**) An autonomous, soft robot that is driven from equilibrium by an incorporated power source. Reproduced with permission from Wehner *et al.* [3]. Copyright © 2016, Springer Nature. (**D**) Motility, chasing, and dissipative self-assembly in a system of two species of droplet. Scale bar, 100 µm. Reproduced with permission from Meredith *et al.* [4]. Copyright © 2020, Springer Nature. Note that only **B** and **D** satisfy all three criteria of PAM.

The most developed printed animate materials are tissue-like, droplet-based constructs built around droplet interface bilayers (**Figure 8.1A**) [1]. These form when phospholipid surfactants adsorb to oil-water interfaces and zip together to form bilayers when two droplets are brought into contact. The resulting droplet-droplet adhesion enables the printing of complex structures comprised of thousands of droplets. Incorporation of nanopores into the bilayers enables size-selective, spatially defined communication between droplets. Incorporating hydrogels into the printed structures leads to remarkable robustness, enabling the fabrication of light-, heat-, or magnetic-field-responsive structures. More impressive still is the incorporation of living material and its interfacing with a synthetic component. Recently, a battery formed from a printed chain of ion-selective hydrogel droplets was connected to another chain of droplets containing neural progenitor cells [5]; the resulting electrical stimulus led to a marked change in cell behaviour.

Given its affordability, ease-of-use, and breadth of commercially available and open-source implementations, extrusion printing is the most popular platform for producing PAM. Using this approach, structures that incorporate diatoms, bacteria, and even mycelium have been made (**Figure 8.1B**) [2], with applications in pollution sensing, cellulose synthesis, and a remarkably resilient living synthetic skin. Extrusion printing naturally lends itself to the production of vascularised structures by writing fibrillar, sacrificial structures into shear-thinning, yield-stress support baths. This approach has been harnessed to produce animate materials ranging from wholly synthetic autonomous soft robots (**Figure 8.1C**) to vascularised, organ-like constructs comprised of aggregates of thousands of living cell spheroids [6].

**Current and Future Challenges**

The biggest barrier to developing the field of PAM is getting the communities of active matter, advanced fabrication, and tissue engineering to speak to one another. The dialogue between 3D printing and tissue engineering has a long and successful history. By contrast, interactions between these areas and active matter are surprisingly infrequent; soft and active matter physicists still work too rarely with complex biological systems. Taking a loose definition of 'printing', the advanced fabrication community has been too slow to adopt holographic optical and acoustic tweezers. The main obstacle to intercommunity collaboration is the perceived expense of bioprinting methods, the perceived difficulty of building and using holographic optical and acoustic tweezers, and the fundamental focus of active matter work. The surge in recent years of cheap and open-source soft matter fabrication methods provides some wonderful opportunities to address this shortcoming.

Active matter systems still lack simple formulation and design rules that would encourage their study by workers in advanced fabrication. There are a wide range of active, artificial cytoskeletons that have been incorporated into droplets and vesicles, although these systems are challenging to work with. Easier-to-use are the wealth of active droplets driven from equilibrium by encapsulated chemical compounds, which have come to the fore in recent years (**Figure 8.1D**). Droplets containing magnetic material, whether confined to a soft interface or encapsulated in the bulk, can exhibit highly controlled response to external fields and complex patterning as a result. A simple question to ask is "what happens when these active droplets are printed into tissue-like arrays?". The absence of simple rules-of-thumb by which such systems can be formulated for a specific printing method limits our ability to answer this question.

Finally, and more challengingly, reaction-structure-function relations for PAM need to be elucidated. Much like biological systems, PAM harnesses reactions at the sub-nanometre scale to produce individual action at the micrometre scale, leading to collective action at the millimetre scale or greater. On an experimental level, this requires spatiotemporal mapping of both chemical reactions and their mesoscopic and macroscopic impact. On a theoretical level, models are required that connect phenomena seen in PAM across length scales; it is important to stress that such models should be simple and insightful rather than complicated and overparameterised.

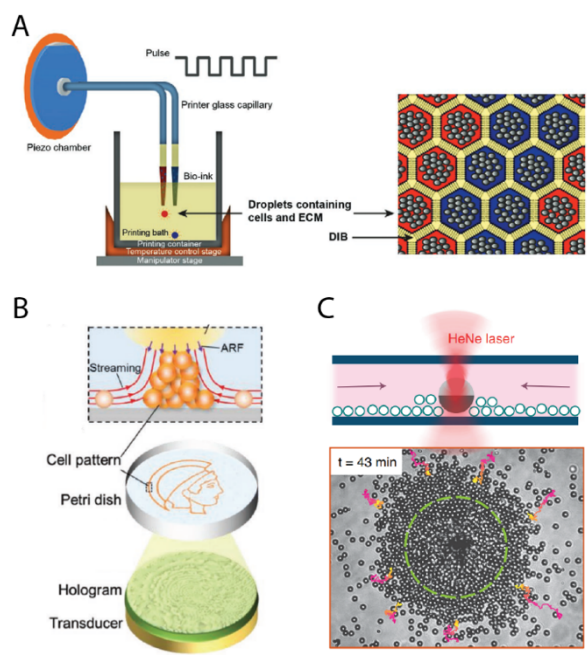

**Figure 8.2 | Methods for fabricating printed animate matter**. (**A**) Droplet deposition printing used to produce DIB-based synthetic tissues. Reproduced with permission from Zhou *et al.* [7]. Copyright © 2020, John Wiley and Sons. (**B**) Complex patterns of living cells assembled using acoustic tweezers. Reproduced from Ma *et al.* [8], use permitted under the CC-BY 4.0 licence. (**C**) A self-building random laser that is nucleated by optical tweezing. Reproduced with permission from Trivedi *et al.* [9]. Copyright © 2022, Springer Nature.

**Advances in Science and Technology to Meet Challenges**

The main advance required for PAM to progress as a field is not the development of new methods but the construction of a community. The proliferation of open-source modifications to fused deposition modelling (FDM) printers in recent years has led to remarkable innovations in chemistry, lab automation, and bioprinting. These advances have happened because of the affordability of FDM 3D printers (a good-enough model currently costs <£300) and the ease with which these setups can be programmed and modified. Most important has been the willingness of the 3D printing community to share its ideas, with knowledge exchange nucleated through hackspaces; the global proliferation of cheap, self-built bioprinters is a stand-out example of what happens when this approach works. While the adoption of self-built devices has been strongest in the extrusion printing community, recent growth in the publication of designs for droplet printers (**Figure 8.2A**) means this area is likely to see a similar surge soon. Such an approach must be taken to the wealth of other advanced fabrication techniques, especially optical methods such as digital light processing and stereolithography. Laudable efforts have already been made to facilitate broader use of holographic acoustic (**Figure 8.2B**) and optical manipulation (**Figure 8.2C**, see also Section 5), however the perceived technical barriers to uptake of these setups, particularly for 3D manipulation and fabrication, remain significant and their use is rather limited compared to that of modified FDM printers.

There are also two technical challenges that need to be overcome to drive the field forward. First, the design principles of active systems and the collective emergent behaviour that they exhibit need to be better understood and reduced to heuristics. This Roadmap documents a great deal of progress in this area, but for translational impact the discipline needs formulation science as well as fundamental science. Automated and high-throughput methods that incorporate algorithms for accelerated exploration of parameter space could prove particularly powerful here. These approaches naturally lend themselves to PAM in which the printer itself plays an active role in the experiment; successful implementations of this concept have led to evolutionary droplet printers that can uncover a rich array of behaviour from a simple model system [10].

Finally, microscopy methods that couple spectroscopic analytical chemistry techniques to multiscale optical characterisation must be applied to provide insight into how processes are coupled across the wide gamut of length scales in PAM. Such techniques are already either in use or under development in the life sciences; the major challenge comes in implementing them without the need for expensive equipment and large-scale facilities. Similarly on a theoretical level, models that connect a systems chemistry approach with the structural evolution of soft cellular systems would enable bottom-up approaches to printing synthetic animate tissues.

**Concluding Remarks**

Printed Animate Matter brings together the principles of active matter and advanced fabrication to produce materials that draw inspiration from biological tissues. These materials are inherently cellular, consisting of individual, energy-converting building blocks that are arranged in an architected structure. The resulting systems are soft, reconfigurable constructs that exhibit complex, emergent, and programmable functionality. To drive this field forward, what is needed is more knowledge exchange and collaboration between the fields of advanced fabrication, active matter, and tissue engineering. Active matter physicists must embrace complex systems; recent efforts to understand the physical principles of confluent cell monolayers, multi-cellular algae, and even tissue spheroids are laudable, but still too rare. Tissue engineers must embrace unconventional fabrication techniques such as holographic optical and acoustic tweezers. Advanced fabrication researchers must embrace dissipative self-assembly to design materials that print themselves. Implementation of many of these ideas is already well underway, but the field has plenty of space for new entrants.

**Acknowledgements**

JF acknowledges start-up support from the University of Liverpool.

## 09 – Self-folding orikata

Christian Santangelo
Syracuse University

**Status**

The origins of origami are now obscured by time but are plausibly as old as paper itself. Recently, however, origami and kirigami, a related art in which one folds and cuts paper, have emerged as a powerful tool to manufacture and actuate three-dimensional shapes [1]. Unlike traditional 3D printing, in which a structure is built layer-by-layer, orikata – meaning "folded shapes" and which we will use to refer to both origami and kirigami – begins with a flat, patterned substrate which subsequently folds up into the target 3D shape [2]. Whereas 3D printing has been touted as a route towards rapid prototyping, self-folding structures can potentially be fabricated in bulk and in sizes that might otherwise be difficult to achieve. The field has advanced rapidly, and numerous examples of self-folding structures can be found in the literature with sizes from the micro- to macroscopic (**Figure 9.1**). The long experience of origami artists has provided a wealth of structures for use in engineering applications. Technological advances have even enabled origami microrobots and devices that can fold and refold into many different shapes [3].

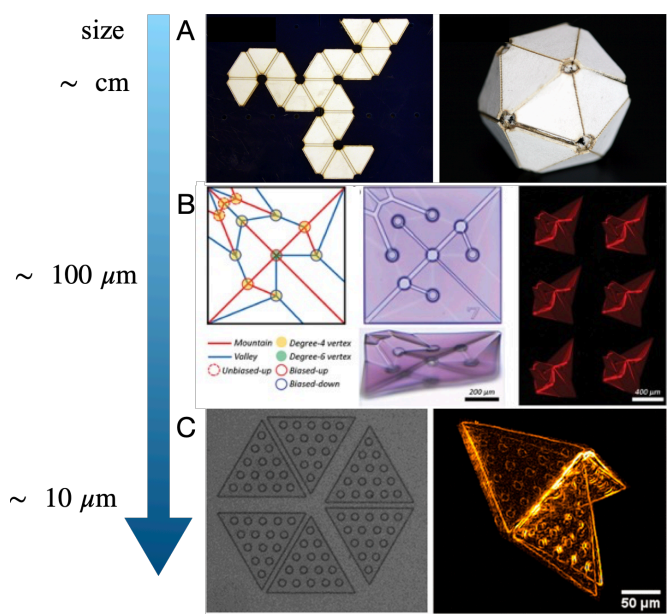

**Figure 9.1 | Self-folding orikata spanning multiple scales**. (**A**) Macroscopic shape memory composite by heating. Reprinted with permission from [3], Copyright © 2014, IOP Publishing Ltd. (**B**) Polymer trilayer folded by swelling. Reprinted with permission from [10], Copyright © 2019, WILEY-VCH Verlag GmbH & Co. KGaA, Weinheim. (**C**) Folding nanofilms. Reprinted with permission from [4], Copyright © 2020, American Chemical Society.

One question one might ask is whether there are any limitations to what shapes could be formed in this manner. Here, the answer is clearly no. First, there are algorithms to generate a fold pattern that can be folded into a given polyhedron from a single sheet of paper, though the results can sometimes be unwieldy [3]. Other algorithms produce origami tessellations [4], or kirigami structures [7] that approximate arbitrary shapes.

In recent years, however, several fundamental limitations of our understanding of self-folding orikata have become apparent. There appears to be a deep connection between the foldability of orikata with rigid faces and computation [8]. The foldability of orikata simplifies when faces can bend, but then the number of folding pathways grows exponentially with pattern complexity. Between these extremes, the energy landscapes of origami exhibit features of glassiness [9] and complexity that can prevent successful folding [10]. This proves a fertile ground for methods of physical learning, in which an origami material can be *trained*, because of material plasticity, to self-fold along a desired pathway. In physical learning as applied to origami, one looks for a local rule that changes the properties of the folds. Subsequently, one is training the origami to self-fold along the desired pathway much as one might train a neural network to recognize patterns [9].

### Current and Future Challenges

The long history of origami means there is no shortage of fold patterns to exploit in new devices. What remains is to expand the range of material systems that can fold and unfold repeatedly, and work on improving control over the fold angle of individual folds. However, there are also fundamental questions about the *physics* of self-folding orikata.

A primary challenge for physics is to understand how to take advantage of kinetic constraints and energetics to program robust folding pathways, especially as complexity is increased. As in protein folding or self-assembly, self-folding orikata must find a target ground state in a high-dimensional, complex energy landscape. Orikata, however, also provide kinetic control over fold geometry and fold stiffness that can be used to hierarchically control the shape and complexity of that landscape. The rigidity of the faces relative to the folds, for example, can be used to tune the nucleation of metastable states and the depth of the deep valleys along which orikata are constrained to fold [10]. Orikata already display many hallmarks of complex physics: the multistability of origami structures can exhibit hysteresis, return-point memory, and other phenomena of complex dynamical response [11]. In greater than one dimension, one expects hysteretic elements to interact to unveil new phenomena, shedding additional light on the connections to computation and physical learning [8, 9].

As origami structures shrink, they will be subject to additional random forces, either thermal motion or motion due to athermal, external driving. Surprisingly little is known about statistical mechanics and phase transitions in branched energy landscapes, but at small scales, the branched structure of the energy landscape [9] suggests that entropy will play a role in determining expected states of self-folding orikata. As one example, thermal fluctuations of structureless, thin sheets are believed to drive a crumpling transition; in the case of origami, whether and how this transition might happen is not clear.

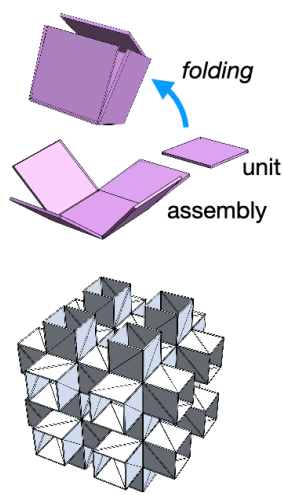

**Figure 9.2 | Self-assembling tiles to generate complex or actuating structures**. (Top) Self-assembly is an alternate pathway to forming complex origakata at the nanoscale. This can produce origami-like structures that cannot be folded from a flat sheet without gluing. (Bottom) A model of a triply-periodic orikata that can be formed with triangular units having the

symmetry cubic symmetry. This cannot be folded from a flat sheet of paper but could be assembled from triangular units following simple rules.

Finally, a third class of challenges will be to extend the range of deformations possible. In addition to folding, one might imagine growing or shrinking the faces so that the vertices develop a small excess or deficit angle. This would be a discrete analogue to a non-Euclidean sheet (or 4D printing), in which the in-plane and out-of-plane stresses are frustrated to buckle into complex 3D shapes (see also Section 12). One can envision orikata that are connected into a non-flat topology either by self-assembly (**Figure 9.2**) or as a multi-layered structure that unfolds from flat to 3D as would a children's pop-up book. The interplay of topology, fluctuations, and folding remains uncharted.

**Concluding Remarks**

Beyond their potential to revolutionize manufacture, self-folding orikata raise critical, and fundamental, questions that highlight the relationship between geometry and elasticity. The basic science has progressed far enough to foresee the use of folding in many engineering applications. Yet, I have argued that there also remaining questions that connect to physics, especially when coupled to actuation and motion. These questions speak to a need to further develop the technology and of continued theoretical development. Orikata constitute a new class of materials and structures for physicists to explore questions of energy landscapes, hysteresis, and fluctuations.

**Acknowledgements**

CDS acknowledges funding through NSF DMR-2217543.

# 10 – Soft robotics

Stefano Palagi

Sant'Anna School of Advanced Studies – Pisa

### Status

A key challenge in robotics is to make robots interact with natural (unstructured) environments and with living beings effectively and safely. To this aim, animals exploit flexible, elastic, and soft tissues, relying on the inherent compliance and adaptability of their body structures. Many animals even have fully soft bodies or use fully soft appendages for extremely dexterous manipulation (e.g., the elephant trunk). This has inspired roboticists to develop a so-called *soft robotics* approach [1] (**Figure 10.1**).

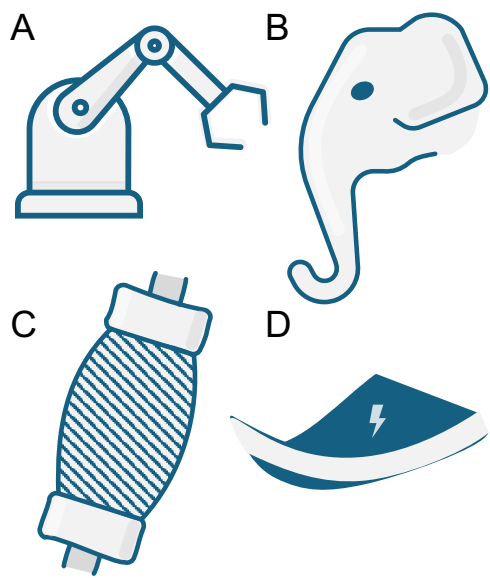

**Figure 10.1 | From rigid to soft robots.** (**A**) Traditional robotic manipulator with rigid links. (**B**) Example of nature as a source of inspiration for soft robots: the elephant trunk. Examples of soft actuators: (**C**) pneumatic artificial muscles and (**D**) dielectric elastomer actuators.

Soft robots can be defined as robotic devices that can interact with the environment with high compliance and deformability [2]. The term initially referred to robots with compliant joints or compliance control and has then identified with the shift from robots with rigid links to continuum robots that are inherently compliant, for instance by consisting of flexible or soft materials (e.g., elastomers and gels). The soft robotics approach thus consists of exploiting compliance and softness of (some of) the robot components to make them more easily and naturally interact with the physical world, without the need for controlling all movements with the highest precision.

In soft robotics, therefore, the distinction between the 'mind' and the body starts to blur. The soft robotics approach can help in making robots inherently adaptable and robust. In other words, it has the potential to make robots more autonomous and intelligent, exploiting the computational features of their bodies rather than just those of their 'brains' (see also Section 16). In this context, materials become key robotic components, not only because of their structural properties and role, but also, and especially, because they implement *functions*. Indeed, smart, stimuli-responsive polymeric materials can be used as soft actuators and/or sensors in soft robots. The design of robots' body and 'mind' thus become intertwined, implementing the principles of *embodied intelligence* observed in living beings [3], leading to the proposal of the term *Physical Artificial Intelligence* (PAI), which involves the co-evolution of body (including morphology, actuation, and sensing) and control [4].

The soft robotics approach could lead to a new generation of life-like (soft) robots. Moreover, it is functional to the development of autonomous and intelligent small-scale robots (e.g., sub-millimetre robots known as microrobots) [5]. This represents a challenging test field as, given their size, small-scale robots cannot be assembled out of standard components and can greatly benefit from implementing robotic functions in the constituent materials. Indeed, in small-scale robots, the material *is* the robot.

**Current and Future Challenges**
Unlike 'traditional' robots made of rigid metallic links and (most often) electromagnetic motors and sensors, which can be accurately designed, fabricated, modelled, and controlled, soft robots are (at least partially) made of soft materials, such as elastomers [6]. Whereas it is relatively simple to fabricate inflatable elastomeric actuators or pneumatic artificial muscles to make soft robots move, doing it with the accuracy and precision needed in many robotic applications is far from trivial. Modelling and controlling the behaviour of such soft robotic parts also presents additional challenges, related to the nonlinear and viscoelastic properties of the adopted soft materials. Model-free, data-based control approaches are thus taking traction in the field [7].

Despite the wide adoption of pneumatic actuators, soft robotics has stimulated (and has been stimulated by) research on smart and active materials that can work as soft actuators and artificial muscles or as soft sensors. They include Electro-Active Polymers (EAPs – e.g., Dielectric Elastomer Actuators) and other stimuli-responsive polymeric materials (thermo-responsive, photo-responsive, or chemo-responsive – e.g., hydrogels and Liquid-Crystal Elastomers). Nonetheless, no smart material is currently able to meet all the specifications required to a general-purpose soft robotics actuator, and the choice of a smart material as actuator depends strongly on the specific implementation.

In addition, if a material responds to environmental stimuli and conditions, rather than to provided control stimuli, they can act as sensors and actuators at the same time, implementing *reactions* to the environment. Indeed, an opportunity and challenge for soft robotics is the progressive decentralization and distribution of 'intelligence' to the different materials and parts of the body (see also Section 16). These reactive behaviours are of particular interest for small-scale soft robots, which must have a simple, electronics-free architecture and can make use of reactive materials to achieve some level of autonomy [5].

Powering soft robots could also prove particularly challenging, especially if untethered operation must be achieved. Indeed, almost no soft power sources or batteries exist, and smart material actuators can have peculiar input requirements (e.g., very high voltages). Moreover, in the most widespread case of soft robots actuated by pneumatic artificial muscles, a source of compressed air must be used, which is difficult to integrate on board and is thus often off board.

Finally, the abilities to grow, self-heal and degrade in the environment are pursued to make soft robots even more adaptable, resilient, and sustainable [8].

**Advances in Science and Technology to Meet Challenges**
Enhancing the actuation (or, in general, functional) performance of stimuli-responsive polymeric materials, as well as their practicality, could greatly benefit soft robotics. Higher forces, deformations, or frequencies might be needed to match the performance of natural muscles. Moreover, lower power consumptions/higher efficiencies, as well as extended durability would significantly enhance the practicality of such materials and make them a standard choice as soft robotics components.

Efforts are also needed to improve the integrability of smart polymeric materials' components, by providing practical and standard ways to interface them with the rest of the robot, e.g., with power supplies and controllers. Developing easy-to-assemble, or even plug-and-play, soft robotic components based on such materials could boost their translation from material science to robotic applications.

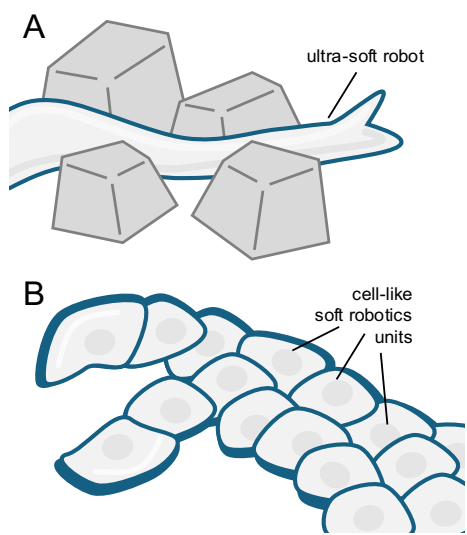

**Figure 10.2 | Future advances in soft robotics.** (**A**) Improvements in responsive materials and new designs can lead to enhanced soft robots able to grow and squish, and to navigate complex environments autonomously. (**B**) Multi-agent soft robots could be developed consisting of basic functional units akin to biological cells.

Improving responsive/smart materials for soft robotics components could indeed enhance (soft) robots. Further benefits could come from new design and fabrication methods, as well as from innovative actuation, sensing, and control approaches, that could lead to even softer (or *ultra-soft*) robots able to dramatically change their body shape, growing or squishing, and thus moving in extremely cluttered environments and narrow openings (**Figure 10.2A**).

Moreover, the field could be revolutionized by the realization of active/autonomous matter units akin to cells – the fundamental units of living beings. These could implement functions such as active deformations and movement, interact with each other and with the external environment, consuming energy already available in the environment or purposely provided. These functional active units should self-assemble and self-organize, while it should be possible for their collective dynamics to be steered or controlled by input signals (**Figure 10.2B**). This will require a major shift in the design, construction, and operation of (soft) robots, with the potential advantage of drastically improving the robots' robustness and adaptability. The units could consist in centimetre-scale electromechanical devices based on standard rigid materials that collectively behave as a soft body robot [9-10], or could themselves be soft and squishy (and perhaps small) as natural cells, possibly consisting of enclosed suspensions of self-propelled particles or other active matter systems (see also Sections 2 and 5).

Finally, responsive and active materials that, in addition to provide functionalities and performance, are not harmful to the environment and could instead degrade safely after the robots' life are highly needed to make (soft) robots sustainable. This will also be necessary to develop biomedical soft microrobots that could operate and then biodegrade inside the human body.

**Concluding Remarks**

Soft robotics has had the merit of providing a new paradigm in robotics, shifting the focus from highly-precisely fabricated and controlled components to the exploitation of the intrinsic compliance of soft materials to achieve high adaptability to the environment. This has required the contribution of roboticists, material scientists and engineers alike. Although soft robotic components (e.g., soft grippers or pneumatic actuators) have become widely used, there is still a long way to go for fully soft robots. These must prove their full potential, with performance and operability better than, or at least on par with, traditional robots. This is likely to happen for specific applications for which soft robots can be better suited than traditional robots. In addition, further evolution of soft robotics based on advances in responsive materials and active matter could lead to completely novel capabilities and open application opportunities that are completely precluded to traditional robots. A specific application field in which a new generation of soft (micro)robots could play a major role is that of minimally invasive medical robotics, which present challenges for the movement of the robot, its intimate interaction with delicate tissues and its end of life.

**Acknowledgements**

We acknowledge funding from the European Research Council (ERC) under the European Union's Horizon 2020 research and innovation programme (project CELLOIDS: Cell-inspired particle-based intelligent microrobots, Grant Agreement No. 948590).

## 11 – Biohybrid robotics
### Ji Min Seok, Victoria A. Webster-Wood
### Carnegie Mellon University

**Status**

Advances in robotic technologies have actively contributed to reducing human labour needs in agriculture, manufacturing, and shipping [1]. New robotic tools have led to advances in medical interventions, including targeted drug delivery and robot-assisted surgery [2]. Cutting-edge mobile robots are being tested for applications in search-and-rescue and reconnaissance [3]. However, there are still many challenges in developing soft, safe robots with the adaptability, behavioural flexibility, and robustness seen in animals (see also Section 10). To overcome these problems, a growing area of research is investigating the concept of biohybrid robots, which are robotic systems that combine living organisms or biological cells and tissues with synthetic components (**Figure 11.1**). Biohybrid robots are composed of a synergistic combination of organic and synthetic systems, with the organic components often taking the role of structure, actuator, sensor, or control [4,5,6].

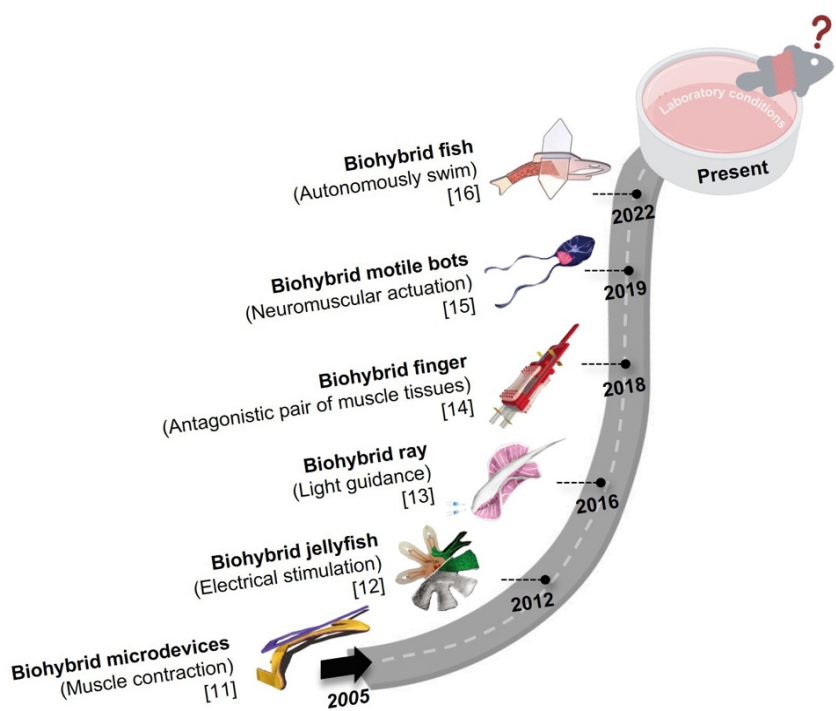

**Figure 11.1 | Time evolution of biohybrid robots**. Schematic overview of a brief history of Biohybrid Robotics. Adapted in part from open access reference [10] under CC-BY 4.0, and with permission from primary sources: biohybrid microdevices reproduced with permission from [11], Copyright © 2005, Springer Nature; biohybrid jellyfish reproduced with permission from [12], Copyright © 2012, Springer Nature; biohybrid ray from [13], reprinted with permission from AAAS, Copyright ©2016, AAAS; biohybrid finger from [14], reprinted with permission from AAAS, Copyright © 2018, AAAS; biohybrid motile bots reproduced from [15] under CC-BY-NC-ND with authors' permission; and biohybrid fish from [16], reprinted with permission from AAAS, Copyright © 2022, AAAS. Additional graphics were created using Biorender.com.

Representative biohybrid robots are mainly categorized into micro-organisms, tissue-based biohybrid robots, and cyborgs [5]. Although biohybrid robotics shows great promise for creating future robotic solutions that are adaptable, sustainable, soft, and biocompatible, many challenges to applying these approaches beyond basic research remain. Each category of biohybrid robots faces its unique challenges. In this roadmap, we will discuss the current and future challenges and directions to overcome the current limitations in tissue-based biohybrid robotics.

### Current and Future Challenges

There are a few challenges worth mentioning in the development of biohybrid robots:

- **Fabrication:** Current fabrication approaches for biohybrid robots can be categorized as top-down and bottom-up methods [7]. In top-down methods, natural tissues are explanted and used as subunits of the robot with the advantage of maintaining the mature tissue hierarchy observed in the source animal. However, it is challenging to maintain tissue functions after explantation from the animal, to resize the pre-designed structure, and to connect the tissues to the robots robustly. In contrast, bottom-up methods start by cultivating small organic units, which can be designed to fit the desired function and shape of the robot. Commonly, flexible elastomers and hydrogels are used as the body, and the biohybrid robot is co-assembled with cast or 3D-printed tissue constructs for the biological component [8]. The biological cells self-assemble based on their surrounding environment to form the final tissue. Therefore, bottom-up approaches allow embedding functional organic components more precisely on the biohybrid robots than top-down methods. However, manufacturing reproducibility, and, therefore, functional performance reproducibility, is difficult to achieve when fabricating systems with biological components, which are apt to remodel surrounding hydrogels and adapt or change in response to environmental cues.

- **Sense-Plan-Act:** True robots exhibit a sense-plan-act cycle in which they sense the world around them, plan motions and tasks, and act in response. Only recently have tissue-based biohybrid robots begun achieving sensing and basic motion control [9]. However, their current capabilities in this realm still fall short of traditional robotic expectations. Therefore, integrated sensing, motion planning, and actuation are important focus areas for enhancing the function of biohybrid robots. Sensing, whether of internal states or external features, remains a substantial challenge in biohybrid robots due to the relative sizes of sensors needed to detect relevant cues and the small payload capacity of existing biohybrid actuators [6]. In addition, sensors that require complex signal processing and transduction are challenging to package appropriately for integration with other biohybrid components. Using biological components directly as sensors, such as optogenetic sensing of light, is one way to overcome these challenges, yet it restricts the types of cells that can be used in the robot due to the need for genetic tools. For autonomous biohybrid robots, motion planning is needed for behaviours more complex than feed-forward walking, gripping, swimming, and pumping. However, controlling cell-based actuation is challenging due to stochasticity in the response and structure of the actuators. To date, external electrical or optical triggers still activate most biohybrid motions [10], which limits translation beyond the lab. Much of the research in tissue-based biohybrid robots has focused on actuation. While small and soft robots have been demonstrated using muscle-based actuators, a key challenge is that the low structural integrity of the muscle tissue can lead to structural instabilities and even tearing of the actuators. This mechanical limitation, combined with low force output, limits the design space available for current biohybrid tissue-based robots.

- **Life support:** The ultimate goal of biohybrid robotics is to perform tasks outside the scope of what traditional synthetic robots can achieve or to improve adaptability, energy efficiency, or sustainability [5]. However, sustaining many of the components within biohybrid robots as they move in the real world is difficult. The highly specific temperature requirements, need for ambient sugars for energy, and need for removal of accumulated wastes, all while maintaining sterility, are substantial challenges facing the field. In addition, it is difficult to accurately check the status of the biohybrid robot during deployment to assess tissue health, making controller design challenging as controllers must adapt to actuator fatigue or damage and manage energy and waste needs.

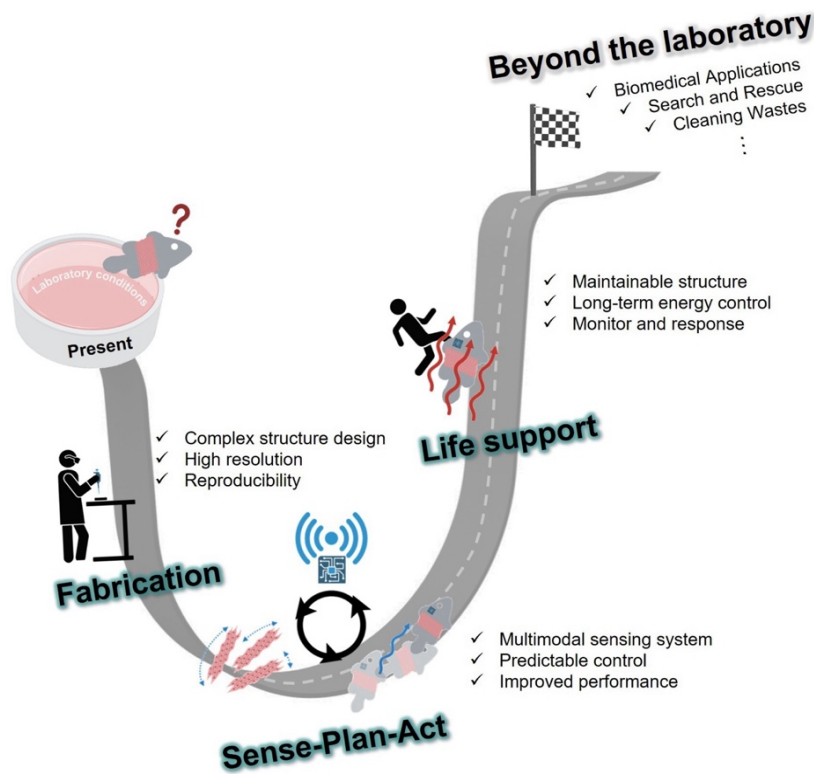

**Figure 11.2 | The future of biohybrid robots**. Roadmap of future research areas needed to overcome existing challenges in translating biohybrid robots beyond the lab. This figure was created in part using Biorender.com.

**Advances in Science and Technology to Meet Challenges**

In creating biohybrid devices, achieving high fabrication resolution and reproducibility is important. Reproducibility of biohybrid robots can be improved by developing platforms for quality control at the fabrication stage, which will be critical for expanding the design and realization of complex biohybrid robots. At the sensing level, new approaches are needed to integrate multimodal sensing into biohybrid robots. Such systems should enable internal state sensing and fusion of multiple external sensing modalities. However, such capabilities will need to be light or biological to not overload bioactuator payload capabilities. Integrated motion planning will require energy-efficient algorithms and lightweight computational platforms [5]. One area for future research is to leverage biological neural networks, or brain-inspired algorithms, to mimic the behaviours and capabilities found in natural systems (see also Section 18). Such approaches may enable self-contained systems that can control more complex movements. To generate autonomous behaviour, once motion planning is performed, that signal must be transmitted to actuators. Although actuation has seen the largest focus in tissue-based biohybrid robotics, substantial research is still needed to achieve autonomy. First, research should focus on improving the robustness and reliability of biohybrid muscle-based actuators for predictable control. Second, the interfaces between bioactuators and the surrounding robot structure must be improved to eliminate stress concentrations. This research can leverage ongoing research in myotendinous repair. Finally, the development of future bioactuators need not be limited by what nature has achieved. For example, novel multi-material muscles could combine conductive materials with cells to improve actuator performance or signal transmission across actuators, or new approaches in cellular engineering could be leveraged to enhance the force capabilities of individual cells.

To move beyond the laboratory, a biohybrid robot must maintain its structure and function, even in non-ideal conditions. There is a clear need for future research to focus on energy harvesting and regulation so that biohybrid robots can explore natural environments over long-term missions. Doing so will also require the development of protective components for maintaining living tissues. The development of novel meta-materials optimized to provide the correct mechanical properties, protection from the elements, energy harvesting, nutrition, or even oxygenation offers promising directions to help biohybrid robots move beyond the laboratory. Finally, the ability of biohybrid robots to monitor and respond to changes in internal state will be critical for complex behaviors. Such capabilities could be used for early-warning systems for hazard detection or to extend device longevity.

## Concluding Remarks

Biohybrid robotics research is growing tremendously across the biomedical, ecological, environmental, and engineering fields. A broad array of current research has focused on biohybrid robots' structural fabrication and functional control. However, several key challenges remain to allow biohybrid robots to move beyond the lab (**Figure 11.2**), including: (i) the realization of heterogeneous biohybrid structures through reproducible and predictable fabrication methods with a high resolution; (ii) complex sensing, motion planning, and actuation using biologically derived materials; (iii) using autonomous cellular systems in complex environments beyond the laboratory. To solve these challenges, continuous interdisciplinary research, multidisciplinary training programs, and collaborative research support are critically needed.

## Acknowledgments

This work was sponsored in part by the National Science Foundation under CAREER award no. 2044785 and by the Army Research Office under Cooperative Agreement Number W911NF-23-2-0138. The views and conclusions contained in this document are those of the authors and should not be interpreted as representing the official policies, either expressed or implied, of the Army Research Office or the U.S. Government. The U.S. Government is authorized to reproduce and distribute reprints for Government purposes notwithstanding any copyright notation herein.

## 12 – 4D printed systems

Shuhong Wang

Morphing Matter Lab, Mechanical Engineering, UC Berkeley, Zhejiang University

Lining Yao

Morphing Matter Lab, Mechanical Engineering, UC Berkeley

### Status

4D printing has been attracting increasing attention since the concept was proposed in 2013 [1]. 4D printing is based on 3D printing technology but involves transformation over time in response to additional stimuli. Initially, it was defined as "4D printing = 3D printing + time" [3]. However, the concept has undergone evolution in recent years: the shape, structure, properties, and functionality of a 3D-printed object can change after being triggered by external stimuli such as temperature, light, pressure, pH, and water.

By designing and preprograming the structure of smart materials, 4D printing allows animate materials to be 3D printed [4] including various shape-changing behaviours, such as bending, folding, twisting, and surface curling [4] (see also Sections 8 and 9). Sequential shape-changing behaviours have also been extensively studied to explore more complex behaviours and functions. These dynamic changes allow the printing of complex and responsive animate materials that can adapt to environmental conditions or user requirements [3].

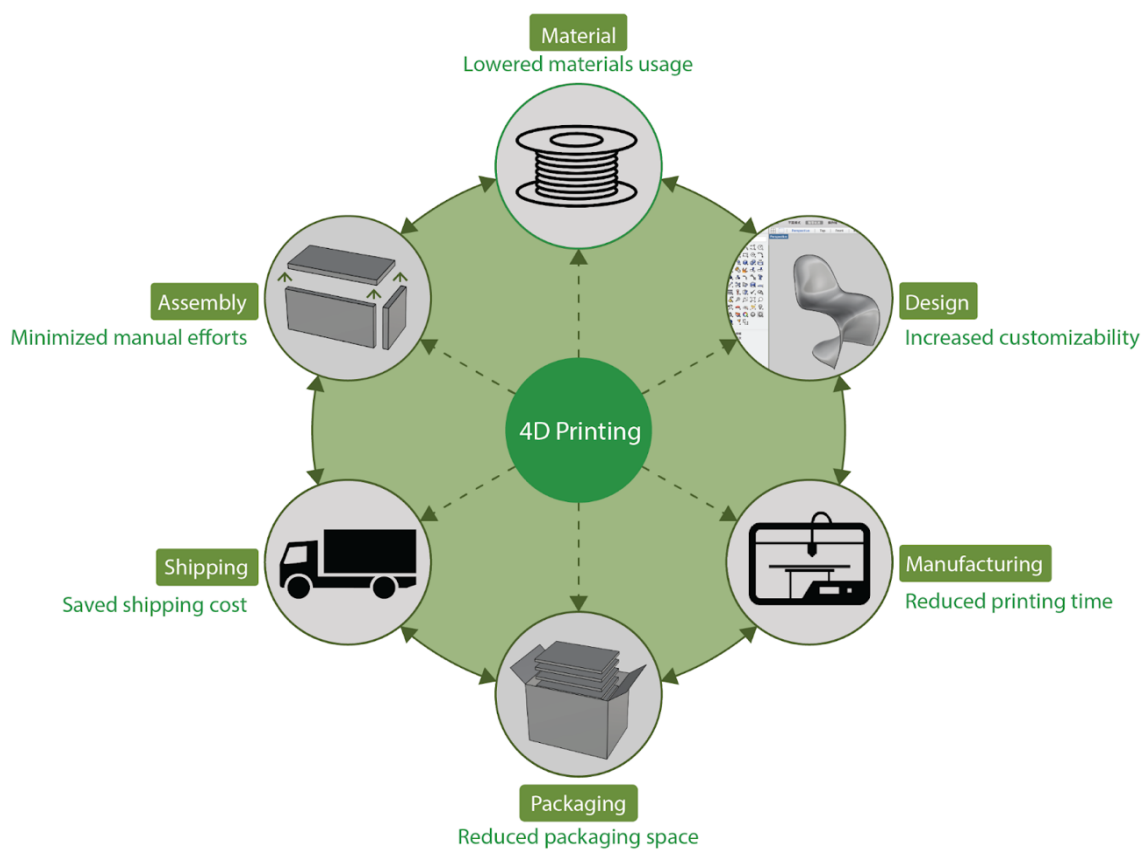

**Figure 12.1 | A schematic envisioning a green and circular manufacturing pipeline enabled by 4D printing**. 4D printing can be used to save costs and time, increase customizability, reduce energy consumption, minimize manual effort, and improve efficiency throughout the entire life cycle of products, including material selection, design, manufacturing, packaging, shipping, and assembly. Initial artwork credited to Guanyun Wang.

The transformative properties of 4D printed materials, including self-assembly, self-adaptability, multi-functionality, and self-repair, exceed those of traditional 3D printed materials [8]. 4D printing has the potential to revolutionize traditional manufacturing, leading to benefits such as reduced manufacturing & assembly, reduced transportation, reduced waste, increased customisation, and more environmentally sustainable products (**Figure 12.1**).

This innovative technology could be utilized to create highly customized and adaptable solutions in various fields such as aerospace, biomedicine, human-computer interaction, soft robotics (see also Sections 10 and 11), and electronics [2]. While 4D printing is still in its early stages as a technology, it is growing rapidly, attracting attention from both interdisciplinary academia and various industries. The technologies of additive manufacturing, stimulus-responsive materials, and related mathematical modelling are under active research and contribute to the growth of 4D printing.

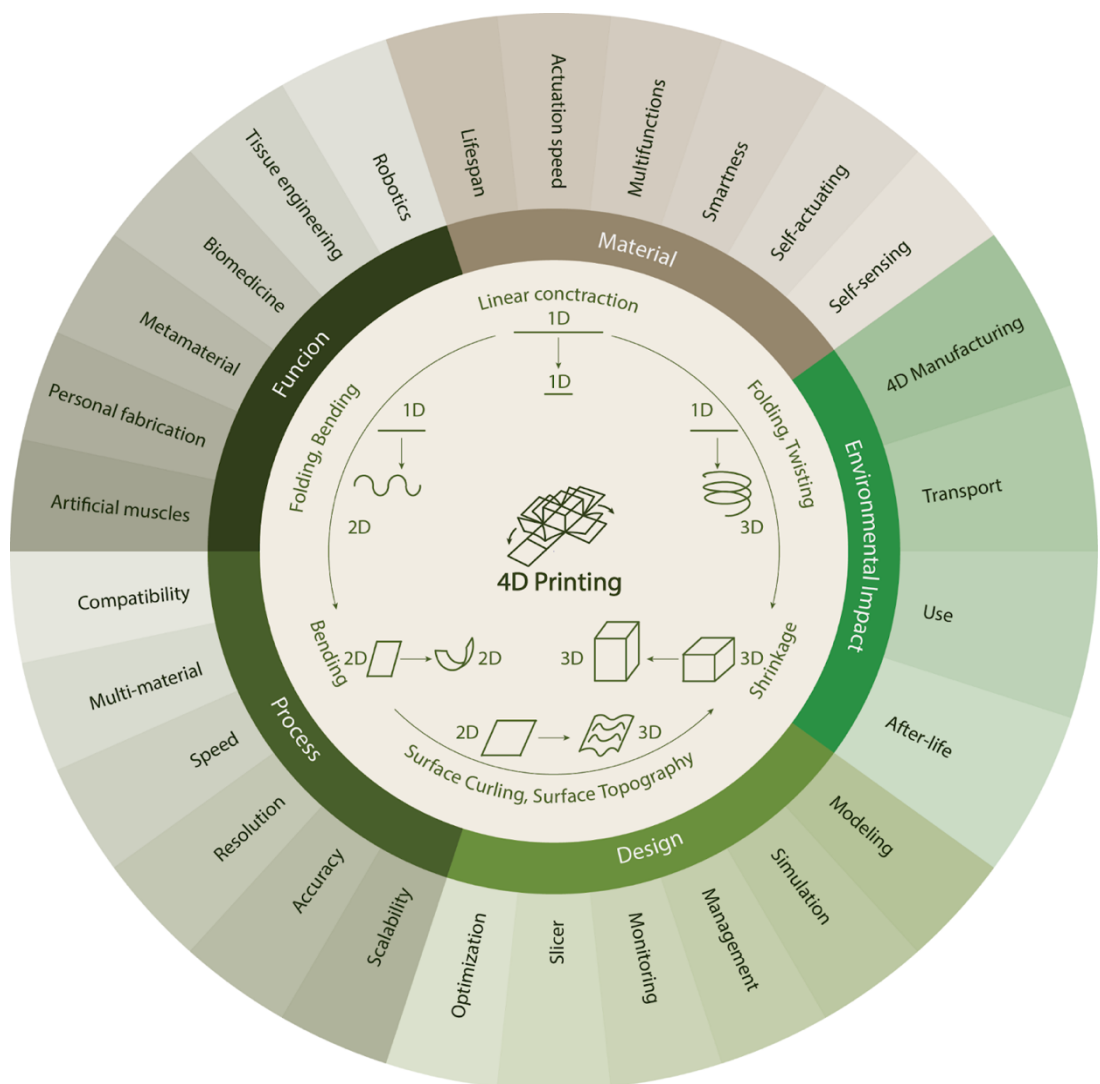

**Figure 12.2 | A schematic showing the important challenges of 4D printing that must be addressed for the technology to reach its full potential**. These cover material, function, process, design space, and environmental impact.

**Current and Future Challenges**
Current 4D printing faces several challenges, including material development [4], process, design, function, and environmental impact (**Figure 12.2**).

- **Material**: The properties of smart materials, such as printability, multifunctionality, multiscale structure, mechanical robustness, actuation speed, and lifespan, all influence their ability to be successfully 4D printed [2]. For most shape-changing polymers, only one-way shape change is allowed. Although there are materials that permit multi-cycle shape changes, most 4D printed structures cannot fully recover their original shapes. Additionally, designing materials that change shape sequentially and respond to multiple stimuli remains a challenging issue.

- **Process**: Material compatibility with 3D printing technology is crucial, requiring directly printable materials that integrate seamlessly with existing technologies. The ability of a printer to accurately and efficiently produce materials with complex structures is essential for the advancement of 4D printing technologies. Regarding 4D printing facilities and technology, factors such as printing accuracy, resolution, speed, multi-material compatibility, and overall material compatibility are crucial [5]. Additionally, current research mostly focuses on layer-by-layer stacking printing. There are still challenges in direct and freeform printing of 3D structures, reducing undesired shape changes during the printing process, and conducting conformal 4D printing on existing non-planar surfaces. Moreover, the portfolio of materials that can be successfully 4D printed is still small.

- **Design**: Developing models and computational design tools increases the efficiency of the 4D printing process compared to manual path planning. The goal of the forward simulation is to describe final shapes from material specifics and stimuli (forward design), while inverse design focuses on deriving print paths or structures from desired outcomes [4]. Simulating the dynamic shape change, often within a multi-physics context, is non-trivial. Moreover, current research falls short in modelling the time-varying behaviour in 4D-printed composite structures. Further development of 4D printing design tools is needed to optimise material properties, printing techniques, and structural considerations. Additionally, the usability of the design tool is important, ensuring that it is accessible for a diverse range of end-users.

- **Function**: Utilizing the advantages of 4D printing techniques, such as self-activation, interactive mechanisms, and customizability, is challenging due to constraints such as triggering time, robustness, and specific stimulus requirements. Furthermore, 4D printing solutions that can cover the full range from the microscale to the human scale are still in their early stages.

- **Environmental impact**: Although crucial, the environmental impact of 4D printing across its life cycle has been insufficiently discussed. Its unique materials, processes, and lifecycle behaviours significantly differ from those of traditional manufacturing methods [9]. However, this also presents opportunities to design a green manufacturing cycle that can lead to greater energy savings compared to conventional manufacturing methods (Figure 12.1). The main hurdles include optimizing manufacturing to reduce waste and energy use, designing for efficient transportation and use, and ensuring end-of-life strategies such as recyclability and biodegradability [9].

**Advances in Science and Technology to Meet Challenges**

The following five advances are needed to develop 4D printing into a technique that can be used reliably to produce animate materials.

- **Material**: Recent studies concentrate on the printing of both single and multiple shape-changing materials as well as composites [6]. Shape memory polymers (SMPs), hydrogels, actuating materials such as liquid crystal elastomers, and composite materials such as magnetic elastomer composites are commonly utilized materials in the field of 4D printing [8]. Multimaterial 3D and 4D printing, utilizing a variety of materials such as polymers, metals, ceramics, and biomaterials, enable shape programmability and enhances part quality by altering materials between layers, thereby surpassing traditional manufacturing techniques [7]. Developing materials with varied properties, durability, and rapid response capabilities will enable the creation of customised objects to meet diverse needs of product manufacturers and withstand various environments. Meanwhile, multi-stimuli responsive and self-computing materials are desirable for making multifunctional objects.

- **Process**: In recent years, various manufacturing methods and facilities have been utilized and further developed to enhance 4D printing, focusing on improvements in printing speed, resolution, and greater material compatibility. These methods encompass extrusion-based techniques such as Fused Deposition Modelling (FDM) as well as inkjet and Direct Ink Writing (DIW) [5] and photopolymerization methods such as Stereolithography (SLA) and Digital Light Processing (DLP). Advanced machines that combine multiple printing parameters are crucial for making heterogeneous printed structures [8]. Furthermore, the customized functions of 4D printed objects can be enhanced by utilizing various printing methods to print multiple material systems. Additionally, 4D printing can be extended to 4D manufacturing, incorporating not only printing process but also diverse manufacturing techniques that can manufacture flat objects which can consequently morph into 3D with additional energy stimuli [9]. Other 4D manufacturing techniques include laser cutting, 2D jetting, layer lamination, etc.

- **Design**: Advancements in predictive methodologies and simulation capabilities are paving the way for the automated design of 4D printed objects, guided by predefined material compositions, attributes, and shape models. Various modelling methods, including simplified geometrical models, the spring-mass system, the 1D elastic rod, the multi-physics finite element model, and data-driven surrogate models, have been developed [4]. Furthermore, advanced machine learning techniques are being utilized to enhance simulation predictions and fast design tools. In the future, advanced software that can be used with different machines and integrated with various hardware should be developed to improve efficiency and coordination.

- **Function**: Applications of 4D printing, leveraging its self-assembly, self-adaptability, multi-functionality, and self-repair capabilities, have been developed across various interdisciplinary fields. 4D printed products find applications in autonomous robotics, personal fabrication [10], human body systems such as artificial muscles, implantable biomedical micro-devices, and wearable sensors [2]. These applications contribute to increased efficiency, reduced human labour, and decreased costs in real-world settings. Future application development needs the convergent mindset of engineering, science, and design thinking. It's crucial to identify the true needs in real-world settings where 4D printed objects can outperform current methods.

- **Environmental impact**: Holistic design guidelines are emerging [9]. These guidelines aim to provide a unified workflow for considering sustainable design practices, addressing the environmental impacts of morphing matter throughout its lifecycle, including 4D manufacturing, transport use, and end-of-life. For example, renewable resources should be used while considering the products' desired functions [9]. Moreover, advances in science and technology, like digital fabrication and AI-based computational design, can improve 4D printing efficiency by optimizing energy consumption and design processes.

**Concluding Remarks**

4D printing, an evolution of 3D printing, introduces dynamic, time-responsive elements to objects, allowing them to change shape, properties, and functionality in response to stimuli. This innovative technology promises significant advancements in fields like aerospace, biomedicine, and robotics by offering customizable, adaptable solutions. However, challenges such as material development, process optimization, design complexity, functional application, and environmental impact remain. Addressing these requires advancements in smart materials, printing technologies, mathematical modeling, and sustainable practices. Despite these hurdles, the potential for self-assembling, adaptive, and multifunctional structures positions 4D printing as a transformative force in manufacturing and design, promising a future where products are more responsive to human needs and environmental sustainability.

## Acknowledgements

The authors acknowledge support from the National Science Foundation Career Grant IIS-2047912 (LY). The authors would like to thank their colleagues at the Morphing Matter Lab for their constant source of insights and inspiration.

# 13 – Active acoustic metamaterials

Amirreza Aghakhani

Institute of Biomaterials and Biomolecular Systems, University of Stuttgart, Pfaffenwaldring 57, 70569 Stuttgart, Germany

**Status**

Acoustic metamaterials have received significant attention since the early 2000s due to advances in materials science, micro- and nanotechnology, and physical modelling. Acoustic metamaterial systems can steer, control, and manipulate sound waves in ways that are not possible with conventional structures. Historically, band gaps in periodically arranged atomic lattices have inspired the development of photonic and phonic crystals, where the lattice constants are on the order of the electromagnetic and acoustic wavelengths, respectively [1]. However, the large wavelengths of acoustic waves in the audible range (centimetres to metres) led to large and bulky phononic designs.

The emerging field of acoustic metamaterials overcame this limitation by allowing lattice sizes much smaller than the acoustic wavelength and expanding their application in acoustic waveguides and acoustic imaging. Since the first experimental realization of acoustic metamaterials, called locally resonant sonic materials [2], several local resonators have been developed, such as Helmholtz-based [3], or membrane-based resonators [4], which provide unconventional, overall bulk modulus and system's mass density. Despite the great progress of these unit-cell resonators, their designs rely on passive and linear systems with fundamentally limited bandwidth, which hinders their applicability in practical time-variant applications.

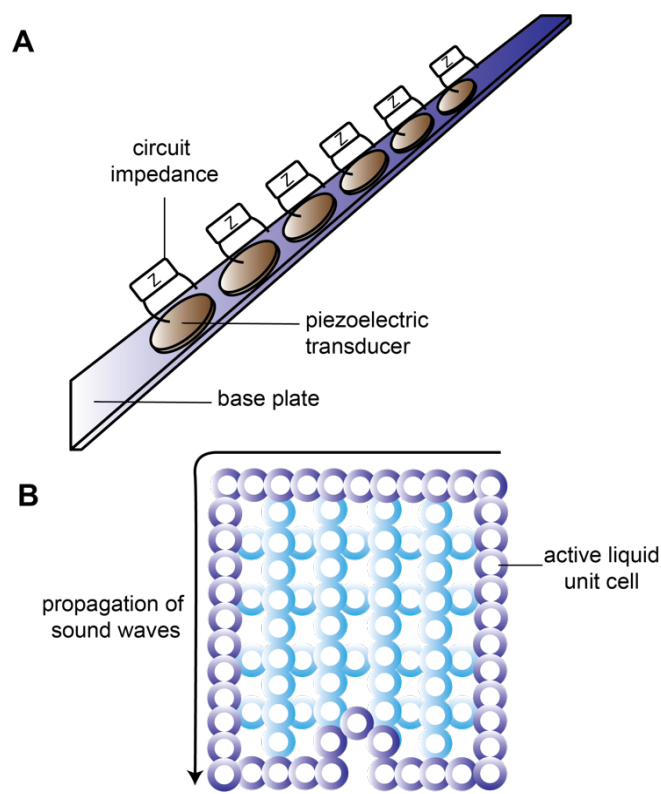

**Figure 13.1 | Schematics of two active acoustic metamaterials**. (**A**) Active acoustic metamaterial using external electric field through piezoelectric unit cells [5]. The piezoelectric local resonators can tune the stiffness of the host structure and tailor the acoustic wave propagation behaviour. (**B**) An animate acoustic metamaterial consisting of active liquid unit cells, enabling topological acoustic property [7]. The liquid unit cells control the propagation of sound and enable topological insulators which are immune to structural defects. The high-density active particles in the unit cells are shown in dark blue and low-density active particles are shown in light blue.

Active acoustic metamaterials (**Figure 13.1**), in contrast to passive acoustic metamaterials, use external control (e.g., electric, light, or magnetic field) or the material's adaptive and active energy absorption properties (e.g., active colloids as animate materials, see Section 2) to tune the effective material properties of the structure, enabling a wide range of on-demand acoustic wave manipulation. The most common active unit cells are electromechanical systems, piezoelectric materials, and electrically loaded loudspeakers [5]. The active elements can dynamically tune their properties in response to changes in the acoustic waves, expanding their applications in rapidly changing noise environments. Another interesting subclass of active acoustic metamaterials is non-reciprocal acoustic metamaterials, which allow unidirectional acoustic propagation and isolation in their medium [6]. By breaking time-reversal symmetry, non-reciprocal acoustic metamaterials allow unique unidirectional propagation of sound waves immune to the backscattering caused by structural imperfections and perturbations in the material. As such, they have enabled the creation of acoustic diodes and isolators, topological insulators, one-way mirrors, and circulators, spanning their application in biomedical ultrasound, targeted therapy, architectural acoustics, and vibration control [6]. Regarding real-world applications, active acoustic metamaterials face challenges regarding their practical applicability. Future developments should include new design, fabrication, and control methods, as well as theoretical modelling developments, to improve their practical use.

**Current and Future Challenges**

The development and application of acoustic metamaterials face several significant challenges that impede their transition from laboratory research to practical real-world applications. One of the primary concerns is the fabrication and scalability of these metamaterials, which are crucial for their widespread adoption. However, these metamaterials currently suffer from high costs and complex production methods. Furthermore, the performance of acoustic metamaterials is frequently constrained by their frequency bandwidth. In particular, those that rely on locally resonant unit cells typically exhibit inherently narrow frequency bandwidth properties, which restricts their applicability across various frequency ranges in noisy and broadband acoustic spectra. Another challenge is the passive nature of many acoustic metamaterials, which suffer from a combination of material's viscous energy loss and thermal losses. Therefore, addressing these inherent material losses is a vital area of ongoing research.

Conversely, active acoustic metamaterials offer enhanced dynamic range and tunability across a broad frequency bandwidth. Nevertheless, their practical implementation remains complex, requiring the development of simpler designs to facilitate their use in real-world settings. Moreover, the design and fabrication of specialized forms, such as topological and non-reciprocal acoustic metamaterials, entail intricate processes that significantly increase costs and hinder scalability. The physical dimensions of some acoustic metamaterials also restrict their suitability for compact environments due to their considerable dimensions and bulk, on the order of metres. Integrating these materials into existing conventional systems presents additional challenges, notably impedance matching issues, which complicate their practical deployment. Moreover, translating theoretical and numerical models into experimental setups is a non-trivial task that requires meticulous adjustment and validation.

In the emerging field of topological acoustics – the study of materials and structures that allow sound waves to propagate along their edges and isolate sound inside them – independent control and mechanical biasing of individual lattice structures present significant engineering challenges. The majority of current research in topological acoustics has focused on airborne sound with longer acoustic wavelengths, while studies on sound waves in solids, especially in three-dimensional structures, are still lacking [8]. This is attributed to the ease of airborne sound manipulation and measurement as well as low acoustic energy loss in certain distances compared to the propagation of sound waves in solid medium. However, for biomedical applications, frequencies in the ultrasound range are preferable, indicating smaller footprints of acoustic metamaterials in the range of microns to millimetres. Furthermore, despite the advances in three-dimensional fabrication technologies, most research has concentrated on two-dimensional designs, such as metasurfaces, likely due to their relative ease of production. Moving forward, a shift towards more extensive research on three-dimensional metamaterial structures, enabling the control of sound propagation at different directions, is crucial to fully leverage the capabilities of acoustic metamaterials in diverse biomedical and environmental applications.

## Advances in Science and Technology to Meet Challenges

The recent advancements in three-dimensional (3D) additive manufacturing have significantly facilitated the fabrication of intricate three-dimensional architectures for acoustic metamaterials. This technological progress has facilitated the integration of theoretical and computational designs with experimental applications, thereby enhancing the translation of complex models into tangible, functional materials. As the ease of use of 3D printing techniques continues to improve, there is a growing necessity for an increased focus on 3D metamaterials research over traditional 2D metasurfaces. This could further expand the potential applications and effectiveness of these materials. To address the inherent limitations of the materials used, current research efforts are focused on refining the unit cell designs in order to minimize losses, thereby maximizing sound absorption and reducing dissipation and damping. This entails the investigation of novel materials and structures with low-loss profiles, which could markedly enhance the functionality of acoustic metamaterials.

Furthermore, the field of active materials represents an intriguing avenue for the advancement of acoustic metamaterials. These materials employ active control mechanisms, such as electromechanical transducers, to facilitate adaptive and reconfigurable properties within the structures, allowing for precise control of mechanical deformations triggered by electrical energy inputs. This approach is exemplified by research on animate matter, where the propagation of topologically protected sound modes in metamaterials made from active liquids—fluids comprising self-propelled particles (see also Section 2)— was explored [7]. Such animate materials can support unidirectional sound waves and establish topological order, enabling sound to travel without backscattering, even in the presence of obstacles or disorder.

Finally, there is a shift in material research towards the design of solid acoustic metamaterials within fluidic media, such as underwater environments, where acoustic contrast is less pronounced. One such innovation is the use of air bubbles as locally resonant unit cells to create promising underwater acoustic metamaterials suitable for medical ultrasound applications [9]. The direct applications of these materials are being explored in fields such as transdermal drug delivery. In this context, pyramidal lattice structures have been used to create localized acoustic streaming around sharp edges, thus enhancing the diffusion of transdermal drugs [10]. These developments illustrate the transformative potential of active acoustic metamaterials in a variety of practical applications.

## Concluding Remarks

In conclusion, acoustic metamaterials stand at the forefront of innovative materials science, offering unprecedented capabilities in sound wave manipulation through their unique design and composition. The development of these materials has progressed from passive, narrowband systems to more dynamic, actively controlled structures, and animate materials that can actively absorb energy and adapt to various acoustic environments. These advancements are largely attributable to significant breakthroughs in micro- and nanotechnology, as well as theoretical and computational modelling. Despite these strides, challenges in scalability, fabrication costs, and integration into existing systems persist, limiting their broader application. Looking ahead, the focus is shifting towards refining 3D additive manufacturing techniques to enhance the practical deployment of these materials. Innovations in unit-cell design and material properties aim to reduce losses and increase functionality, facilitating their use across a wider array of applications, from architectural acoustics to biomedical ultrasound. Additionally, the exploration of non-reciprocal and topologically protected acoustic waves promises new dimensions in sound control, potentially revolutionizing how active acoustic metamaterials are utilized in everyday environments. The ongoing research and development in this field are poised to not only overcome existing limitations but also to unlock a myriad of new possibilities for acoustic manipulation and application.

## Acknowledgments

-

## 14 – Active granular matter

Thomas Barois, Hamid Kellay

University of Bordeaux, CNRS, LOMA, UMR 5798, F-33400 Talence, France

**Status**

Granular matter denotes a very large ensemble of similar small elements that are however not small enough to be ruled by Brownian motion and equilibrium statistical mechanics. In active granular matter, each granular element of the ensemble can convert some input energy into motion in which friction and shocks are the main sources of dissipation. The development of active granular matter can be traced back to the 2000's with first active grains powered by vertical vibrations [1,2] and then individually powered granular robots [3]. The dominant studied situation was directional motion, which relies on various self-propelling mechanisms for granular particles with some head-tail asymmetry. Because of aligning interactions, assemblies of self-propelled grains typically show transitions between disordered gas-like states at low densities and ordered collective states at higher densities.

In the context of animate matter, the specificity of active granular matter is the simplicity of the granular elements that contrasts with the complexity of the collective regimes observed. If we put aside the active mechanism, granular particles are easy to find in nature and laboratories. The granular interaction phenomenology at the macroscale is however not so simple due to the non-smooth properties of contact interaction. First, the collisions are very short-time events with sudden change in the velocity of the particles. Second, friction between grains or walls can induce transitions between sticking and sliding regimes.

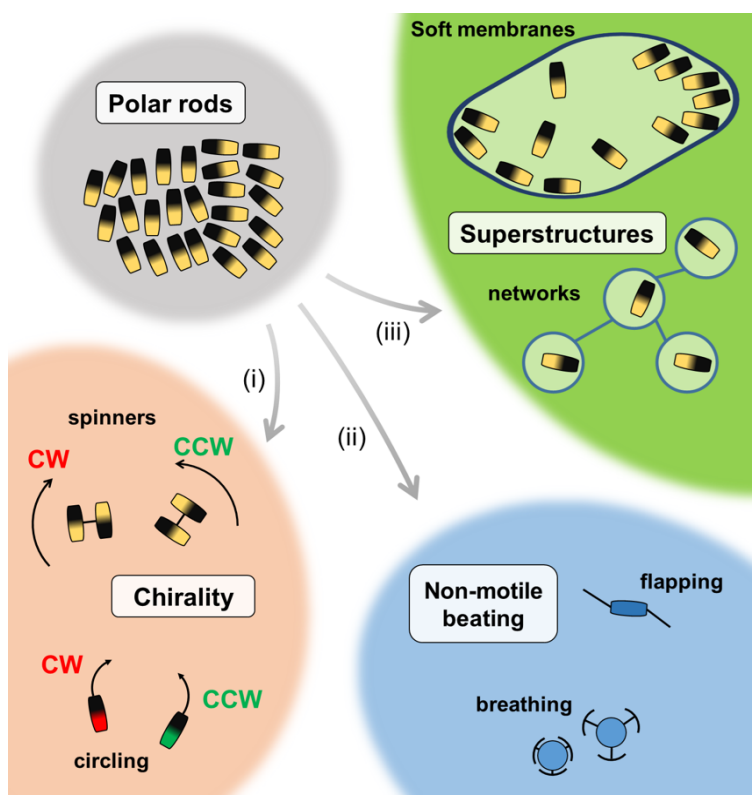

**Figure 14.1 | Perspective for active granular matter as animate matter.** At first, active was mostly referring to self-propelled elements such as polar rods and disks with a persistent directional motion. More recently, activity beyond translational motion was studied with (i) rotational activity for chiral active matter and (ii) non-motile active particles with body

deformation decoupled from the centre-of-mass motion for the individual particles. Structures using granular active elements as building blocks are also proposed (iii) to form more complex animate objects.

Because the granular interactions are repulsive, the collective states typically appear and persist for sufficiently large densities under the influence of steric effects. Elongated particles are more likely to have strong aligning interactions but, even with self-propelled particles of circular shapes, the interplay between collisions and friction of the moving particles also leads to velocity alignment and flocking.

The interaction of active granular particles with walls is an important aspect of the individual and collective particles' dynamics. Because of the aligning interaction, and depending on confining geometries, self-propelled particles can be guided, jammed or trapped. The collisions with boundaries can also enable the formation of clusters of particles even for moderate densities.

**Current and Future Challenges**

Granular matter offers the possibility to explore activity beyond linear motion – a topic of great interest with further potential for the future. Recently, chirality has attracted much attention in active granular matter [4] (**Figure 14.1**). A chiral active particle can be a spinner, i.e. a particle self-propelled into a rotating motion or a particle moving on a circular trajectory, which can be viewed as a combination of directional motion and body rotation. Rotation necessarily implies the existence of two-particle types, clockwise (CW) and counter-clockwise (CCW). Consequently, chirality brings a new set of specific pair interactions CW-CW, CCW-CCW, and CW-CCW. This leads to situations of segregation with particle clusters only composed, or dominantly composed, of particles of type either CW or CCW. The interaction of chiral particles with walls is also type-specific, which leads to edge currents with a parallel transport at the walls depending on the sign of the rotation. This surface flow can be used to sort particles depending on their sign of rotation. Surface flow is also observed at the free surface of chiral particle ensembles and the peculiar dynamics of interacting rotating particles leads to exotic phases with odd properties [5].

Another illustration of novel type of individual motion is the volumetric oscillation of disk-like particles with T-shaped arms [6] (Fig. 14.1). In such systems, the activity of the individual elements has no action on their centre of mass. Collectively, however, the particles can influence each other and self-organize into static or dynamic clusters. In this situation, one of the new physical ingredients is the possibility to choose the relative phase of the oscillating particles.

A last approach consists of embedding active granular particles into more complex structures to form compound objects with peculiar dynamical properties (Fig. 14.1). With active particles placed inside a flexible closed membrane [7], the coupling between the particles' motion and the membrane deformation leads to a cell-like dynamics of the membrane with locomotion, large deformation, and exploration of confined environments. Active particles are also used to realize active elastic crystals and therefore explore the effect of activity on mode dynamics in periodic structures. Finally, it is possible to combine the approaches of activity beyond directional motion and superstructures [8]: flapping particles composed of three articulated segments placed in a circular enclosure form an entity capable of diffusive motion under the collective contact interaction of the active flapping particles. If the activity is made dependent of a light intensity threshold, the superstructure demonstrates phototaxis, which is here a rectified motion towards a light source.

Another question with active granular systems is the role of inertial regimes in the emergence of collective states. In the Vicsek model, the dynamics of the particles' orientation is of first order, which means that two particles close to each other will align in an overdamped regime. For experimental physical systems, this overdamped regime is valid for, e.g., microswimmers with relatively large viscous forces (see, e.g., Section 2). With dry granular elements, there are almost no viscous forces at play and the inertial regime dominates. Inertia can be responsible for non-aligning collisions of elongated particles, which prevents the formation of aligned phases in the absence of walls. Inertial regimes can also lead to unusual thermodynamics [9].

The future challenges concern both the control of the trajectory, the different types of motion, and the configurational dynamics of the particles. So far only very few examples have been considered and it is not clear how to optimize different collective behaviours by modifying a simple property whether for the trajectory or the configurational change. Further, as encapsulating particles into scaffolds opens new possibilities, the challenge is again how to optimize the behaviour and function of superstructures by designing intelligent scaffolds and new particles with well-adapted properties.

**Advances in Science and Technology to Meet Challenges**

The field of active granular matter is going through a diversification phase with new types of assemblies performing new collective tasks. In Fig. 14.1, we have sketched an overview of recent strategies implemented in active granular matter. By modifying either the particles' activity properties or environment, new types of functions and collective phases can be observed.

A direct challenge is not only how to implement different types of motions or configurational changes but how they can be optimized to generate a particular function or behaviour. The advent of 3D printing is certainly a step forward, however, optimization in view of a task or function is not straightforward and may need accompanying simulations and numerical work. Using enclosures can benefit from the advent of new materials with, for example, sensitivity to local stresses or forces so that one can generate more deformation or resist deformation depending on the forces acting on the enclosures; this may give rise to possible feedback loops between collective formation near the walls and the scaffold itself.

The essence of granular systems is the simple structure of their constitutive elements. This is both a strength and a limitation in the possible achievements of active granular systems. The recent results in the field of active granular matter have shown that it is possible to foresee active granular schemes as part of complex animate entities as it is for robotic systems (see, e.g. Section 17). In robotics, the design of a specific robot is usually via reverse engineering approaches in which the structure of the robot is designed from the task it is expected to perform. In the field of granular systems, the research approach is via exploration with usually no specifically targeted function. The challenge is to bridge the two approaches and be able to gauge when active granular elements are interesting compared to conventional approaches in robotics with higher levels of programming. Introducing some form of elementary communication between granular agents is a way to explore new collective properties: introducing some simple communication between agents in collision-dominated crowds can for example lead to collective learning schemes [10].

**Concluding Remarks**

Active granular matter is a surprisingly rich domain. Although granular particles are simple entities, making them active, using different schemes, reveals the complexity of granular interactions. This activity can consist of simple locomotion, rotation, or simply shape shifting. These different modes give rise to a multiplicity of behaviours. This complexity opens the way towards the emergence of functional entities capable of deformation, complex exploration, foraging, etc. Besides, a variety of collective dynamics emerge, going from cluster formation, to flocking, to interfacial flows. Finally, adding a layer of basic communication between agents can give rise to learning crowds and even enhanced functionalities: while improved phototaxis has been demonstrated, other tasks, e.g. cargo transport or searching, still needs testing with learning strategies.

**Acknowledgements**

-

# 15 – Active mechanical metamaterials

Corentin Coulais

University of Amsterdam, Institute of Physics, Science Park 904, 1098 XH, Amsterdam, the Netherlands

Martin van Hecke

Huygens-Kamerlingh Onnes Laboratory, Universiteit Leiden, PO Box 9504, Leiden, 2300 RA, the Netherlands

AMOLF, Science Park 104, 1098 XG, Amsterdam, the Netherlands

### Status

Animate materials interact with their environment in a dynamical and adaptive manner. This entails the combined ability to (1) change shape, (2) do work and (3) process information. Over the last few years, mechanical metamaterials that exhibit these three abilities are starting to appear. We consider each in turn.

First, a range of design methods have been developed for realizing metamaterials that morph between arbitrarily complex shapes [1]. These methods include spatially grading the material such that the local kinematics add up to a shape-change over many unit cells, using combinatorial techniques to control the local kinematics and texture, or exploiting geometrical frustration to obtain large out-of-plane deformations. Moreover, the building blocks of these metamaterials can be either monostable, leading to a single shape, or multi-stable, leading to the integration of many shapes and memory [2,3].

Second, while most active matter systems to date fall into the class of active fluids that lack a reference state, recently active solids have emerged where elastic restoring forces to a reference state compete with active forces [4,5,6]. These materials consist of discrete units – thus falling in the class of active metamaterials – and their response typically fundamentally differs from that of ordinary elastic solids, viz. whereas ordinary passive solids are described by symmetric dynamical matrices and elastic tensors, activity can turn dynamical matrices and elastic tensors asymmetric [4,5]. As a result, when the active driving components are sufficiently strong, geometric feedback between deformations and direction of forcing may lead to active waves, synchronized oscillations, and other complex spatiotemporal patterns [6]. These may allow the active metamaterial to do work on its environment and to locomote autonomously [4].

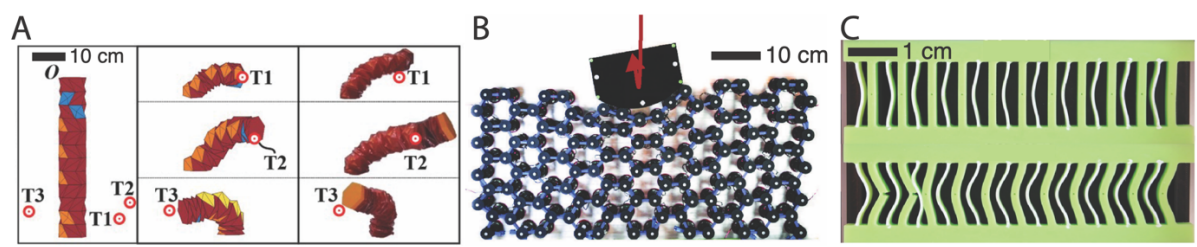

**Figure 15.1 | Examples of active mechanical metamaterials**. (**A**) Example of a shape-changing metamaterial made from an origami unit cell. The metamaterial consists of a pneumatically actuated origami, which is only actuated via a single pressure source. It can bend and twist such that the top of the column can reach points T1, T2 and T3 only by adjusting the history of the pressure loading. Left panel: metamaterial at rest; middle panel: simulations; right panel: experiments. Reproduced with permission from [11], Copyright © 2022, Wiley-VCH GmbH. (**B**) Example of an active metamaterial deflecting a projectile. The metamaterial consists of 120 robotic hexagons made from building blocks that break reciprocity. The metamaterial as a whole is described by a generalization of elasticity called odd elasticity. Odd elasticity leads to non-reciprocal coupling between the two shear modes of the material when perturbed by the projectile and, in turn, to deflect the projectile [4]. (**C**) Example of a multistable metamaterial performing a counting task under cyclic compression. The metamaterial consists of a

series of thin and thick beams. The thick beams are designed to be asymmetric: they buckle initially towards the left, but once nudged by the thin beams, they snap towards the right, in turn making the adjacent thin beam on their right snap. The metamaterial at the smallest (top panel) and largest (bottom panel) compression of cycle 2 [9].

Finally, the geometric nonlinearities of mechanical metamaterials have recently been explored to realize mechanical logic, storage of information, and elementary information processing *in materia* [7,8,9,10]. Crucially, the input and output here are via specific physical channels, distinct from the symbolic, general-purpose functionality of silicon-based computing. The simplest form of information processing has no memory, but directly translates (a number of) inputs to one or more outputs, with Boolean logic the paradigmatic example; such logic gates have now been realized in metamaterials by using buckling and snapping in, e.g., origami and slender elements [7]. More advanced forms of information processing require memory, in which the output depends on both the current input and the internal state, and where the latter stores information on past input and computations. This requires developing strategies to write, erase, and read information, and to design multistable metamaterials [2,3,8,10].

**Current and Future Challenges**

Shape-changing, activity and computation have thus far been achieved in separate platforms. Yet, to achieve animacy, they would need to be integrated in a single metamaterial. We first discuss what we see as the key challenges in each field, and then present a future strategy – based on realizing non-reciprocal cycles in active, multistable metamaterials – to integrate these key elements of animacy.

*Shape-changing metamaterials* (**Figure 15.1A**) with a single shape change are by now well established. The next challenges lie in the design of multiple on-demand shapes into a single metamaterial and in the controlled actuation of such metamaterials. There is an inherent trade-off between freedom in the shape-change and complexity of actuation: metamaterials with a single shape-change are easy to robustly actuate, but their single shape can be limiting, whereas metamaterials with many shape-changes can be more versatile but require more advanced actuation.

One key challenge for *active metamaterial* (**Figure 15.1B**) is to do useful work on their environment in an adaptive fashion: think of an active metamaterial that can autonomously roll or crawl depending on the terrain. We see several important questions:

- We need efficient design methods for active metamaterials that combine the flexibility to efficiently do work on their environment with the rigidity to carry a load.
- We need robust and scalable approaches to embed activity within a flexible metamaterial. One strategy may be to leverage rapid advances in materials such as liquid crystal elastomers, graphene actuators or shape memory alloys, and use their responsiveness to continuously harvest energy from the environment and convert this energy into purposeful work.
- We need strategies to control the behaviour of active metamaterials - as their dynamics are inherently far-from-equilibrium and nonlinear, they are hard to predict and to design.

*Computing metamaterials* (**Figure 15.1C**) should be able to perform sequential computational tasks. Such materials materialize finite state machines, the paradigm of computing that describes finite sequential computations. Strategies that use a distribution of bistable or hysteretic material bits - realized by, e.g., buckling and snapping beams or bistable origami elements - form an attractive route towards controllable and scalable multistability [8,9,10]. The first examples of metamaterials capable of sequential computations (specifically, counting and recognizing a sequential sequence of inputs) are appearing [8,9,10]. The complexity of mechanical metamaterials also allows other forms of *in-* *materia* computing, inspired by artificial neural networks, reservoir computing and neuromorphic computing.

**Advances in Science and Technology to Meet Challenges**

Realizing the next generation of active metamaterials that integrate autonomous, life-like functions (such as shape changes, locomotion and intelligence) relies on fundamental breakthroughs in each of these fields, as well as integrative work.

Scientifically, there is an enormous open space for advanced design techniques for metamaterials, building on recent advances in our understanding of geometric nonlinearities, topology optimization, and combinatorial design as well as integrating recent advances in machine learning and AI (see Section 18). In addition, distributed control strategies for such advanced metamaterials need to be developed. This may require revising tools of continuum mechanics, dynamical systems and control theory so that they can be applied in active systems, i.e., when reciprocity is broken; alternatively, model-free methods such as physical reinforcement learning may be helpful once they are generalized for multi-agent schemes. Both design and control suffer from the combinatorial explosions of the design and action space; efficient computational and optimization techniques need to be developed to navigate these vastly under-sampled, rugged parameter spaces and discover the rare metamaterial designs that are functional.

Technologically, we see great potential for improving the manufacturing of advanced active metamaterials. 3D printing techniques allow for an ever-increasing palette of materials as well as for increasingly fine resolutions. However, these fall far short of the complexity seen in living systems, or more modestly, in most advanced metamaterial designs; we currently cannot (routinely) print complex metamaterials with $1000^3$ voxels – which would not be unreasonable if one wants a linear dimension of at least 50-unit cells, and each unit cell with a reasonable resolution of $20^3$ voxels. Another crucial challenge is to combine different materials classes, which is needed to combine multiple functions (e.g. combining passive elastomers for deformations and responsive materials for driving them). Finally, integrating chemical, electrical or electromagnetic sources of energy with sufficient power density and practical actuation mechanisms in metamaterials is a daunting challenge.

## Non-reciprocal cycles of shape changes

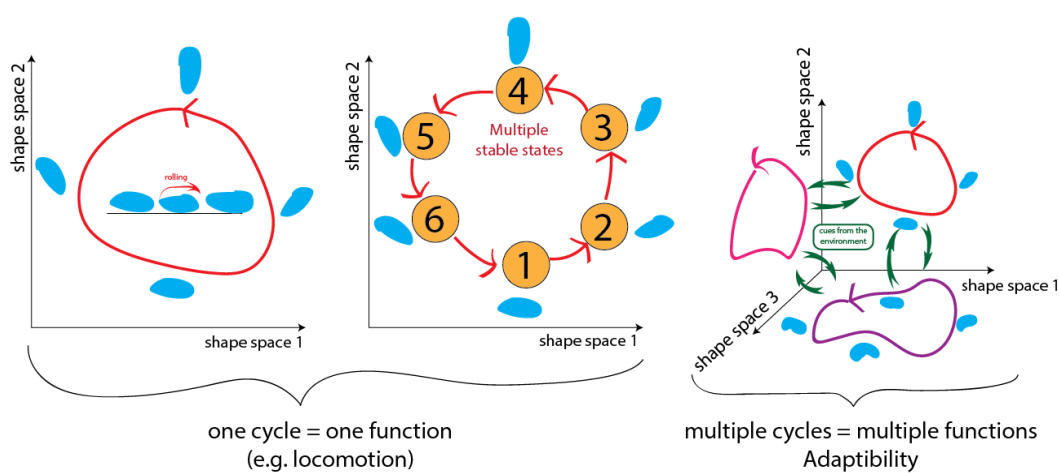

**Figure 15.2 | The notion of non-reciprocal cycles of shape changes**. Non-reciprocal cycles of shape changes through configuration or shape space break time-reversal symmetry and enable energy conversion into work (e.g., locomotion). (Left) In active metamaterials, limit cycles or travelling patterns are often emerging from a Hopf bifurcation. (Middle) Multistable metamaterials can feature periodic pathways through a sequence of metastable states. (Right) An important challenge is to

embed multiple non-reciprocal cycles to integrate multiple functionalities (e.g., multiple modes of locomotion) and with adaptability between cycles (e.g., swap between rolling and crawling when the terrain demands it).

Finally, conceptual advances are required to unify the three subfields of shape-changing, active and multistable metamaterials. We put forward the notion of non-reciprocal cycles: time-ordered cycles through configuration or shape space (**Figure 15.2**). In the context of active metamaterials, such cycles imply a sequence of shape changes that break time-reversal symmetry. These sequences are key to: (i) convert input energy into mechanical work for e.g. locomotion [4] or controlled gripping and release; (ii) perform advanced computation, store and process information in multistable metamaterials that sequentially hop between many stable states in sequence [8,9,11] (Fig. 15.1B), e.g. to achieve multiple shape-changes [3] yet using a single input [8,11] (Fig. 15.1A). Questions remain open: How do we embed such non-reciprocal cycles in a metamaterial? How do we efficiently couple these to large shape-changes to optimize their interaction with the environment? How do we embed multiple such cycles to achieve multiple functionalities (e.g., rolling, crawling and jumping)? How do we ensure that a target cycle is selected on the fly when the environment demands it?

**Concluding Remarks**

Animate matter should be able to perform purposeful actions in an autonomous and adaptable fashion. Mechanical metamaterials have been introduced over the past decade that display the individual functionalities of animacy: the ability to shape change, to do work and locomote and to perform *in-materia* information processing. Interweaving these functions via the notion of non-reciprocal cycles provides a unique opportunity to elevate metamaterials towards synthetic animate matter, to push forward the fields of rational and computational design, soft and active matter and alternative computing, and to reach impactful applications. These applications may include intelligent soft robots (see also Section 10), non-invasive medical devices, autonomous infrastructure capable of self-adaptation to environmental fluctuations (see also Section 20), and efficient distributed information processing systems (see also Section 18).

**Acknowledgements**

CC acknowledges funding from the Netherlands Organisation for Scientific Research under grant agreement Vidi 213-131-3.

# 16 − Mechanically intelligent devices: Mechanical intelligence in animate systems

Christopher J. Pierce, Tianyu Wang, Baxi Chong, Daniel I. Goldman
Georgia Institute of Technology

**Status**

Animate matter is not merely active, it responds dynamically, adaptively, and intelligently to its environment. Environmentally adaptive systems, be they biological or robotic, are often described in terms of active sensing and neuronal/electronic computation. However, passive physical processes that occur downstream of sensing, computation and locomotor commands can profoundly impact the dynamics of animate systems, functioning as an auxiliary control system. When these passive mechanical processes are tuned such that they spontaneously facilitate environmentally adaptive responses, we call this system *mechanically* [1] or *physically* intelligent (**Figure 16.1**) [2]. Roboticists have begun to identify situations where mechanical intelligence can supplement and simplify computational control to improve locomotion (see also Sections 10 and 14), and in biology the importance of passive mechanical processes in controlling adaptive responses has become increasingly clear. However, limited progress has been made in developing descriptions of mechanical intelligence that apply across different systems. Where computational intelligence is governed by the laws of logic and discrete mathematics, mechanical intelligence is governed by the dynamics of driven and damped mechanical systems. Hence, physics may provide a bridge between biological and engineered systems, by identifying fundamental principles that describe how mechanical and computational intelligence interact to produce behaviour across the diversity of animate systems.

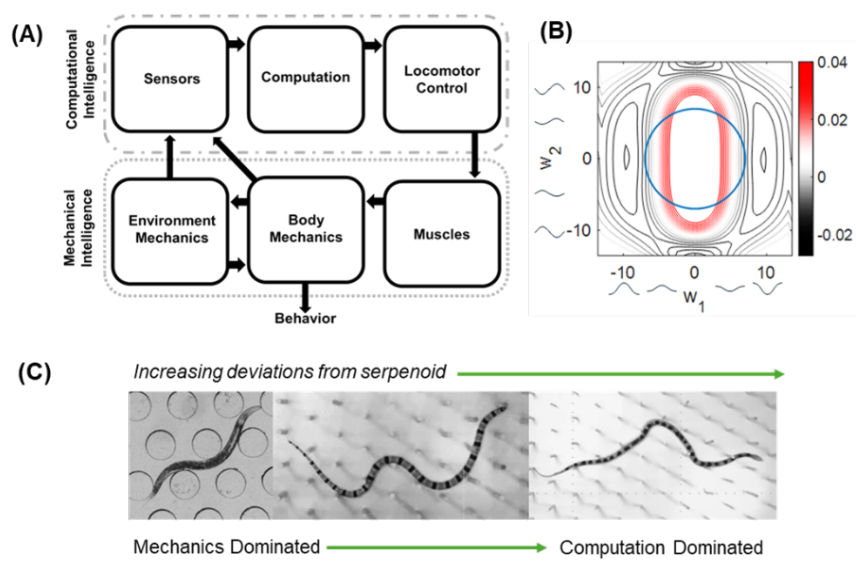

**Figure 16.1 | Defining mechanical intelligence**. (**A**) Schematic illustrating the interaction of computational and mechanical intelligence. Sensors take information from the environment in the form of mechanical, optical and other information. Algorithms perform computations based on those inputs to create commanded locomotor patterns, these then activate muscles or actuators, which filter the commands based on their dynamic properties. The forces generated by the muscles induce response forces in the passive material of the body and surrounding environment that feedback to modulate the final output behaviour. (**B**) A serpenoid curve (blue circle) in the shape space defined by the weights $\alpha_1$ and $\alpha_2$ which indicate the amplitude of the shape basis functions. The height functions can be used to calculate speed in the lab frame. Curves which enclose more of the positive parts of the height function (red region) move faster. (**C**) Images of *C. elegans*, *C. occipitalis*, and *P. guttatus* in lattices.

One successful approach to discerning principles of mechanical intelligence in locomotion is the use of "robophysical" models [3]. In this approach, robotic systems embodying drastically simplified models of organismal mechanics and 'neural' (electronic) feedback control are used to model the emergently simple, low-dimensional dynamics of organisms. This technique has its intellectual origins in the work of early cyberneticists, who first recognized that animals and machines could be described in a common language of feedback and control. Starting in the 1980s, robots as models were used to identify the role of mechanical control in legged locomotion, including hopping, bipedal and hexapodal legged locomotion, later in flapping flight and recently in limbless and myriapod locomotion. These models demonstrate how mechanical intelligence can simplify and augment active neural feedback control processes for diverse organismal morphologies. However, these approaches are often hard to generalize. Integrating robophysical models, which act as experimental tools, with theoretical tools taken from other branches of physics, offers an attractive path towards fundamental principles that describe how mechanical and physical intelligence interact to create behaviour across systems.

## Current and Future Challenges

Mechanically intelligent systems are complex, involving many degrees of freedom and multiple length scales. However, such systems typically display emergent simplicity. For example, longitudinal behaviour recordings of the nematode *C. elegans* have been described with data-driven dimensionality reduction techniques, such as PCA and other embedding schemes [4]. Subsequently, theoretical models of the transition probabilities between the low-dimensional behavioural states have invoked concepts like Markov models, and free energy landscapes – using ideas from non-equilibrium statistical physics and dynamics to rationalize behaviour in a generic way suggestive of principles [4]. Because these models focus on describing *kinematics,* they are fundamentally limited in their ability to discriminate between the computational and mechanical origins of animate behaviours.

In contrast, mechanically oriented models offer an alternate path to dimensionality reduction that may point to principles of mechanical intelligence. Many organisms, including *C. elegans,* move in a regime where inertia is unimportant (e.g., viscous fluids), and consequently, body-environment interactions are analytically tractable. This facilitates the connection of *geometric phase* (initially developed within particle physics) to locomotion [5], which *mechanically* rationalizes the low-dimensional kinematics of undulatory gaits. This approach (referred to as "geometric mechanics") has helped to explain the gaits of worms, lizards and snakes [6], all which converge to a common gait template called a 'serpenoid curve' [7] – a sinusoidal traveling wave of curvature, which, when mapped into a two-dimensional 'shape space' creates a circular orbit that can be used to calculate the centre-of-mass motion using the enclosed area of a 'height function' (Fig. 16.1B). This technique describes the role of mechanics in creating motion *outside* the animate agent (in the environment) and does not capture the role of mechanical feedback in producing the serpenoid gait in the first place. To complete the rich feedback system in Fig. 16.1A requires delving into the dynamic material properties of the body itself, and understanding how these properties control the mechanical coupling to the environment.

Internal mechanical control is especially important when encountering environmental heterogeneities, which produce large mechanical effects. *C. elegans* [8] and the sand-specialist snake *Chionactus occipitalis* [9] (Fig. 16.1C) display small deviations from serpenoid curves, thought to arise from mechanically controlled dynamic buckling. Alternatively, the corn snake *Pantherophis guttatus* displays large deviations from serpenoid behaviours suggestive of complex mechanosensory feedback [9] (Fig. 16.1C). Dissecting the relative roles of these modes of control in response to mechanical perturbations requires a new class of robophysical models and corresponding theory.

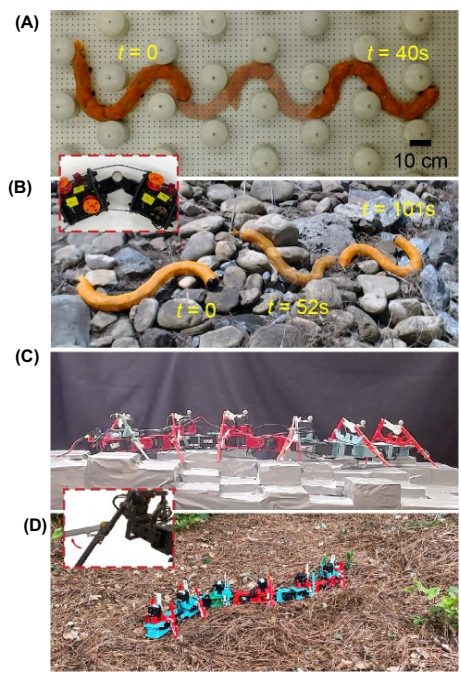

**Figure 16.2 | Examples of mechanically intelligent devices**. (**A**-**B**) A robophysical model of a bilaterally actuated undulating robot leverages the redundancy of its multi-joint architecture and the asymmetric compliance to confront (**A**) lab and (**B**) natural terrain through mechanical intelligence. From [9]. Reprinted with permission from AAAS. The inset shows a close-up on the multi-joint architecture. (**C**-**D**) A centipede robot similarly leverages both the redundancy of its legs and the asymmetric compliance its limb joints in (**C**) lab and (**D**) natural settings. From [10]. Reprinted with permission from AAAS. The inset shows a close-up on the limb joints. Reprinted with permission from [11], Copyright © 2020, IEEE.

### Advances in Science and Technology to Meet Challenges

To describe the role of mechanical feedback *within and outside of* the body, a recent robophysical model (**Figure 16.2A,B**) was developed to capture aspects of the bilateral muscle activation patterns of worms, snakes and other undulating organisms [8]. This model goes beyond previous work by revealing how mechanical feedback can *create* kinematics. An open-loop serpenoid gait (derived from the geometric phase approach) is commanded by the robophysical 'nervous system'. Downstream mechanical feedback mediated by collisions with environmental obstacles and the dynamic material properties of the body then conspire to modify the commanded gait kinematics to produce effective locomotion in highly complex terrain. Specifically, the dynamic, asymmetric compliance of the body, acts like a mechanical diode – would-be drag-inducing collisions spontaneously lead to body yielding and turning around the obstacle, while thrust-producing collisions, like hooking behaviours, result in body rigidity and forward movement. This example illustrates how external perturbations to optimal gaits derived from geometric mechanics can be mediated from mechanical feedback alone, demonstrating mechanical intelligence.

Beyond undulation, recent work on the limbed locomotion of centipedes has revealed another case where a combination of theory borrowed from fields outside biology, and robophysical models of mechanical control (**Figure 16.2C,D**) have shed light on mechanical intelligence. Centipedes exploit limb redundancy, as well as the asymmetric compliance of each leg to rectify perturbations to their commanded stepping gaits from obstacles in the environment. This effect was formally described using an analogy from Shannon's information theory [10], where discrete foot contact patterns were mapped onto binary bit sequences, and message transmission was analogized to the robot's self-transportation through the terrain.

Both cases illustrated above offer examples of how robophysical models of mechanical feedback, combined with theoretical models of computational control, can point to principles of mechanical intelligence that transcend morphology. In both cases asymmetric compliance (either of the limbs or the body) along with the redundancy (either of legs or body joints) helped to buffer the optimal, physics-derived gait commanded by the nervous system to unexpected environmental perturbations. Elevating these observations to the level of principles will require further experimental and theoretical development. On the experimental side, a proliferation of novel soft actuation schemes analogous to the ones above and testing in a greater variety of environmental regimes will produce a rich foundation of phenomenology which can be organized and explained with theory.

**Concluding Remarks**

Developing principles of mechanical intelligence will lead to major changes in the way engineered animate systems are designed and controlled. The emphasis on kinematics in the study of biological animate matter reflects a similar focus on body-shapes – and not body-environment physical dynamics – in the design of existing robotic systems. Focusing on connecting behaviour (defined as kinematics) to neurobiology (defined as neural activity) as the goal of neuro-ethology, while neglecting mechanical control, produces a fundamentally incomplete picture of animate behaviour in general. To develop engineered animate matter that can reckon with the full complexity of the external environment efficiently and robustly will require both mechanical and computational intelligence. However, the overwhelming paradigm of contemporary robotics is oriented towards computational solutions to often intrinsically mechanical problems. Shape control either relies upon the environment being simple or requires high bandwidth sensing and computation. This approach limits animate matter in two ways: 1) at microscopic length-scales, the relative cost of computation may increase rapidly; 2) for macroscopic matter, at the scale of large swarms (see also Section 17), the cost of computation may too become prohibitively significant. Principles of mechanical intelligence could lead to a paradigm shift in the control of animate matter to overcome these challenges.

**Acknowledgements**

# 17 – Swarm robotics

Andreagiovanni Reina

Centre for the Advanced Study of Collective Behaviour, Universität Konstanz, Germany

Department of Collective Behaviour, Max Planck Institute of Animal Behavior, Konstanz, Germany

Vito Trianni

Institute of Cognitive Sciences and Technologies, National Research Council, Rome, Italy

### Status

Swarm robotics stands as a beacon of innovation in the realm of robotics, embodying the principles of collective intelligence, resilience, and scalability [3]. This emerging field harnesses the power of collective behaviour exhibited by groups of simple robotic agents (e.g., see a swarm of miniature robots in **Figure 17.1**), which resemble an animate material composed of many interacting, autonomous units like cells of an organism. More precisely, robot swarms take inspiration from collective animal behaviour, where animal societies display coherent, cognitive behaviour as a group, such as ant colonies or flocks of birds. Self-organisation imbues swarm systems with the ability to exhibit emergent behaviours. Hence, at its core, swarm robotics revolves around the concept of decentralised control, where individual robots collaborate to achieve common objectives. Unlike traditional robotics, where centralised control prevails, swarm robotics embraces autonomy at the local level, enabling adaptability to dynamic environments and robustness against failures.

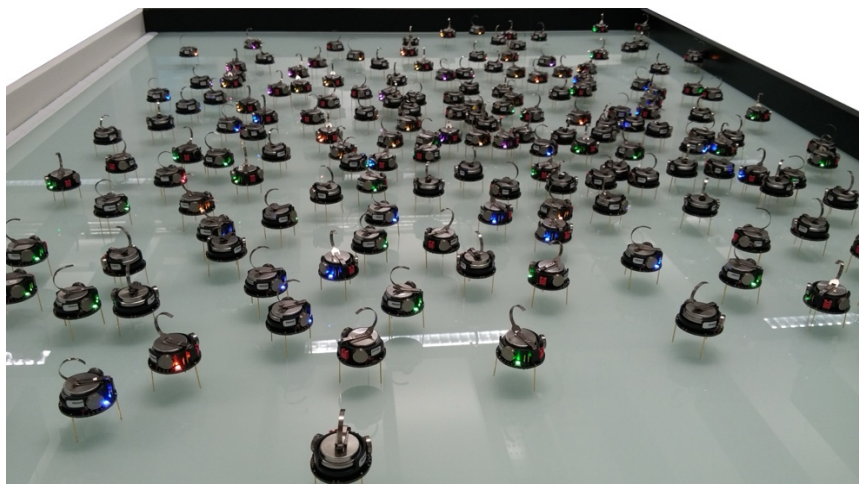

**Figure 17.1 | A swarm of Kilobots, a *de facto* standard platform for swarm robotics research**. This is a small robot of 3.3-cm diameter, designed to move on a flat surface and communicate efficiently with its immediate neighbours. Several design choices have been made to simplify the electromechanical components and experimental procedures like recharging and downloading control programs—which become cumbersome with large swarms—enabling experimentation with thousand robots lasting several hours [8].

Research has focused on two main problems. On the one hand, specialised robotic platforms have been developed to enable coordination through communication and physical interactions [8]. For instance, self-assembly of autonomous swarms in large functional structures constitutes a distinctive paradigm of swarm robotics [4]. On the other hand, engineering methodologies have been proposed to address the design of the individual control rules that can lead to a desired swarm behaviour. However, general-purpose solutions to this fundamental problem are still missing.

The versatility of swarm robotics promises exploitation in numerous application domains, spanning from disaster response and environmental monitoring to precision agriculture and space exploration [3]. For instance, in disaster scenarios such as search-and-rescue operations, swarm robots could efficiently explore hazardous environments, locate survivors, and deliver essential supplies. Similarly, in agricultural settings, they could collaborate to perform tasks like crop monitoring and pest control, enhancing agricultural productivity and sustainability. Moreover, swarm robotics could find applications in automated warehouses, infrastructure maintenance, surveillance, and many other fields, showcasing its potential to revolutionise diverse industries.

However, despite its promising applications, achieving the full potential of swarm robotics remains an ongoing challenge. There are still no commercial applications of swarm robotics. The main limitations to demonstrating robot swarms at scale are flexibility and adaptability to dynamic and uncertain environments, cost-effectiveness and security. Overcoming these limitations requires addressing various prospects and challenges in swarm robotics, including miniaturisation and control, heterogeneity, ensuring long-term autonomy, achieving scalability, and enhancing robustness.

**Current and Future Challenges**
**Miniaturisation** and **control** are critical aspects of swarm robotics, offering both opportunities and challenges [10]. Shrinking robotic components enables navigation of confined spaces and execution of precise tasks, e.g., in precision medicine. However, it also poses challenges like limited computational power, sensing, actuation and communication capabilities, which constrain the autonomy of the individual robots. Novel concepts in hardware, energy efficiency and control algorithms are necessary to advance the field. Integrating soft robotics principles introduces deformable robots capable of seamless interaction with the environment [9], also enhancing adaptability and cooperation within swarms. Indeed, soft materials introduce novel possibilities for physical interaction within the swarm that challenge current ways of designing cooperation and control. Future challenges should address robot interactions within soft-bodied aggregates where boundaries get blurred, making continuous structures adaptable to unpredictable situations, hence offering versatility in various applications (see also Section 10).

**Scaling up** the complexity of the tasks performed by robot swarms likely requires diversifying the functions and roles that robots can perform. Heterogeneity in swarm robotics, be it morphological or behavioural, enhances adaptability and task performance, as specialised robots can provide different functions to fulfil task requirements at hand [7]. Integrating robots with different hardware and behaviours can lead to efficient resource allocation and adaptive responses to environmental changes. Further challenges derive from scalability of swarm systems, managing task execution effectiveness as swarm size increases. This is even more compelling in the case of heterogeneous swarms. Scalability limitations are well known and sometimes unavoidable [5]. A current challenge stands in guaranteeing effective scalability at least within the range of swarm sizes requested by the problem at hand.

Enhancing **robustness** involves fault-tolerant mechanisms and resilience against adversarial conditions, ensuring mission success despite uncertainties and disruptions. This is intimately related to safety in the deployed system, ensuring that it can sustain internal failures or external attacks. Addressing technological challenges together with ethical considerations is crucial for building trust and acceptance of swarm technologies. This also intersects with the need to achieve long-term autonomy, so that functions can extend over long timeframes without the need for human intervention. This would unlock several application scenarios in remote environments, such as underwater or in space, where the possibility to rely on cooperation and collective intelligence would be a game changer.

**Advances in Science and Technology to Meet Challenges**

There are two main paths that need to be followed to meet the above challenges, one that is more science-oriented, and one that is rather technology-oriented. Clearly, these two paths should not develop in parallel but intersect and cross-fertilise.

The science-oriented path leads to the exploration of novel challenging directions for understanding swarm systems and enabling miniaturisation and control. First of all, we need to produce better theoretical models that include spatial and topological dynamics, which can be used to explain collective behaviour and support design, especially tailored to predict and control heterogeneous swarm systems. Further modelling efforts should also guide the development of novel mechanisms able to exploit internal and external perturbations—e.g., malfunctioning or noise—as an asset rather than a hindrance, for example, to allow a collective system poised at a critical state to transit between different states, or to enable exploration and adaptation to environmental changes by exploiting stochastic fluctuations. Bio-inspiration has been one of the main approaches to designing swarm systems but should be extended beyond the development of control systems to embrace also new materials and hardware concepts. Physical computing reduces requirements from control, especially if soft physical interactions are considered. Bio-inspired morphologies can support collective behaviour, making robots more similar to living organisms. Pushing this idea to the extreme, organic swarm robots can be considered, obtained through genetic engineering and controlled growth [1].

The technology-oriented path leads to the implementation of swarm robotics principles for real-world applications. Recent research has shown that the integration of blockchain technology is a promising solution to address several of the outstanding critical problems for real-world deployment including security, resilience, accountability, and data integrity [2]. Human swarm interaction needs to be considered, as swarm robots do not operate in isolation but must blend within our societies and enable efficient user control, with guarantees of no harm and unintended emergent effects [6]. Finally, real-world applications should not be limited to battery lifetime. While material science contributes with novel designs for solar cells and ultra-low-power sensors, swarm robotics research needs to focus on energy-awareness, harmonising swarm functionalities with sustainable usage of resources. Hence, minimalist approaches that achieve coordination and control with little or no computation should be considered, which also fit well with miniaturised hardware [11]. Orchestration of swarm activities and heterogeneity can enable managing continuous operations as well as collective energy management.

**Concluding Remarks**

Swarm robotics promises to effectively produce a new form of animate material, a kind of artificial super-organisms that displays adaptive and resilient behaviour in the face of dynamic task requirements. While such large-scale robotics systems can have a disruptive impact on our society and industries, current research has shown that controlling a large number of autonomous robots—which influence non-linearly each other's behaviour—is a particularly challenging problem. We envision substantial progress can be achieved thanks to an interdisciplinary scientific effort, as the deployment of swarm robotics requires addressing challenges in several fields other than robotics, including material science, microelectronics, computer science, biology, and nonlinear dynamics. Once swarm robotics starts being employed in real-world applications, addressing ethical, political and economic concerns also becomes critical.

**Acknowledgements**

A.R. acknowledges financial support from the Deutsche Forschungsgemeinschaft (DFG, German Research Foundation) under Germany's Excellence Strategy – EXC 2117 – 422037984. VT acknowledges support from the projects FAIR (Future Artificial Intelligence Research, PNRR MUR Cod. PE0000013 - CUP: E63C22001940 006) and BABOTS (Horizon Europe, PathFinder Open EIC Work Programme, GA N. 101098722).

# 18 − Machine Intelligence: Integrating Artificial Intelligence and Animate Matter

Giovanni Volpe

## Department of Physics University of Gothenburg, Sweden

**Status**

The integration of Artificial Intelligence (AI) into the development of bio-inspired synthetic systems heralds a new era with materials and agents that are self-sufficient and capable of complex behaviours mirroring the complexities of living systems. Historically, the field of AI has evolved from rudimentary algorithms simulating basic cognitive functions to sophisticated neural networks capable of learning and making decisions with minimal human oversight [1]. **Figure 18.1** presents an overview of current AI techniques. This progression has paralleled advancements in soft and active matter physics and chemistry, where the focus has increasingly shifted towards systems that can dynamically respond to their environment [2].

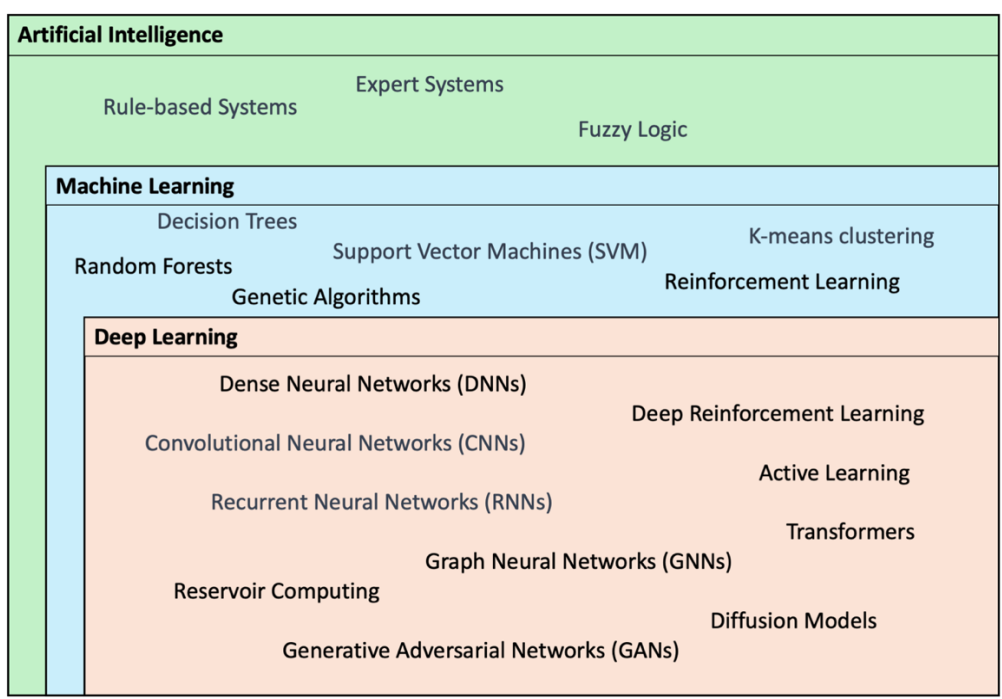

**Figure 18.1 | Overview of AI techniques**. Hierarchy within artificial intelligence (AI), breaking down AI into its core components: Machine Learning (ML) as a subset of AI that focuses on the ability of machines to learn from data, and Deep Learning (DL) as a further specialization of ML employing neural networks with multiple layers. Key techniques under each category are highlighted, showcasing the progression from general AI principles to specific computational models and algorithms in ML and DL.

By utilizing AI, researchers can endow synthetic materials and agents with the ability to adapt, evolve, and perform tasks autonomously. Such capabilities are inspired by living systems, where active matter constantly adjusts to stimuli through an interplay of mechanical, chemical, and biological processes. Current AI techniques facilitate this by enabling the design and control of agents and materials at an unprecedented level of precision and functionality. Machine learning algorithms, for instance, can predict these system's behaviours under various conditions, optimize their performance, and even discover new agents and materials with desired properties [2].

The importance of AI combined with animate matter extends beyond the laboratory into practical applications promising revolutionary advances. In healthcare, the integration of AI with bio-inspired systems could enhance the development of responsive and adaptive drug delivery systems, for example, by using geometric deep learning to identify spatiotemporal fingerprints of microscopic motion of drug carriers engineering their diffusion properties [3]. Similarly, there is a huge potential to advance diagnostics and patient care through improved imaging techniques relying on deep learning to improve robustness against noise [4, 5] or to decrease the need for training data [6]. In environmental sustainability, AI has been shown to contribute to the understanding of microplankton life histories [7], which could inform the design of materials for cleaning pollutants or converting waste into energy. Finally, AI can be used to create biohybrid microrobots using biological cells to create microscopic active agents [8] (see also Section 11).

The ongoing advancements in AI promise to further bridge the gap between synthetic materials and living systems. As AI algorithms become more sophisticated, their integration with bio-inspired materials will likely yield systems with levels of autonomy and adaptability that closely mimic, or even surpass, those of natural organisms. The potential gains from further advances in this interdisciplinary field include more efficient energy use, enhanced resilience and durability of materials, and the creation of entirely new functionalities that today remain unimaginable.

### Current and Future Challenges
The ambitious vision of creating bio-inspired synthetic systems and materials that are active, adaptive, and autonomous through the integration of AI presents significant challenges. These challenges span technical, ethical, and scalability issues:

- **Complexity of integration**: One of the foremost challenges lies in the complexity of integrating AI with the inherently unpredictable and dynamic nature of bio-inspired systems. The stochastic behaviours observed in living systems, influenced by countless variables, present a significant hurdle for AI models that thrive on predictability and quantifiable data. Developing algorithms capable of accommodating, predicting, and effectively responding to the dynamics of these systems remains a critical task—especially algorithms that can be trained with small amounts of data.
- **Data limitations and model accuracy**: The scarcity of high-quality, comprehensive datasets that accurately represent the complexities of bio-inspired materials limits the ability of AI to learn and make reliable predictions. Furthermore, the models themselves must balance accuracy and generalizability, ensuring that they are robust enough to handle real-world variability without being overly specialized to specific datasets.
- **Ethical and societal implications**: As these materials gain autonomy, ethical considerations come to the forefront. Questions regarding the decision-making processes of AI-driven systems, their impact on privacy, safety, and the environment, and the potential for unforeseen consequences of autonomous action pose significant challenges. Addressing these concerns requires a multidisciplinary approach, involving ethicists, policymakers, and the public in the development process [9].
- **Scalability and deployment**: Transitioning from laboratory prototypes to widely deployed, scalable solutions is another challenge. Issues related to manufacturing, energy consumption, longevity, and maintenance of AI-integrated bio-inspired systems must be addressed. Furthermore, ensuring these materials can be produced and operate sustainably and economically on a large scale is essential for their widespread adoption.
- **Interdisciplinary Collaboration:** The cross-disciplinary nature of this field demands unprecedented levels of collaboration between material scientists, biologists, AI researchers, and engineers. Bridging the knowledge gaps and fostering effective communication among these diverse groups are crucial for tackling the complex problems at the intersection of AI and bio-inspired materials.

Looking ahead, overcoming these challenges will require technological advancements and a concerted effort to address ethical considerations and foster interdisciplinary collaboration. The future success of integrating AI with bio-inspired systems will hinge on our ability to navigate these multifaceted challenges, pushing the boundaries of what is possible in materials science and artificial intelligence.

### Advances in Science and Technology to Meet Challenges

Addressing the challenges of integrating Artificial Intelligence (AI) with bio-inspired synthetic systems requires significant scientific and technological advancements, matched with the challenges described above:

- **Enhanced machine learning models**: Progress in machine learning algorithms is vital to handle the unpredictable nature of bio-inspired systems, while exploiting the intrinsic symmetries and properties of these systems to enhance the model performance also with minimal amounts of training data. Physics-inspired deep learning, deep reinforcement learning, and unsupervised learning offer promising pathways to develop models that can learn complex patterns, adapt to new environments, and make decisions in real-time. Advances in these areas will enable more accurate predictions and control of material behaviours, leading to systems that can autonomously adapt to their surroundings.

- **Synthetic data generation and simulation technologies**: To circumvent the limitations of scarce or incomplete datasets, advances in synthetic data generation and simulation technologies are essential—as already successfully done in the field of microscopy [10-12]. These technologies can create high-fidelity, diverse datasets that simulate the vast range of conditions that bio-inspired materials and agents might encounter. This approach improves the training of AI models, while facilitating the testing of materials and agents under virtual conditions that are difficult or impossible to replicate in the lab.

- **Ethical AI frameworks:** Developing frameworks for ethical AI is critical to ensure that autonomous systems operate within predefined ethical boundaries, for example, in biomedical and ecological applications. These frameworks should include guidelines for transparency, accountability, and privacy protection, incorporating feedback from diverse stakeholders. Furthermore, research into explainable AI will enhance our understanding of AI decision-making processes, making them more interpretable to humans.

- **Scalable manufacturing and integration techniques:** Technological advances in scalable manufacturing processes and integration techniques are required to transition bio-inspired materials from prototype to production. Innovations in 3D printing, bioprinting, and nanofabrication will facilitate the creation of complex, multi-functional materials. Additionally, the development of energy-efficient, durable systems that can self-repair or degrade safely after their lifecycle is crucial for sustainable deployment.

- **Interdisciplinary collaboration platforms:** Finally, the establishment of platforms and frameworks that promote interdisciplinary collaboration is essential. These platforms should facilitate knowledge sharing and joint research efforts, leveraging cutting-edge tools in data sharing, computational resources, and collaborative software. By breaking down the barriers between disciplines, these platforms will accelerate the pace of innovation and the application of AI in developing adaptive, autonomous materials. An effective way to enhance these interdisciplinary collaborations is through open challenges, such as the Anomalous Diffusion Challenge to characterize the motion of microscopic particles [13, 14].

These advancements are the keystones for realizing the full potential of autonomous, adaptive materials and agents, paving the way for ground-breaking applications across various domains in materials science and beyond.

**Concluding Remarks**

The integration of AI with bio-inspired synthetic systems represents a frontier in materials science that promises to redefine our interaction with technology. By drawing inspiration from the adaptability, autonomy, and complexity of living systems, and harnessing the analytical and predictive power of AI, we stand on the cusp of creating materials and devices that can dynamically respond to environmental stimuli, self-organize, and even mimic biological processes. The challenges in this interdisciplinary endeavour are substantial, spanning technical, ethical, and scalability concerns. However, the ongoing advancements in AI and materials science, coupled with a growing emphasis on interdisciplinary collaboration, provide a solid foundation for overcoming these obstacles. As we navigate these challenges, we must recognize the ethical implications of autonomous systems and strive for sustainable, beneficial applications. The journey towards fully realizing the potential of AI-driven animate matter is complex and fraught with unknowns, but it is a journey that holds unparalleled promise for the future of technology, robotics (see also Sections 16 and 17), built environment (see also Section 20), medicine, and environmental sustainability.

**Acknowledgements**

Funding from the Horizon Europe ERC Consolidator Grant MAPEI (Grant No. 101001267) and the Knut and Alice Wallenberg Foundation (Grant No. 2019.0079).

## 19 – Probiotic Materials

Richard Beckett, Sean P. Nair
UCL

### Status

Probiotic materials are a class of biologically active materials for the built environment containing benign microbes that are harmless to humans, but that are beneficial for health through their ability to shape the indoor microbiome and the microbiomes of building occupants towards a healthy state [1].

Recognition of the importance for limiting harmful microbial transmissions in buildings in relation to human health is long standing, and over the last 100 years, material strategies in buildings have broadly operated under the assumption that healthy buildings should contain fewer microbes. Materials in buildings serve as a significant source of microbial transmission indoors where they act as 'fomites', defined as surfaces or other inanimate objects onto which a microorganisms can deposit and from which they can be transferred to a host. Early material strategies of whitewashing and glazing were partly concerned with limiting the presence of germs, but since the development of antibiotics in the mid-20th Century, the use of antimicrobial materials and antibiotic chemicals has become commonplace across a range of building typologies. Their use rose drastically during the COVID-19 pandemic and is predicted to continue to increase in the coming years.

While there is a clear need to target sterile surfaces in certain buildings such as hospitals, antimicrobial materials and chemicals have various limitations that are problematic for human health. As well as being harmful for human exposure, chemical disinfectants are time-dependant and are subsequently ineffective against recolonisation [2] — a key factor in building acquired infection. More worryingly, their overuse appears to be creating conditions of stress which are actively selecting for antimicrobial resistance in buildings [3]. As well as removing harmful microbes, they also remove other benign microbiodiversity which act as a kind of shield against pathogens through principles of microbial competition. In line with the emerging understanding of microbiome health, this lack of microbial diversity in buildings is also being associated with the observed rise in immunoregulatory illnesses described as diseases of missing microbes [4].

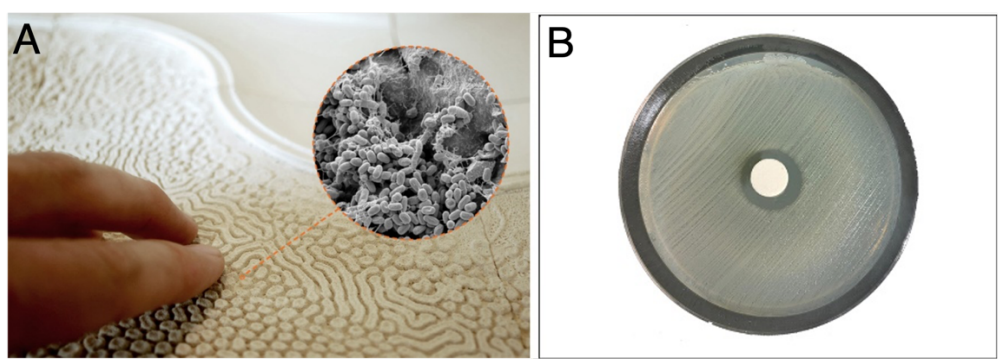

**Figure 19.1 | Examples of probiotic surface tiles**. (**A**) Photograph of probiotic surface tiles embedded with *B. subtilis*; inset with SEM image of bacterial cells within the material matrix. (**B**) Growth inhibition of Methicillin-resistant *Staphylococcus aureus* (MRSA) by a probiotic disc. The creamy yellow/white background on the plate is MRSA and the zone of clearing around the disc shows inhibition of MRSA.

These positions suggest that healthy materials could be ones that support and maintain a degree of benign microbial diversity. This can serve to outcompete pathogens and potentially support healthy immune function through their impact on the human microbiome. This animacy of competitive exclusion, and the need to implement strategies to restore microbial diversity in certain built environments forms the basis for Probiotic Materials. These approaches align with the emergence of other 'probiotic' strategies for the built environment including cleaning products [5] and efforts to create microbiome inspired landscape designs [6], but with a specific focus on architectural application (see also Section 20). Work by the authors to date has developed methodologies for the creation of probiotic ceramics, concretes and plastics, and is exploring their use towards the engineering of surface tiles (**Figure 19.1**), furniture, or even entire structures (**Figure 19.2**) out of microbially-active materials [7].

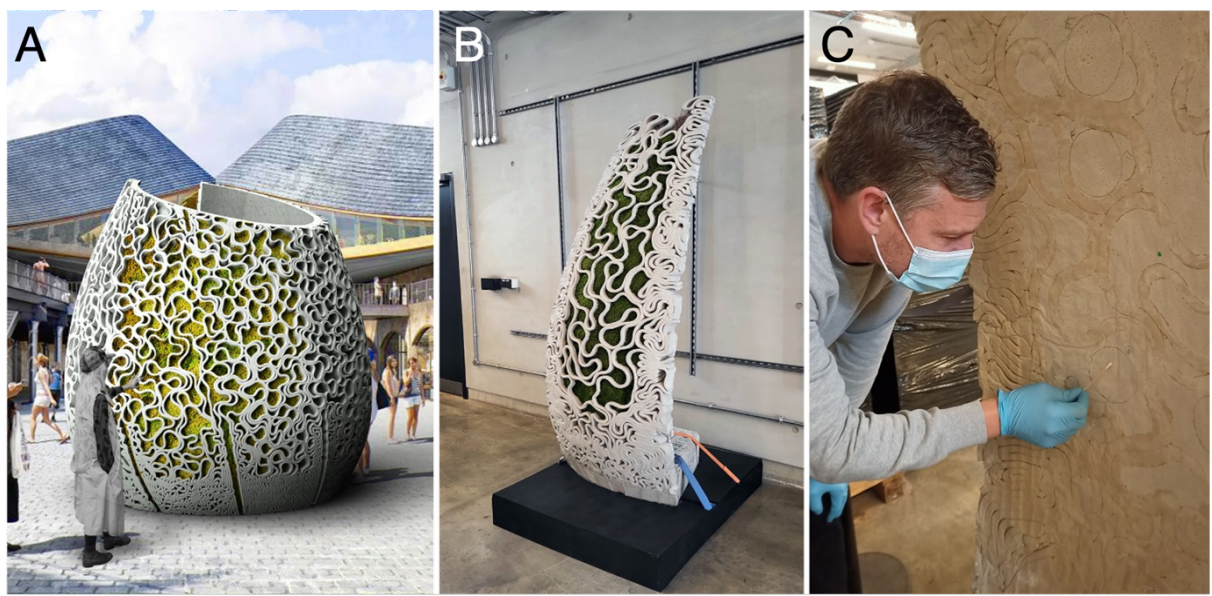

**Figure 19.2 | Probiotic building protypes**. (**A**) Rendering of an imagined probiotic pavilion structure. (**B**) 1:1 scale 3D printed, microbially active building prototype, embedded with bryophytes (outer facing) and soil microbes (inner facing). (**C**) Photograph of prototype undergoing a microbiome monitoring study using 16S DNA sequencing.

**Current and Future Challenges**

Probiotic materials have emerged to include the microbial scale as a design parameter in shaping materials for buildings that are sustainable, healthy and resilient. Such approaches involve the intentional integration of living organisms, typically microorganisms or microbial communities, into the design and function of materials, that utilise the beneficial biological agencies of the respective organisms (see also Sections 8, 12 and 20). Within this hybrid condition, more research is needed to understand how the physical and chemical properties of the material substrate can augment the beneficial biological mechanisms of benign microbes. Important properties might facilitate adhesion of microbes to the material surface, modulate biochemical pathways and ensure their ability to survive and remain viable indoors over time. Building materials have been described as microbial wastelands, due to the typically dry and nutrient poor nature of contemporary indoor environments [8]. Materials might provide nutrient sources for specific strains or facilitate spore formation to offer long term survival with limited maintenance. Spore of *Bacillus* spps have been directly integrated into hydrogel materials [9], yet such materials are unlikely to be able to survive the rigours of daily building life. The use of harder materials for buildings (ceramics, concretes, hard plastics, etc.) is preferable. Methods of manufacture will be important to reliably produce material properties such as porosity, pH and hydrophobicity so they can serve as a resilient source of beneficial microbes, but also perform as part of the building fabric.

Probiotic materials can be fundamental in shaping what we envisage as healthy beneficial human-microbe interactions in buildings. In line with this, we highlight two key outstanding challenges:

- **Probiotic materials to limit AMR**: Biologically active materials and surfaces could be applied in buildings to limit the emergence and spread of AMR in the built environment. Researchers recently developed novel porous ceramic materials, inoculated with strains of the *Bacillus subtilis* bacterium, a Gram-positive soil microbe which exhibits known antimicrobial mechanisms. The study demonstrated the efficacy of these probiotic material to inhibit the presence and persistence of the AMR superbug Methicillin-resistant *Staphylococcus aureus* (MRSA) and were shown to remain viable over time under normal building conditions without the need for nutrient or water supply [1]. These hybrid materials can reduce adverse selection pressures and offer potential for application across a range of building types including hospitals, homes and other public buildings. New networks of materials scientists, architects and microbiologists will be required to make this happen.

- **Probiotic materials to promote immunoregulation**: The hypothesis that probiotic materials could modulate or inform the microbiome of the built environment (MoBE) towards a condition that is directly beneficial for the human microbiome is an attractive yet currently understudied. Probiotic materials can serve as a source of a diverse range of immunoregulatory relevant microbiota for both the building and the human microbiome. Such approaches have been explored for biodiverse outdoor landscaping materials that aim to enhance the immunoregulation of children in urban daycare centres [10]. Similar material strategies for the indoor condition will require new collaborations between immunologists, material scientists, architects and engineers. These might inform new planning policies to promote microbial diversity in buildings in dense urban areas where the surrounding environmental microbiome is likely to be low.

**Advances in Science and Technology to Meet Challenges**
A series of key advances are needed moving forward:

- **Probiotic Microbes**: Probiotic microbial strains can vary widely in their biocontrol potential, with dramatic shifts in activity against pathogens with different materials surface chemistries and under different environmental conditions relevant to buildings. To better understand the underlying mechanistic interactions between probiotic microbes and pathogens that influence competition outcomes, the development of Metabolic and Gene Expression Models will be required. This will require longitudinal competition studies using multi-omics methodologies to generate gene expression and metabolite data. Advances in this knowledge may then permit the possibility to genetically modify specific probiotic strains, to engineer traits to improve their persistence and competitive activity on specific materials.

- **Microbiome**: To engineer probiotic materials that can provide human health benefits through their impact on the human microbiome, scientific advances are needed to better understand exactly what constitutes a healthy microbe, a healthy human microbiome or indeed what a healthy indoor microbiome might constitute. This will require real-world trials to identify and determine the effectiveness of probiotic materials in buildings that would need to take into account many of the complexities relating to different populations, geographical regions and building uses etc.

- **Understanding data**: The field of the MoBE has been informed through the sciences and technologies of metagenomics, and to date, its characterisation has been led through the field of the building sciences. While further advancements relating to the sciences of the microbiome are required to address the challenges described above, the clear role that architectural design plays in shaping the indoor microbiome, highlights the need for interdisciplinary knowledge and workflows that can span the fields of science and design. Key to this is the development of computational technologies and software's that will enable architects and designers to understand and engage with the complex data sets that emerge from metagenomic studies and use them as a tool to inform, test and iterate design strategies.

- **Broader challenges**: Finally, the further development of probiotic materials also faces broader cultural and societal challenges related to people's perception of such materials. Their use in hospitals may be particularly controversial and would require significant shifts in existing cleaning practices to avoid inhibiting probiotic agency. There is also likely to be challenges in relation to regulatory oversight for how to govern probiotic microorganisms in building materials. Their long-term use is likely to result in other unknowns or knock-on effects which will need to be monitored.

## Concluding Remarks

Probiotic materials offer a new line of investigation for designing healthy buildings which can radically transform how we engineer building materials, construct buildings and maintain our built environments in relation to microbes (see also Section 20). Although the field is relatively new and studies in practice are limited, the potential for applying probiotic materials in hospitals, and other buildings such as homes, schools and workplaces in which people spend most of their day is fascinating, necessary and should be investigated further.

## Acknowledgements

Authors gratefully acknowledge funding support from the AHRC (AH/R001987/1) and EPSRC (EP/X026892/1) and The Bartlett School of Architecture, UCL.

## 20 – Towards living architectures

Rachel Armstrong

KU Leuven

### Status

Living architecture operates at the intersection of life and technology to offer a vision for next-generation sustainable buildings by applying 'living technology' to the practice of the built environment (see also Section 19). Approaches are diverse range and integrate living organisms, such as plants, microorganisms, and fungi, into built environments to enhance sustainability, improve air quality, and promote ecological resilience. Spanning all scales from construction materials, to building services, and urban systems, its animate character confers our homes and cities with life-like properties. These demand ethical consideration, inviting our care and consideration of how they generate life-promoting impacts.

The pioneering project "Future Venice" held significant implications for the field of living architecture. Based on laboratory experiments, it proposed the growth of an artificial reef beneath Venice through a chemically mediated biomineralization process, catalysed by dynamic droplets [1]. Designed to protect the wooden piles of the city from rotting, it demonstrated that human-designed dynamic material transformations could positively impact the overall ecosystem in innovative ways to address the environmental challenges associated with construction, maintenance, and the end-of-life phases of buildings. Although the 'living technology' was at the earliest stage of commercial readiness, it pioneered the concept that an animate platform could uniquely counter the negative environmental impacts of the building industry that globally accounts for 40% of total carbon emissions. **Figure 20.1** shows a computer rendering of the concept.

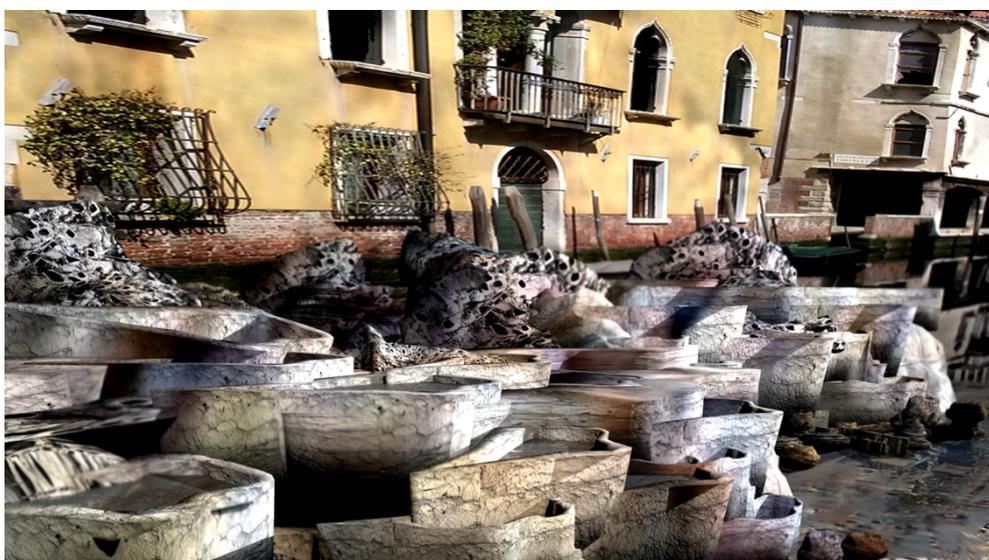

**Figure 20.1 | Future Venice.** Rendering by Christian Kerrigan, concept by Rachel Armstrong, 2009. Conceptual illustration: Here, buildings along the Venetian waterways have been fortified by applications of an animate, protocell-mediated mineralization process that also potentiates biomineralization in the native lagoon ecosystem. Collectively, the organisms and protocell technology form a protective reef-like structure around foundations, reinforcing damaged brickwork, and sealing woodpiles to prevent rotting, providing a sustainable solution for Venice's challenges with rising waters.

### Current and Future Challenges

Today living architecture uses a range of living technologies to make buildings much more sustainable by addressing two main energy-based challenges: firstly embodied energy, which is used to mine and make the materials for our buildings, and secondly the energy for building operations to run as well as maintain and manage our buildings. These two are responsible for the high carbon emissions of the construction industry.

Rather than extracting materials from the ground, living architecture considers microorganisms as a technical platform that can grow sustainable materials. Fungal roots [2] can penetrate and bind organic wastes together to make a single, solid material with excellent thermal and acoustic insulating properties called mycelium biocomposites [3]. Several Small and Medium Enterprises (SMEs), such as Ecovative, MOGU and Grown.bio, now provide these materials for packaging, home insulation, soundproofing, and as attractive interior panels for sustainable interior design.

Microbes like *Sporosarcina pasteurii* bacteria [4] can also be cultured in biomass, aggregate, nutrients, and minerals like sand, to generate a sandstone analogue material through a biomineralizing process. This does not require firing and is expected to make a significant reduction in embodied carbon emissions in the brick making process. Microbial products are also the active ingredient for bioconcrete that lasts longer than ordinary concrete as it biomineralizes to produce carbonate crystals that self-heal microfractures in the concrete in the presence of moisture [5].

While inventing new materials that can reduce embodied energy is exciting, the even bigger challenge is in reducing emissions from building operations. Microbial fuel cells (MFCs) are living technologies in the form of a specialized kind of anaerobic bioreactor that provide electrical energy without combustion. By forming an anaerobic biofilm in the apparatus, the microbes metabolize organic substrates in the anode, converting this chemical energy through an external electron transfer process into action potentials that are then harvested by electrodes [6]. For over a decade, MFC-based PeePower® urinals have been used in refugee camps and schools to turn urine into electricity, cleaned water, sanitation, biomass, and provide basic lighting that keeps people safe at night. The same technology has also been used at the pop-up Glastonbury festival where revellers can exchange their urine, e.g., for charging their mobile phones, playing computer games, and powering some of the festival screens.

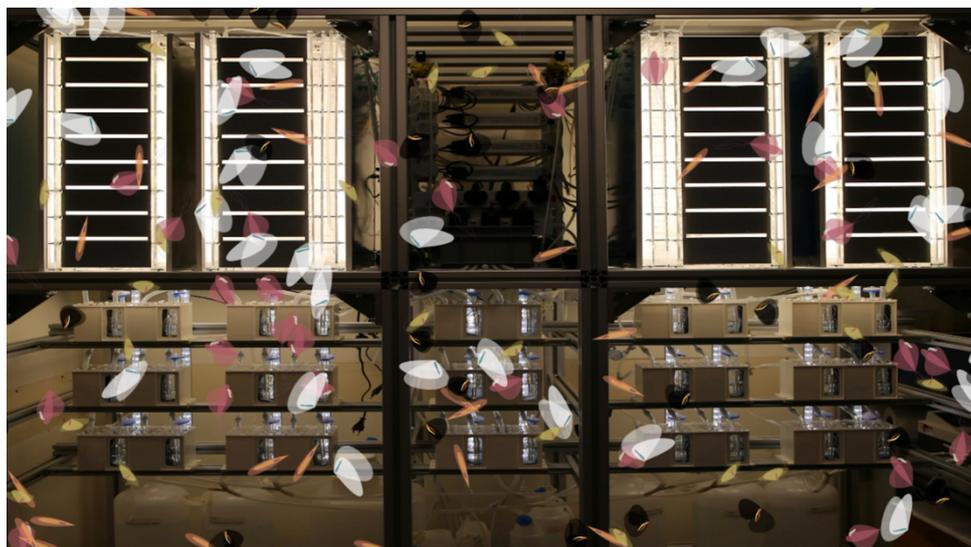

**Figure 20.2 | Living Architecture with Augmented Reality microbial animations**. Courtesy of the Active Living Infrastructure: Controlled Environment (ALICE) demonstrator, 2020. Here, the Living Architecture series of bioreactors: photobioreactor with algae, microbial fuel cells with anaerobic biofilms and a synthetic bioreactor with modified organisms are generating a circular system of metabolic exchange that transforms household grey water into valorised substances that can be re-used within the home: bioelectricity, cleaned water, nutrient-rich biomass, and high-value biomolecules. The bioelectricity generated by the platform directly correlates with the metabolisms of the biofilm activity, giving an overall reading of the health and performance of the platform, where action spikes can be considered as 'data' that is then transformed by software

to generate readable animations. The augmented reality image poetically shows the animate nature of the system generated by microbes via its real time metabolic activity.

Bioreactors (MFCs, synthetic bioreactors, algae photobioreactors) and hydroponics, can even be sequenced to turn household liquid wastes into electrical energy and generate biomolecules while bioremediating the effluents. This concept was demonstrated by the EIC Pathfinder Challenges Microbial Hydroponics (Mi-Hy) project, which builds on the H2020 FET Open Living Architecture project to incorporate hydroponics and biosynthesis for the first time (2024-2027). The maximum power output for this system of 15 MFCs has reached up to 6.5 mW (32.5 W m$^{-3}$) [7], which makes it applicable for low-powered electronic applications in households. Importantly, software can transform the electrical signals of microbial data from the electrode into characterful animations to render microbes more relatable (**Figure 20.2**) Just like a pet, the apparatus can be fed, or gently warmed with LEDs, encouraging the homeowner to keep the electrons flowing via this empathetic connection and establishing the foundations for a bioremediating Green Digital Revolution.

**Advances in Science and Technology to Meet Challenges**

The evolution of living architecture raises new challenges for the future with continued scientific and technological advancements needed in 'gardening' living microbes. The field of biofilm engineering is of particular interest, where new insights will enable increased complexity and scale of microbial metabolism.

Such approaches are being explored in the development of bioreceptive surfaces [8], which are patterned cladding materials that can also be thought of as bioreactors without walls that distribute water across the surface. This enables complex interactions between various living organisms maintaining a balanced ecosystem that benefits humans and urban nature. Within the home environment and perhaps in hospital wards, probiotic panels [9] are being prototyped that are colonized by 'friendly' *Bacillus subtilis* bacteria that ward off antibiotic resistant strains (see also Section 19). These developments suggest the possibility of gardening the whole microbiome of the built environment to ensure we are healthy and safe at the microbial scale. Communities of organisms can also be combined to create new kinds of materials, whose performance exceeds the sum of the ingredients. For example, BioMason (which makes microbially produced sandstone bricks) is working with Evocative (which is known for its mycelium biocomposites) to grow a new kind of stone from agricultural waste, mushrooms, bacteria, and sand. Living architecture composed of heterogenous combinations of living materials will enable new functionalities of buildings, for instance to dynamically respond to environmental cues such as by healing cracks, metabolising dangerous toxins from the air, and even glowing on command [10].

Unconventional computing methods will also integrate machine intelligence and microbial intelligence to optimise spatial practices by performing complex calculations relating to real-world problems (see also Section 18). Microbes like slime moulds and mycelium networks have been widely used as unconventional computers [11]. These operate via components that are in different subpopulations of the community to find the shortest routes around cities or proposing a network of blue-green path systems. Presently, these modes of unconventional computing are generally understood via single overarching parameters such as pattern production. But advances in biofilm engineering, metabolic engineering and machine learning may converge via neural networks to solve the complex urban challenges in sustainable planning, construction, and design.

Research into the computational abilities of animate materials have also demonstrated the potential for harnessing biological processes and developing emergent physical behaviours to solve complex challenges in materials science, such as healing microcracks in concrete by activating biomineralizing bacterial spores. As our understanding of these systems deepens, we can expect further innovation in fields ranging from smart construction materials that can respond to environmental changes to cyber-physical systems (CPS) that are physical systems interconnected with computing and communication technologies, enabling them to monitor, coordinate, control, and integrate their operations in real-time. CPS can facilitate the integration of living organisms or biological processes with engineered systems, creating hybrid systems with enhanced functionalities. For example, CPS could enable the monitoring and control of biological processes in real-time, allowing for the optimization of bioreactors or the regulation of biological functions in biohybrid robots. By incorporating sensors, actuators, and communication interfaces, CPS can enable the dynamic interaction between animate matter and its environment, opening new possibilities for applications in healthcare, agriculture, environmental monitoring, and addressing grand challenges across various domains.

**Concluding Remarks**

Living architecture's visionary approach paves the way for next-generation sustainable solutions to reduce and even reverse the construction sector's current huge environmental impact. While it faces various challenges, such as scalability and regulation, pathways to overcoming these hurdles are promised through rapidly developing advances in 'living technology'. There is still some way to go before all these solutions can be integrated in a single 'living' house or building within the next 10 to 20 years. Around 2035, we may start to see more 'living' features incorporated into new construction projects, such as green roofs, bioreactors for wastewater treatment, and bio-based building materials. By 2040 or beyond, the vision of coming home to a fully integrated 'living' residence equipped with digital interfaces for personalized comfort and sustainability could become a reality for many households. Advances in artificial intelligence, sensor technology, and smart building systems will play a crucial role in realizing this vision, enhancing the functionality and efficiency of 'living' spaces. By 2070, we will come home to a 'living' residence of our own that will greet us through digital signals, to make sure that our living space is *just the way we like it*.

**Acknowledgements**

***Living Architecture*** is Funded by the EU Horizon 2020 Future Emerging Technologies Open programme (2016-2019) Grant Agreement 686585 a consortium of 6 collaborating institutions—Newcastle University, University of Trento, University of the West of England, Spanish National Research Council, Explora Biotech and Liquifer Systems Group.

***The Active Living Infrastructure: Controlled Environment (ALICE)*** project is funded by an EU Innovation Award for the development of a bio-digital 'brick' prototype, a collaboration between Newcastle University, Translating Nature, and the University of the West of England (2019-2021) under EU Grant Agreement no. 851246.

***Microbial Hydroponics: Circular Sustainable Electrobiosynthesis (Mi-Hy)*** is Funded by the European Union under Grant Agreement number 101114746, which is a collaboration between *Beneficiaries,* KU Leuven (Belgium), the University of Southampton (UK), SONY Computer Science Laboratory (France), BioFaction KG (Austria), Spanish National Research Council (Spain), and *Associated Partners*, the University of the West of England (UK) and University of Southampton (UK). Mi-Hy is also supported *through the interdisciplinary KU Leuven Institute for Cultural Heritage (HERKUL).*